\documentclass[a4paper,11pt]{article}
\usepackage{jheppub} 
\pdfoutput=1
\usepackage{amsmath,amssymb,amsfonts}
\usepackage{graphicx}
\usepackage[normalem]{ulem}
\usepackage{slashed}
\usepackage{color}
\usepackage{braket}
\usepackage{tikz}
\usepackage{url}
\usepackage{subcaption}
\usepackage{amsmath}

\usepackage{hyperref}
\newcounter{qnumber}

\newcommand{\beq}{\begin{equation}}
\newcommand{\eeq}{\end{equation}}

\newcommand{\order}[1]{\mathcal{O}\left(#1 \right)}
\newcommand{\inn}[2]{#1 \cdot #2}
\newcommand{\dfk}{\frac{d^4k}{(2\pi)^4}}
\newcommand{\dtk}{\frac{d^3\vec{k}}{(2\pi)^3}}
\newcommand{\nn}{\nonumber}

\usepackage{stackrel}

\makeatother

\begin{document}
\title{Bound-Unbound Universality and the All-Order Semi-Classical Wave Function in Schwarzschild}

\author[a,b]{Majed Khalaf,}
\emailAdd{khalafmajed@gmail.com}
\affiliation[a]{Racah Institute of Physics, Hebrew University of Jerusalem, Jerusalem 91904, Israel}
\affiliation[b]{Laboratory for Elementary Particle Physics,
 Cornell University, Ithaca, NY 14853, USA}

\author[c,d,e]{Chia-Hsien Shen,}
\emailAdd{chshen@phys.ntu.edu.tw}
\affiliation[c]{Department of Physics and Center of Theoretical Physics, National Taiwan University, Taipei 10617, Taiwan}
\affiliation[d]{Leung Center for Cosmology and Astroparticle Physics, Taipei 10617, Taiwan}
\affiliation[e]{Physics Division, National Center for Theoretical Sciences, Taipei 10617, Taiwan}

\author[a]{and Ofri Telem}
\emailAdd{t10ofrit@gmail.com}

\abstract{
We present a systematic method for analytically computing time-dependent observables for a relativistic probe particle in Coulomb and Schwarzschild backgrounds. 
The method generates expressions valid both in the bound and unbound regimes, namely \textit{bound-unbound universal} expressions. To demonstrate our method we compute the time-dependent radius and azimuthal angle for relativistic motion in a Coulomb background (relativistic Keplerian motion), as well as the electromagnetic field radiated by a relativistic Keplerian source. All of our calculations exhibit bound-unbound universality. Finally, we present an exact expression for the semi-classical wave function in Schwarzschild. The latter is crucial in applying our method to any time-dependent observable for probe-limit motion in Schwarzschild, to any desired order in velocity and the gravitational constant $G$.
}

\maketitle
\section{Introduction}
With the influx of gravitational wave signals from inspiralling black holes \cite{Abbott2016}, the precise calculation of radiation-reacted bound motion has become a major focus point for theoretical and computational development. While many useful analytic approaches deal with bound motion directly \cite{Blanchet:2013haa,Porto:2016pyg,Goldberger:2004jt,Damour:2014jta,Jaranowski:2015lha,Bernard:2015njp,Bernard:2017bvn,Marchand:2017pir,Foffa:2019rdf,Foffa:2019yfl,Blumlein:2020pog,Blumlein:2020pyo,Blumlein:2021txj}, some of the most cutting-edge computational strategies involve calculations in the \textit{unbound} regime \cite{
Damour:2016gwp,Bini:2017wfr,Damour:2017zjx,Bjerrum-Bohr:2018xdl,Cheung:2018wkq,Kosower:2018adc,Bern:2019nnu,Bern:2019crd,Cristofoli:2019neg,Damgaard:2019lfh,Bjerrum-Bohr:2019kec,Damour:2019lcq,DiVecchia:2020ymx,Bjerrum-Bohr:2021vuf,
Damour:2020tta,Herrmann2021,Herrmann2021a,DiVecchia:2021bdo,
Bern2022a,Bern2022,Bern2023,Bini:2021gat,Manohar:2022dea,
Kalin:2020mvi,Kalin:2020fhe,Dlapa:2021npj,Kalin:2022hph,Dlapa:2022lmu,
Mogull:2020sak,Jakobsen:2022psy,Jakobsen:2023ndj,Bern:2024adl,Dlapa:2024cje,Driesse:2024xad,Driesse:2024feo}, which are then extrapolated to bound motion, for example using the effective-one-body approach \cite{Buonanno:1998gg,Damour2016a,Bini:2019nra,Bini:2020wpo,Bini:2020nsb,Bini:2020hmy,Bini:2020uiq,Bini:2020rzn,Buonanno:2024vkx}, or the boundary-to-bound map \cite{Kalin:2019rwq,Kalin:2019inp,Cho2022a,Dlapa:2024cje,Adamo:2024oxy}. 
Importantly, within the framework of Post-Adiabatic (PA) perturbation theory (c.f. \cite{Poisson:2011nh,Barack:2018yvs,Pound2021}), it is enough \cite{VanDeMeent2018,Miller2021,LISAConsortiumWaveformWorkingGroup:2023arg} to compute the radiation-induced force on osculating orbits \cite{Pound:2007th,Hinderer2008} -- conservative orbits that are momentarily tangential to the physical trajectory. Once this radiation-induced force, or self-force, is computed throughout the phase space and tabulated, one can solve the dynamical differential equation for an inspiralling trajectory and obtain the emitted gravitational waveform efficiently \cite{Hughes:2021exa,Wardell2023,Katz:2021yft}. Consequently, the computation of self-force on osculating orbits is a central element of modern inspiral computations, whose state-of-the-art is second-order in the mass ratio of the inspiral \cite{Pound:2015tma,Pound:2017psq,Barack:2018yvs,Pound:2019lzj,Pound2020,Upton2021,Warburton2021,Albertini2022,Albertini2022a,Spiers2023,Spiers2023a,Meent2023,Wardell2023,Bini:2024icd,Long:2024ltn}.  

This raises the question: for conservative orbits, what is the \textit{most natural quantity} to compute in the unbound regime in order to learn about the bound regime? 
In this paper, we propose a definite answer to this question; given a time-dependent observable $\mathcal{O}(t)$ (e.g. position on trajectory, radiated field, etc), the natural quantity to compute is its Laplace transform $\hat{\mathcal{O}}(s_L)$ in the unbound regime. We refer to the latter as a ``\textit{Laplace observable}" for short\footnote{We use the term ``observable" in a liberal way to mean time-dependent quantities and their Laplace transform. In GR these quantities can be coordinate-dependent, i.e. not strictly observables.}.
In very special cases this quantity can be computed by brute force, while generically it can be calculated analytically using the Quantum spectral Method (QSM), developed by two of the present authors \cite{Khalaf:2023ozy}.
Once $\hat{\mathcal{O}}(s_L)$ is reconstructed as a function of the complex Laplace variable $s_L$, it can be used to reproduce the time-domain observable $\mathcal{O}(t)$ both in the unbound and bound regimes, via an inverse Laplace transform. The latter leads to qualitatively different results depending on the sign of $E^2-\mu^2$, the difference between the squared energy and the squared rest mass; for $E^2>\mu^2$ the inverse Laplace transform gives unbound motion; while for $E^2<\mu^2$, the contour integral of the inverse Laplace transform localizes to the poles of $\hat{\mathcal{O}}(s_L)$, and we get a Fourier series for the periodic observable $\mathcal{O}(t)$ for bound motion.

We emphasize that the bound-unbound universality of the Laplace observable $\hat{\mathcal{O}}(s_L)$ for conservative motion is a fundamental property independent of the particular method used to compute it in the unbound regime. Nevertheless, the analytical continuation to the bound regime requires the knowledge of its complex structure. For this purpose, a bonafide \textit{Post-Minkowskian} (PM) or \textit{Post-Lorentzian} (PL) computation of $\hat{\mathcal{O}}(s_L)$ is insufficient, as it fails to reconstruct its poles in the bound regime. In this paper, we lay the foundations for a new generic method for computing any $\hat{\mathcal{O}}(s_L)$ exactly for Keplerian motion, and perturbatively\footnote{The word perturbative here has a very specific meaning -- we can expand the Laplace observables for Schwarzschild as a sum of Keplerian Laplace observables with $\mathcal{O}(\beta^iG^j)$ coefficients. Each Keplerian Laplace element is still non-perturbative in $G$ and manifestly bound-unbound universal.} for a Schwarzschild background. The essence of the method is as follows:
\begin{enumerate}
\item Compute any Laplace observable $\hat{\mathcal{O}}^{Kep}_i(s_L)$ \textit{exactly} for bound/unbound relativistic Keplerian motion, using the QSM.
\item For a probe particle in a Schwarzschild background, expand each Laplace observable as a sum of Keplerian Laplace observables with $\mathcal{O}(\beta^iG^j)$ coefficients, $\hat{\mathcal{O}}^{Sch}(s_L)=\sum\,a_{ij}\,\beta^iG^j\,\hat{\mathcal{O}}^{Kep}_{ij}(s_L)$.
\end{enumerate}
In this work, we complete part (1) of this method, and derive the exact semi-classical wave function in Schwarzschild,  required for the QSM computation of (2) to any desired order in $\beta$ and $G$. In a follow-up work, we will use this semi-classical wave function to compute geodesic motion in Schwarzschild as well as 1st order self-force, analytically and to any desired order in $\beta$ and $G$.

The paper is structured as follows.
Section~\ref{section:units} is a summary of the units and key quantities used in the paper. In Section~\ref{section:univstat} we show that time-dependent conservative motion can indeed be analytically continued between the bound and unbound cases. While our results for Keplerian motion are known, our analysis of the Schwarzschild case is new. In particular, we derive a novel implicit analytical solution for the time-dependent radius and azimuthal angle in Schwarzschild. In Section~\ref{section:univt}, we introduce Laplace observables and explain how they can be used to compute both bound and unbound time-dependent motion. Section~\ref{section:glance} is a recap of relativistic Keplerian motion, which is the main setting for this work, as well as the backbone of the Schwarzschild strategy presented in the last section. The section also contains a classical computation of $\hat{r}(s_L)$, the Laplace observable corresponding to the time-dependent radius of the motion. 

Section~\ref{section:QSMKep} is the main technical part of our paper, in which we show how to utilize the QSM to compute any Laplace observable for relativistic Keplerian motion. We demonstrate the method by computing two additional Laplace observables that have not been computed analytically before: 
(a) the azimuthal angle $\hat{\varphi}(s_L)$; and (b) $\hat{A}_\mu(s_L)$, the electromagnetic (EM) radiation field emitted by a relativistic Keplerian electron. These results, together with the time-dependent radius, are presented in Figs.~\ref{fig:Unbound_Bound},\,\ref{fig:Unbound_Bound_phi} and~\ref{fig:Unbound_Bound_At}, and compared with explicit numerical solutions for a perfect match. To emphasize, all of our computations are in the probe limit and to all orders in the coupling $\alpha_{EM}$. To make contact with the standard PM/PL expansion, we expand our result for the radiated EM field $\hat{A}_\mu(s_L)$ to first order in $\alpha_{EM}$ in Section~\ref{section:pert}. This provides us with a very non-trivial analytical benchmark between the QSM result and its 1PL counterpart.
Finally, in Section~\ref{section:Schwarz} we derive an all-order expression in $G$ for the Schwarzschild wave function, in the $\hbar\rightarrow 0$ limit. This allows to present any Schwarzschild Laplace observable as a sum over Keplerian Laplace observables. We conclude in Section~\ref{section:Conc}, where we outline the immediate prospects of our method to calculate geodesic motion and self-force in Schwarzschild, to any desired order in velocity and in $G$.
\newpage
\section{Units and Definitions}\label{section:units}
In this paper, we set the speed of light to be $c=1$, but not $\hbar=1$. This means that all dimensionful quantities have units $[\rm {(distance)}^a{\rm (mass)}^b]$. We summarize the different parameters defined in this table and their units in the following table:
\vspace*{10pt}
\begin{table}[ht]
\centering
\begin{tabular}{|c|c|c|}
\hline
{~~\textbf{Symbol}~~}          & {\textbf{Description}}                 & {\textbf{Units}} \\ \hline
$t$        & time                       & {[}distance{]}               \\ \hline
$r,\,x$        & distance                   & {[}distance{]}               \\ \hline
$k$    & wave number          & ~~{[}${\rm distance}^{-1}${]}~~ \\ \hline
$\Omega_r,\,\Omega_\varphi$    & fundamental frequencies          & ~~{[}${\rm distance}^{-1}${]}~~ \\ \hline
$\mu$        & mass                       & {[}mass{]}                   \\ \hline
$E$        & energy                     & {[}mass{]}                   \\ \hline
$p$        & momentum                   & {[}mass{]}                   \\ \hline
$L$        & angular momentum           & ~~{[}distance $\times$ mass{]}~~ \\ \hline
$L_z$        & azimuthal angular momentum           & ~~{[}distance $\times$ mass{]}~~ \\ \hline
$\mathcal{N}$      & ~~effective angular momentum~~ & ~~{[}distance $\times$ mass{]}~~ \\ \hline
$J_r$      & ~~radial action variable~~ & ~~{[}distance $\times$ mass{]}~~ \\ \hline
$\hbar$    & Planck's constant          & ~~{[}distance $\times$ mass{]}~~ \\ \hline
$K=-Qq/4\pi$    & EM constant$\times$ ${\rm (charge)}^2$          & ~~{[}distance $\times$ mass{]}~~ \\ \hline
$G M \mu$    & gravitational constant$\times$ ${\rm (mass)}^2$          & ~~{[}distance $\times \,\text{mass}${]}~~ \\ \hline
$\alpha_r,\,\alpha_\varphi$    & action-angles          & ~~{[dimensionless]}~~ \\ \hline
$\gamma(t),\,\beta(t)$    & momentary boost and velocity          & ~~{[dimensionless]}~~ \\ \hline
$\gamma_\infty,\,\beta_\infty$    & asymptotic boost and velocity          & ~~{[dimensionless]}~~ \\ \hline
$s_L$    & Laplace variable          & ~~{[dimensionless]}~~ \\ \hline
$j_r,\,\ell,\,m,\,\nu,\,n$    & quantum numbers          & ~~{[dimensionless]}~~ \\ \hline
\end{tabular}
\caption{In this table, we summarize the different parameters used in this paper, along with their units.}\label{tab:units}
\end{table}

\section{Bound-Unbound Universality in Kepler and Schwarzschild}\label{section:univstat}
It is well known that the trajectories for Keplerian motion have a well-defined analytical continuation between bound and unbound motion. Here we show that a similar relation holds for (special) relativistic Keplerian motion as well, and, perhaps surprisingly, to the orbiting motion of a probe mass in a Schwarzschild background. We begin with a quick presentation of the standard Keplerian case, followed by an easy relativistic generalization. Our results for Schwarzschild, however, are new to the best of our knowledge; in particular, we present a novel implicit analytic solution for the time-dependent radius and azimuthal angle, valid both for bound and unbound motion.
\subsection{Non-Relativistic Keplerian Motion}
Consider the non-relativistic Keplerian motion of a body of mass $\mu$ is a potential $V(r)=-K/r$. The resulting trajectories have the functional form
\begin{eqnarray}\label{eq:rtnonrel}
r(\varphi)=\frac{p}{1+e\cos(\varphi)}\,,
\end{eqnarray}
where $p$ and $e$ are the semi-latus rectum and eccentricity, respectively, and they are given by
\begin{eqnarray}\label{eq:penonrel}
p=\frac{L^2}{K\mu}~~,~~e=\sqrt{1+\frac{2EL^2}{K^2\mu}}\,,
\end{eqnarray}
where $E$ is the conserved (non-relativistic) energy (excluding the rest mass) and $L$ is the conserved angular momentum. Famously, for $E<0$ we have $0\leq e< 1$, and the body undergoes bound, elliptic motion. Conversely, for $E\geq 0$ we have $e\geq 1$ and the body undergoes unbound, hyperbolic motion. The continuation between bound and unbound trajectories carries over to time-dependent motion. For bound motion, we can implicitly solve for the time-dependent radius $r$ via the bound \textit{eccentric} anomaly $\kappa^b_r(t)$ as \cite{Goldstein2011}
\begin{eqnarray}\label{eq:rtnr}
r(t)=\frac{p}{1-e^2}\,\left(1-e\cos\kappa^b_r\right)~~({\rm bound})\,.
\end{eqnarray}
where 
\begin{eqnarray}\label{eq:kbnr}
\Omega^b_r t=\kappa^b_r-e\sin\kappa^b_r\,,
\end{eqnarray}
is Kepler's equation. Here $\Omega^b_r$ is the fundamental frequency given by
\begin{eqnarray}\label{eq:fundnr}
\Omega^b_r=\frac{\mu}{K}\,\left(-\frac{2E}{\mu}\right)^{\frac{3}{2}}\,.
\end{eqnarray}
The explicit time dependence of $r(t)$, and especially its generalization to relativistic Keplerian motion, will be given in Section~\ref{section:glance}. Here we will show how this time dependence has a natural analytic continuation to the unbound case. For the latter we have $E>0$, and so $\Omega^b_r$ is imaginary\footnote{We choose the square root so that ${\rm Im}(\Omega^b_r)>0$.}. It is then useful to define a positive real fundamental frequency
\begin{eqnarray}\label{eq:fundnrub}
\Omega_r=\frac{\mu}{K}\,\left(\frac{2E}{\mu}\right)^\frac{3}{2}=-i\Omega^b_r\,.
\end{eqnarray}
Defining $\kappa_r=i\kappa^b_r$, we get Kepler's equation for unbound motion,
\begin{eqnarray}\label{eq:kbnrub}
\Omega_r t=e\sinh(\kappa_r)-\kappa_r\,,
\end{eqnarray}
as well as
\begin{eqnarray}\label{eq:rtnrub}
r(t)=\frac{p}{e^2-1}\,\left(e\cosh\kappa_r-1\right)~~({\rm unbound})\,,
\end{eqnarray}
which is indeed the solution to the equations of motion (EOM) in the unbound regime.
\subsection{Relativistic Keplerian Motion}
We will analyze in detail the special relativistic version of Keplerian motion in Section~\ref{section:glance}. Here we simply present the relativistic generalization of~\eqref{eq:rtnonrel}-\eqref{eq:fundnr}. First, \eqref{eq:rtnonrel} is slightly modified to 
\begin{eqnarray}\label{eq:rtrel}
r(\varphi)=\frac{p}{1+e\cos(\frac{\mathcal{N}}{L}\varphi)}\,,
\end{eqnarray}
where $\mathcal{N}^2=L^2-K^2$, which accounts for the precession of the relativistic motion. The semi-latus rectum and eccentricity are given in the relativistic case by 
\begin{eqnarray}\label{eq:perel}
p=\frac{\mathcal{N}^2}{\gamma_{\infty} \mu K}~~,~~e=\sqrt{1+\frac{\mathcal{N}^2(E^2-\mu^2)}{K^2E^2}}\,,
\end{eqnarray}
where $E$ is the total energy. We see that for $E^2<\mu^2$ the motion is bounded, while for $E^2\geq \mu^2$ it is unbounded. As for the time-dependent radius, \eqref{eq:rtnr} is still valid for the time-dependent radius, while the Kepler equation is modified to
\begin{eqnarray}\label{eq:kbrrl}
\Omega^b_r t=\kappa^b_r-e\gamma^2_{\infty}\sin\kappa^b_r\,,
\end{eqnarray}
where $\gamma_{\infty}\equiv E/\mu$ and
\begin{eqnarray}\label{eq:fundrl}
\Omega^b_r=\frac{\mu}{K}\left({1-}\gamma^2_{\infty}\right)^{\frac{3}{2}}\,.
\end{eqnarray}
One can see immediately that the analytical continuation to the unbound case holds in exactly the same way as in the non relativistic case. In particular, for $E^2>\mu^2$ we define $\Omega_r=-i\Omega^b_r$ and $\kappa_r=i\kappa^r_b$ so that
\begin{eqnarray}\label{eq:rtnrubrel}
r(t)=\frac{p}{e^2-1}\,\left(e\cosh\kappa_r-1\right)~~({\rm unbound})\,,
\end{eqnarray}
with
\begin{eqnarray}\label{eq:kbnrubrel}
\alpha_r\equiv\Omega_r t=e\gamma^2_\infty\sinh(\kappa_r)-\kappa_r\,,
\end{eqnarray}
which indeed gives the time-dependent radius for unbound motion.
\subsection{Probe Mass in Schwarzschild}
\subsubsection{Trajectory}
The EOM for the shape $r(\varphi)$ of the trajectory in Schwarzschild is given by \cite{Carter:1968ks}
\begin{eqnarray}\label{eq:rphischeq}
\frac{dr}{d\varphi}=\pm\frac{1}{L}\sqrt{P^{Sch}_r(r)}\,.
\end{eqnarray}
where $\Delta(r)=r(r-2GM)$ and
\begin{eqnarray}\label{eq:Psch}
P^{Sch}_r(r)&=&E^2r^4-\Delta(r)\left(L^2+\mu^2r^2\right)\nonumber\\[5pt]
&\equiv& (\mu^2-E^2)\,r\,(r-r_{b})\,(r-r_{min})\,(r_{*}-r)
\end{eqnarray}
is the characteristic radial polynomial in Schwarzschild. The $+\,(-)$ sign is for the first (second) part of each period. The radii $r_b, r_{min}$ and $r_*$ are defined in~\eqref{eq:Psch} so that $0<r_b<r_{min}\leq r\leq r_*$ for bound motion, and $r_*<0<r_b<r_{min}\leq r$ for unbound motion (we do not consider infalling trajectories in this work). We begin by analyzing the bound case, for which $E^2<\mu^2$ and $r_{b}<r_{min}<r_{*}$, and consider the bound motion between $r_{min}\leq r\leq r_{*}$.
The solution to this equation is (see \cite{Scharf:2011ii,Gibbons:2011rh} for equivalent formulations)
\begin{eqnarray}\label{eq:rphischeqsl}
r(\varphi)=\frac{p}{1+e\,\cos\left[2{\rm am}(A\,\varphi,k)\right]}\,,
\end{eqnarray}
where ${\rm am}(\phi,k)$ is the Jacobi amplitude function \cite{NIST:DLMFE3}, and 
\begin{eqnarray}\label{eq:csn}
p=\frac{2r_{min}r_{*}}{r_{min}+r_{*}}~~,~~e={\rm sign}(L)\,\frac{r_{*}-r_{min}}{r_{*}+r_{min}}\,,
\end{eqnarray}
as well as
\begin{eqnarray}\label{eq:Aphi}
A\equiv\frac{p e}{(1-e^2)(r_{*}-r_{min})}\sqrt{\frac{(\mu^2-E^2)r_{*}(r_{min}-r_{b})}{J^2\mu^2}}~~,~~k^2=-\frac{r_{b}(r_{*}-r_{min})}{r_{*}(r_{min}-r_{b})}\,.
\end{eqnarray}
We put ${\rm sign}(L)$ in the definition of the eccentricity to correctly reproduce the boundary-to-bound map \cite{Kalin:2019inp} in the next subsection.
Note that the ${\rm{am}}$ function satisfies
\begin{eqnarray}\label{eq:deltaphi}
&&{\rm am}(A(\varphi+2\pi+\delta\varphi),k)={\rm am}(A \varphi,k)+\pi\nonumber\\[5pt]
    &&\delta\varphi=\frac{2}{A}K(k)-2\pi\,,
\end{eqnarray}
where $K(k)$ is the complete elliptic integral of the first kind. Consequently, $r$ is periodic under $\varphi\rightarrow \varphi+2\pi+\delta \varphi$. In other words, $\delta\varphi$ is the exact result for the precession of the pericenter in a Schwarzschild background.

\begin{figure}[t]
     \centering
     \begin{subfigure}{0.455\textwidth}
         \centering
         \includegraphics[width=\textwidth]{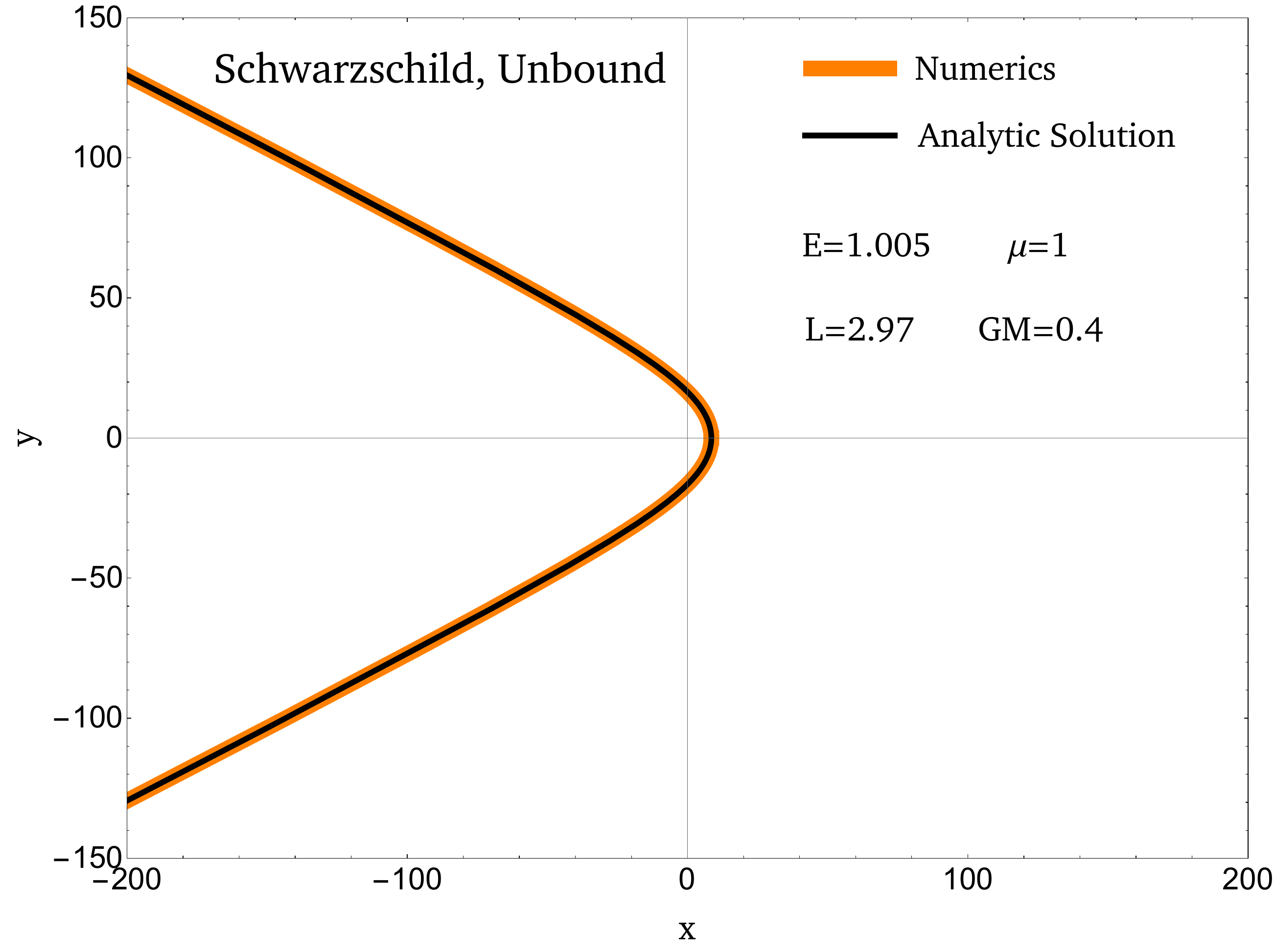}
         \caption{Unbound}
\end{subfigure}
     \hspace{25pt}
     \vspace*{-15pt}
        \hspace{-2pt}  \begin{subfigure}{0.46\textwidth}
         \centering \includegraphics[width=\textwidth]{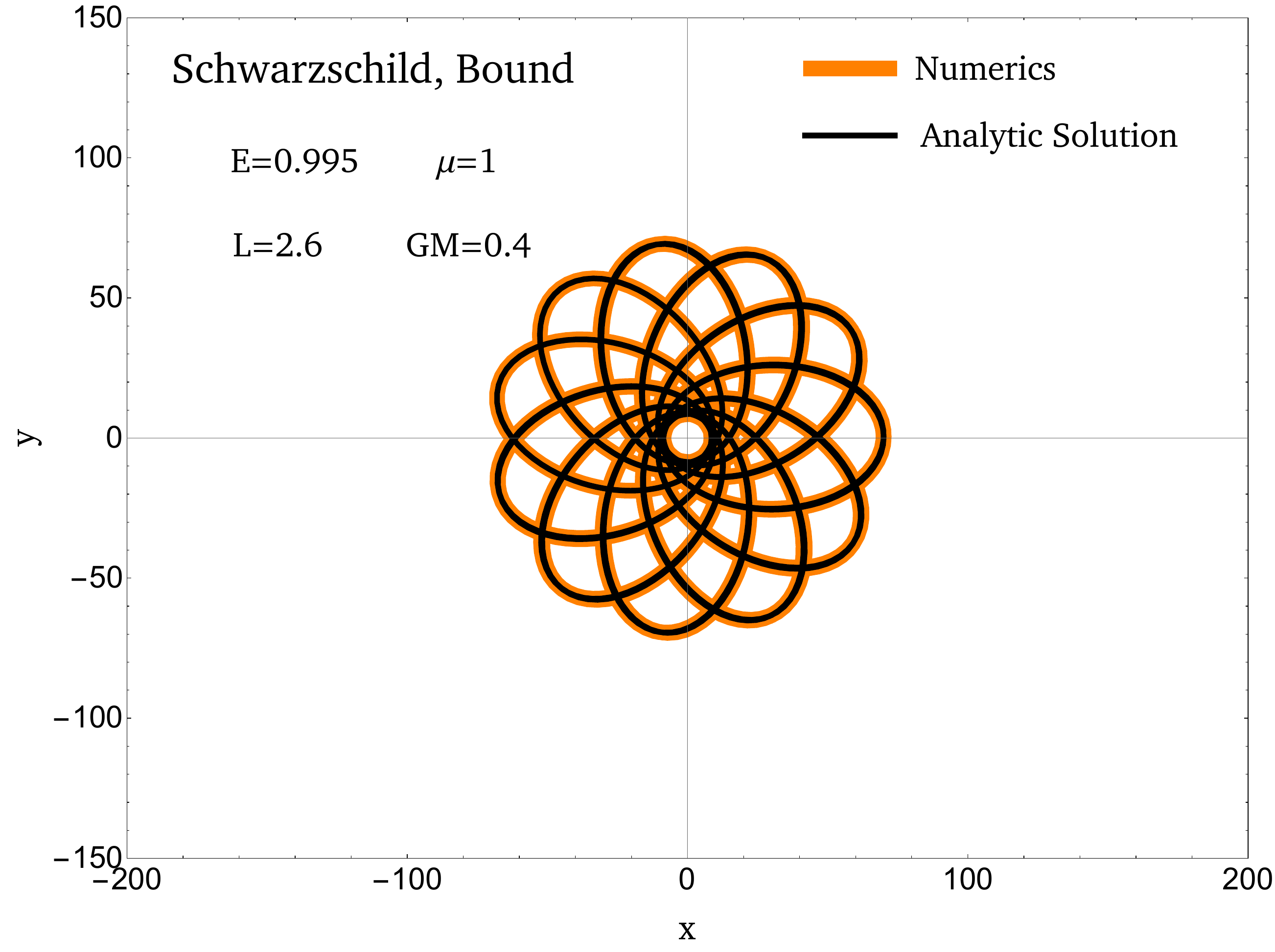}
         \caption{Bound}
     \end{subfigure}
     \vspace*{15pt}
     \caption{Trajectory in XY plane for geodesic motion in Schwarzschild: comparison of analytical solution \eqref{eq:rphischeqsl} with numerical solutions. The units for the parameters $E,L,\mu,G,M$ are given in Table~\ref{tab:units}. Note that the bound motion involves precession of the pericenter.}\label{fig:Sch_traj_Unbound_Bound}
     \end{figure}

Similarly to the Keplerian case, the solution \eqref{eq:rphischeqsl} is valid both in the bound and unbound regimes. This is explicitly shown in Fig.~\ref{fig:Sch_traj_Unbound_Bound}, where we compare it with numerical solutions both in the bound and unbound regimes. 
In the unbound regime $e>1$ and so $r(\varphi)$ diverges at $\varphi=\frac{1}{2}\left(\chi_{scat}+\pi\right)$ where $\chi_{scat}$ is the scattering angle,
\begin{eqnarray}\label{eq:scatsch}
\chi_{scat}=-\frac{2}{A}F\left[-\frac{1}{2}{\rm arccos}\left(-\frac{1}{e}\right),k\right]-\pi\,.
\end{eqnarray}
Here $F$ is the incomplete elliptic integral of the first kind. Using \eqref{eq:deltaphi} and \eqref{eq:scatsch}, one can explicitly check the boundary-to-bound map \cite{Kalin:2019rwq} for the scattering angle,
\begin{eqnarray}\label{eq:b2bsch}
\chi_{scat}(L)+\chi_{scat}(-L)=\delta\varphi~~,~~E^2<\mu^2\,,
\end{eqnarray}
which holds exactly. Note that when we take $L\rightarrow -L$, the sign of the eccentricity flips by definition, and $r_{*}$ is exchanged with $r_{min}$.

\subsubsection{Time Dependence}
The EOM for the time-dependent azimuthal angle is given by \cite{Carter:1968ks}
\begin{eqnarray}\label{eq:dotssch}
\dot{\varphi}=\frac{L\Delta(r)}{Er^4}\,,
\end{eqnarray}
where the overdot denotes a derivative with respect to coordinate (Boyer-Lindquist) time. Since we know $r(\varphi)$ we can directly integrate \eqref{eq:dotssch} to get $t(\varphi)$, with the result given in Appendix~\ref{KepSch}. The result is depicted in Fig.~\ref{fig:Sch_Unbound_Bound}, where we plot the parametric curve $\left(t(\varphi),r(\varphi)\right)$, with a perfect match to numerical results in both bound and unbound regimes. The $t$ axis in these plots is normalized by the radial fundamental frequency
\begin{eqnarray}\label{eq:Omegabr}
\Omega^b_r\equiv\frac{2\pi}{T^b_r}\,,
\end{eqnarray}
where $T^b_r$ is the radial period given by
\begin{eqnarray}\label{eq:Tbr}
T^b_r=t(\varphi=2\pi+\delta\varphi)\,.
\end{eqnarray}

\begin{figure}[t]
     \centering
     \begin{subfigure}{0.455\textwidth}
         \centering
         \includegraphics[width=\textwidth]{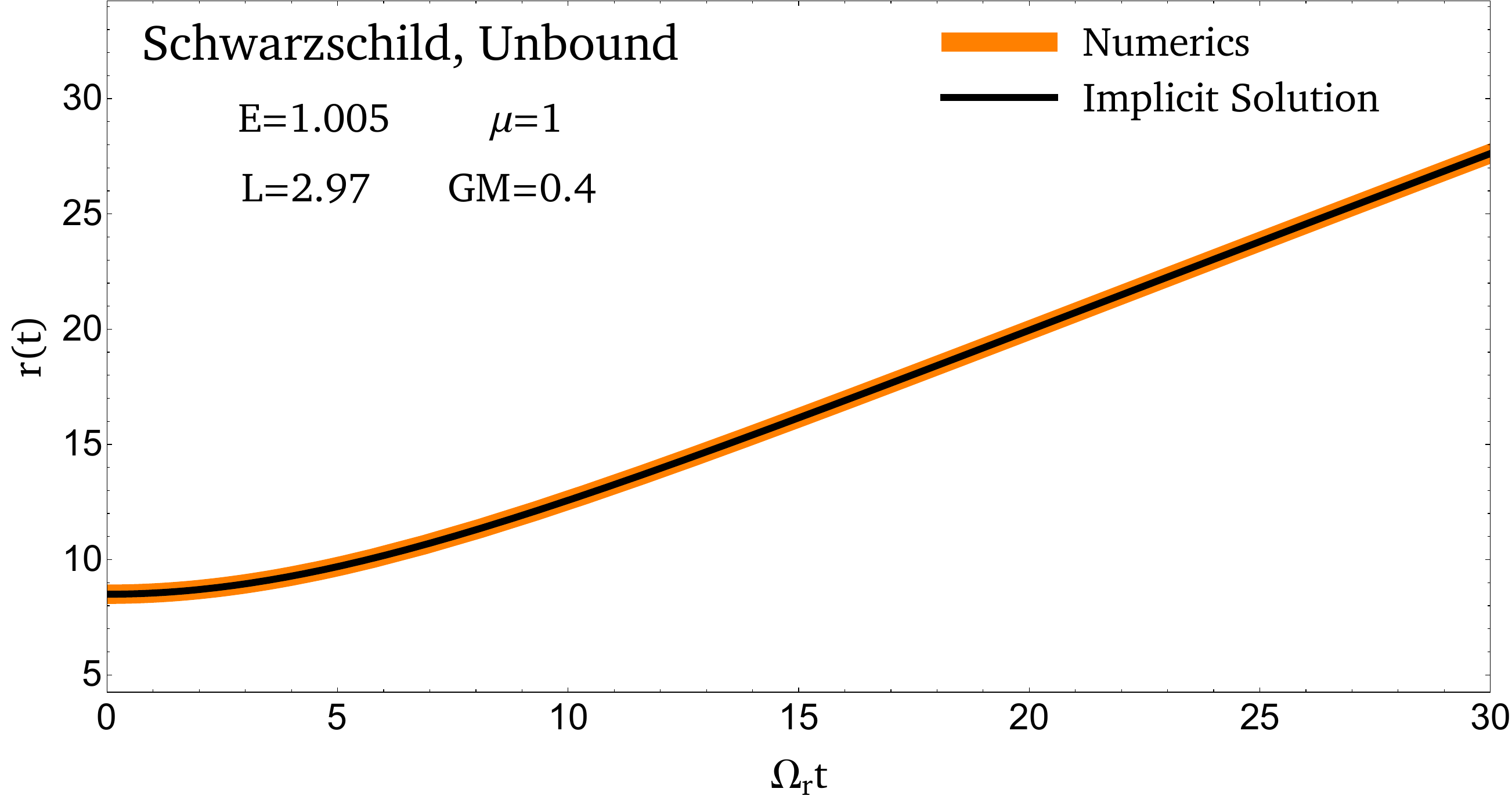}
         \caption{Unbound}
\end{subfigure}
     \hspace{25pt}
     \vspace*{-15pt}
        \hspace{-2pt}  \begin{subfigure}{0.46\textwidth}
         \centering \includegraphics[width=\textwidth]{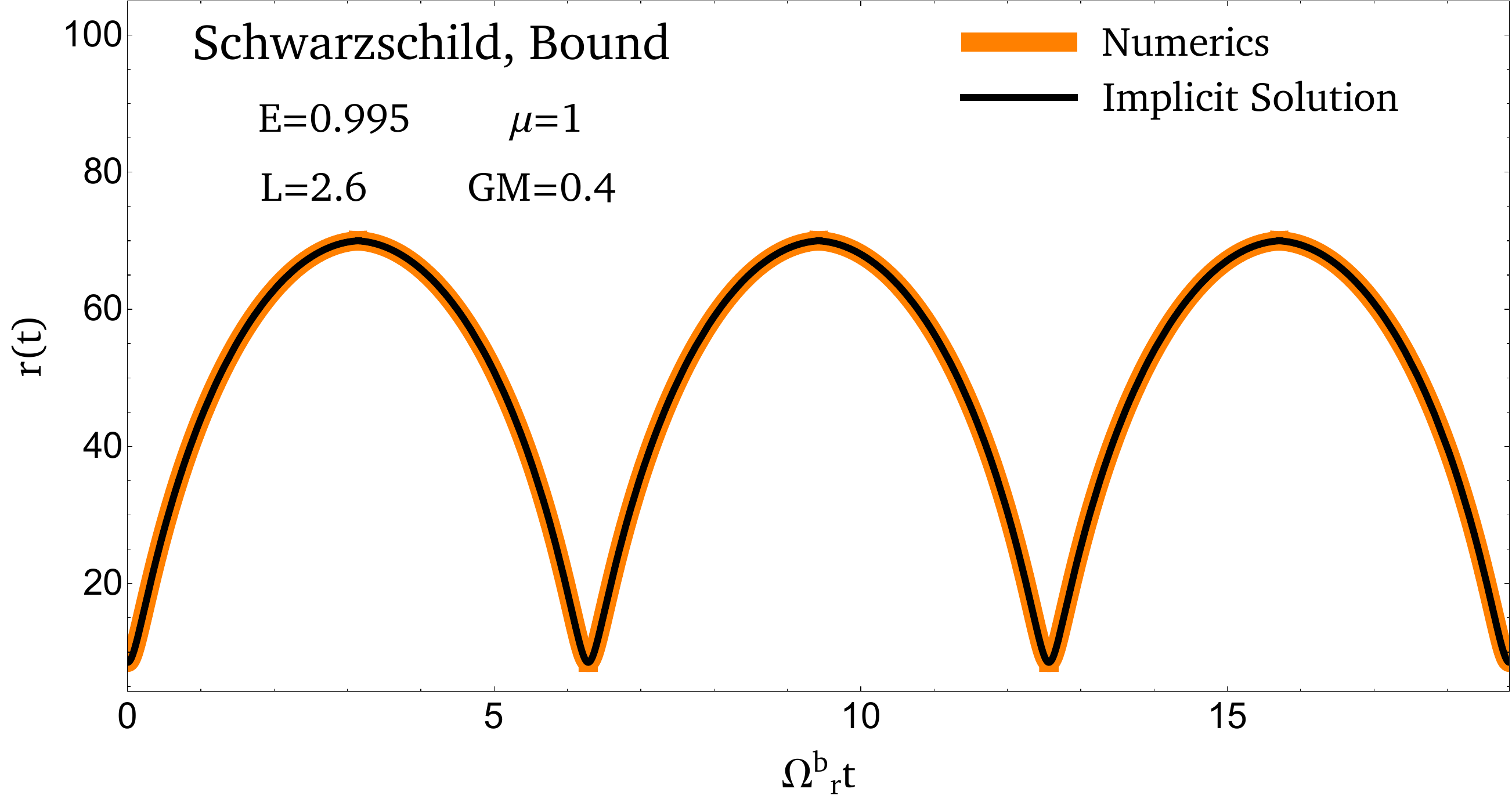}
         \caption{Bound}
     \end{subfigure}
     \vspace*{15pt}
     \caption{$r(t)$ for geodesic motion in Schwarzschild: comparison of implicit analytical solution (\eqref{eq:rphischeqsl} and the solution for \eqref{eq:dotssch} in \eqref{eq:Fexp}) with numerical solutions. The units for the parameters $E,L,\mu,G,M$ are given in Table~\ref{tab:units}. Note that for bound motion, $r(t)$ is periodic.}\label{fig:Sch_Unbound_Bound}
     \end{figure}

\section{Laplace Observables and Bound-Unbound Universality}\label{section:univt}

It is tempting to view the implicit analytical solutions \eqref{eq:rtnr}, \eqref{eq:kbrrl} (for relativistic Kepler) and \eqref{eq:rphischeqsl}, 
 \eqref{eq:Fexp} (for Schwarzschild) as an indication that all time-dependent observables can be computed analytically for these systems. In fact, this statement is true for relativistic Keplerian motion, and holds for Schwarzschild order-by-order in a PM expansion. In practice, however, the analytical computation of more generic observables (such as the emitted field) requires more robust technology than the involved differential equations of Section~\ref{section:univstat}. The development of this robust technology is the main contribution of this current work.

The central point of our paper is quite simple: knowing the Laplace transform of time-dependent observables (Laplace observables for short) allows us to reproduce their time-domain values both for unbound and bound motion. For unbound motion, the inverse Laplace transform gives unbound observables, for bound motion, the inverse Laplace transform becomes a Fourier series and leads to (multi-) periodic observables.

The point above is most vividly illustrated in the context of \textit{conservative, spherically symmetric motion} of a probe particle in an ambient radial potential. We use this limited scope as a laboratory to demonstrate our ideas, though the use of Laplace observables is by no means limited to this special case. For the system in question, the motion is characterized by a conserved energy $E$ and conserved angular momentum $\vec{L}$, which we can assume without loss of generality to be directed along the $z$ axis. Furthermore, all time dependent observables for the system can be recast as functions of the \textit{angle variables} of the system \cite{Goldstein2011} $\alpha_r,\,\alpha_\varphi$, which grow linearly with time as
\begin{eqnarray}
\alpha_i=\Omega_it\,,\,i\in\{r,\varphi\}\,,
\end{eqnarray}
where $\Omega_i$ are the \textit{fundamental frequencies} of the theory\footnote{In a relativistic theory, these are the fundamental frequencies with respect to \textit{coordinate time}.}. The results in this section are independent of the particular definition of these angles and only hinges on one property of the fundamental frequencies:
\begin{eqnarray}
\Omega_i&\in&{\rm Positive~Reals}~~~~{\rm for~unbound~motion}\nonumber\\[5pt]
i\Omega_i&\in&{\rm Positive~Reals}~~~~{\rm for~bound~motion}\,.
\end{eqnarray}
\begin{figure}[t]
     \centering
     \begin{subfigure}{0.455\textwidth}
         \centering
         \includegraphics[width=\textwidth]{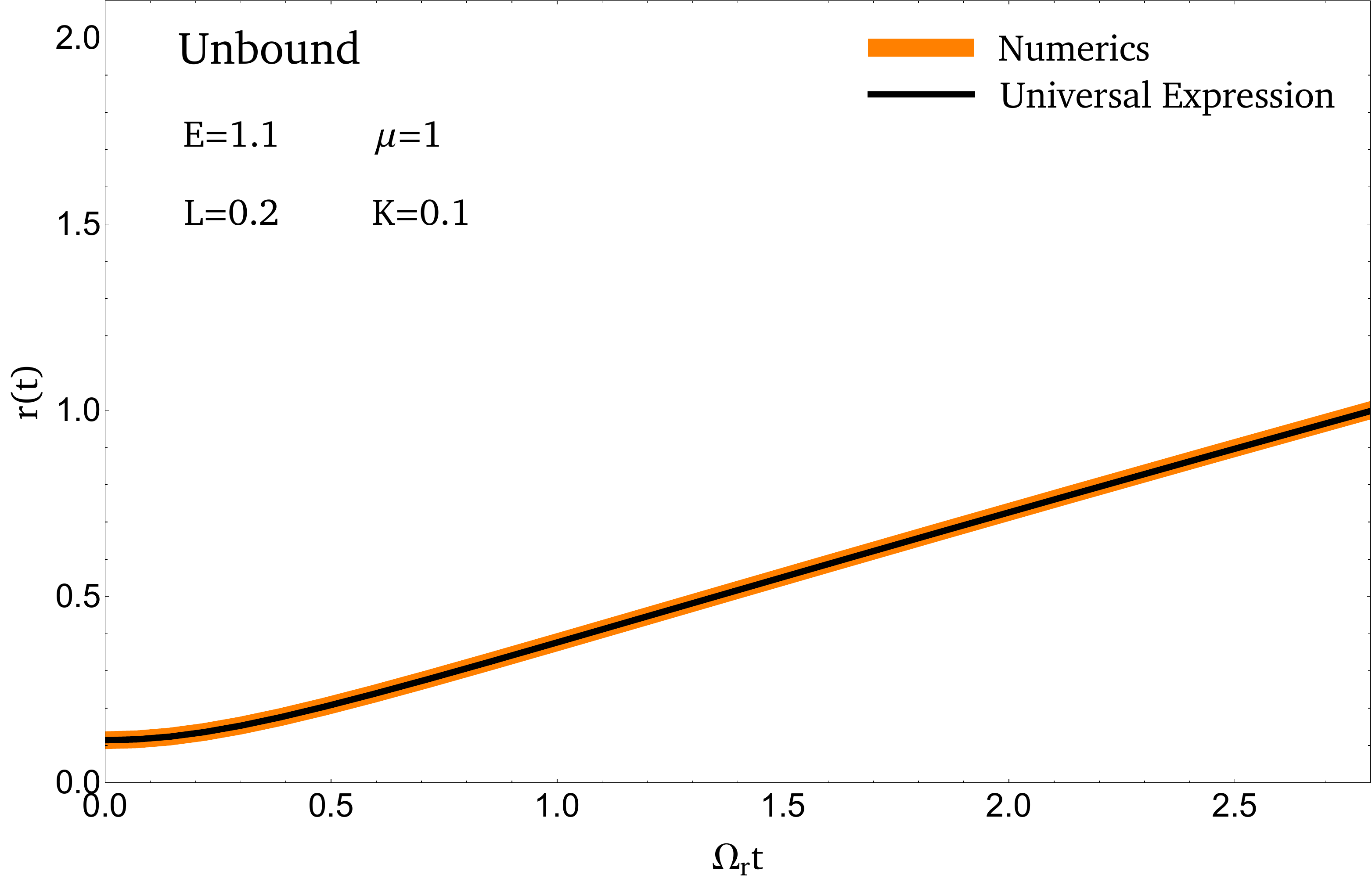}
         \caption{Unbound}
\end{subfigure}
     \hspace{25pt}
     \vspace*{-15pt}
        \hspace{-2pt}  \begin{subfigure}{0.46\textwidth}
         \centering \includegraphics[width=\textwidth]{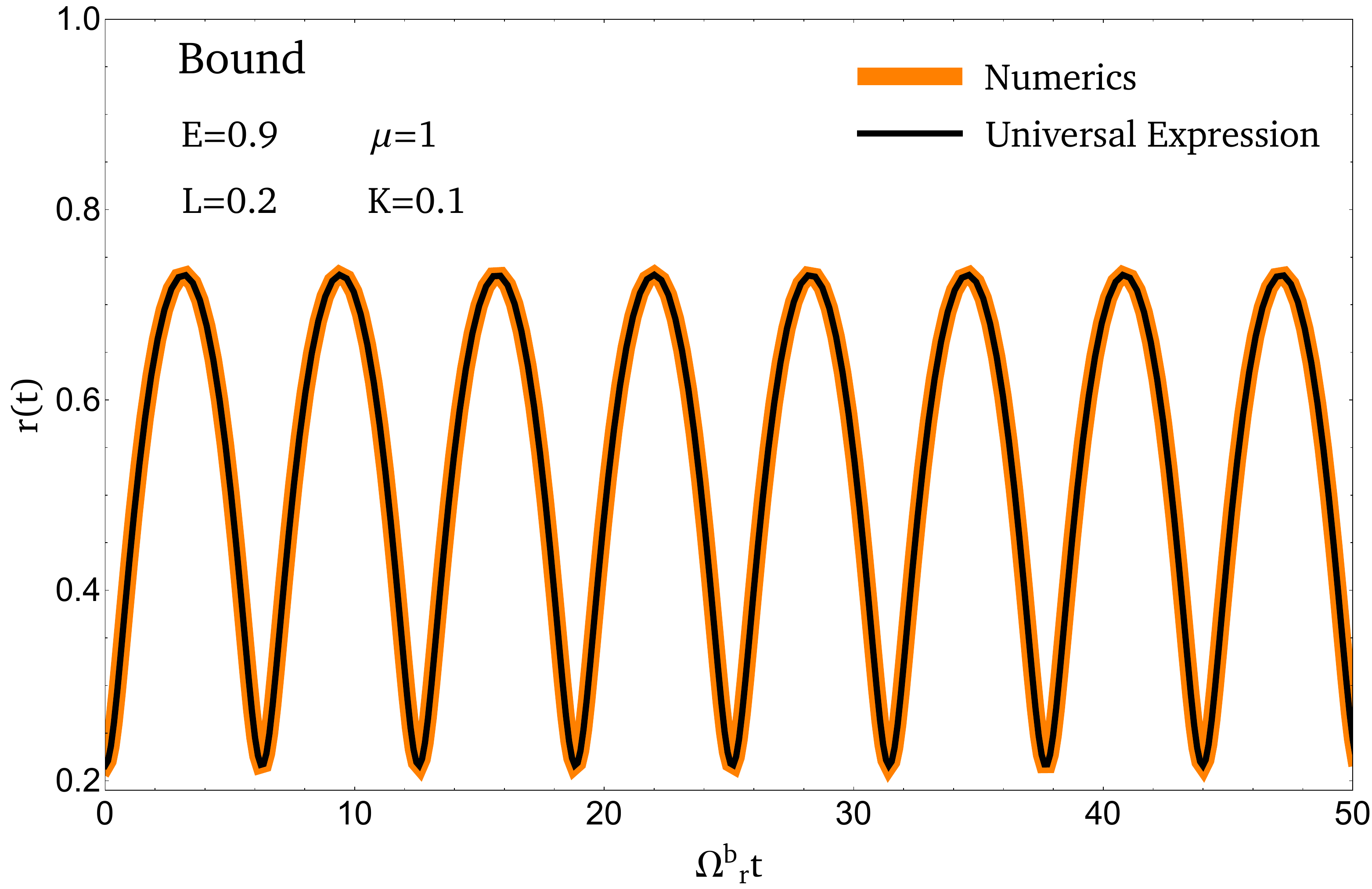}
         \caption{Bound}
     \end{subfigure}
     \vspace*{15pt}
     \caption{$r(t)$ for relativistic Keplerian motion: comparison of universal expression (inverse Laplace/Fourier transform of Laplace observable) with numerical solutions. The units for the parameters $E,L,\mu,K$ are given in Table~\ref{tab:units}. Note that for bound motion, $r(t)$ is periodic.}\label{fig:Unbound_Bound}
     \end{figure}
     \begin{figure}[t]
     \centering
     \hspace{1pt}
     \begin{subfigure}{0.44\textwidth}
         \centering
          \includegraphics[width=\textwidth]{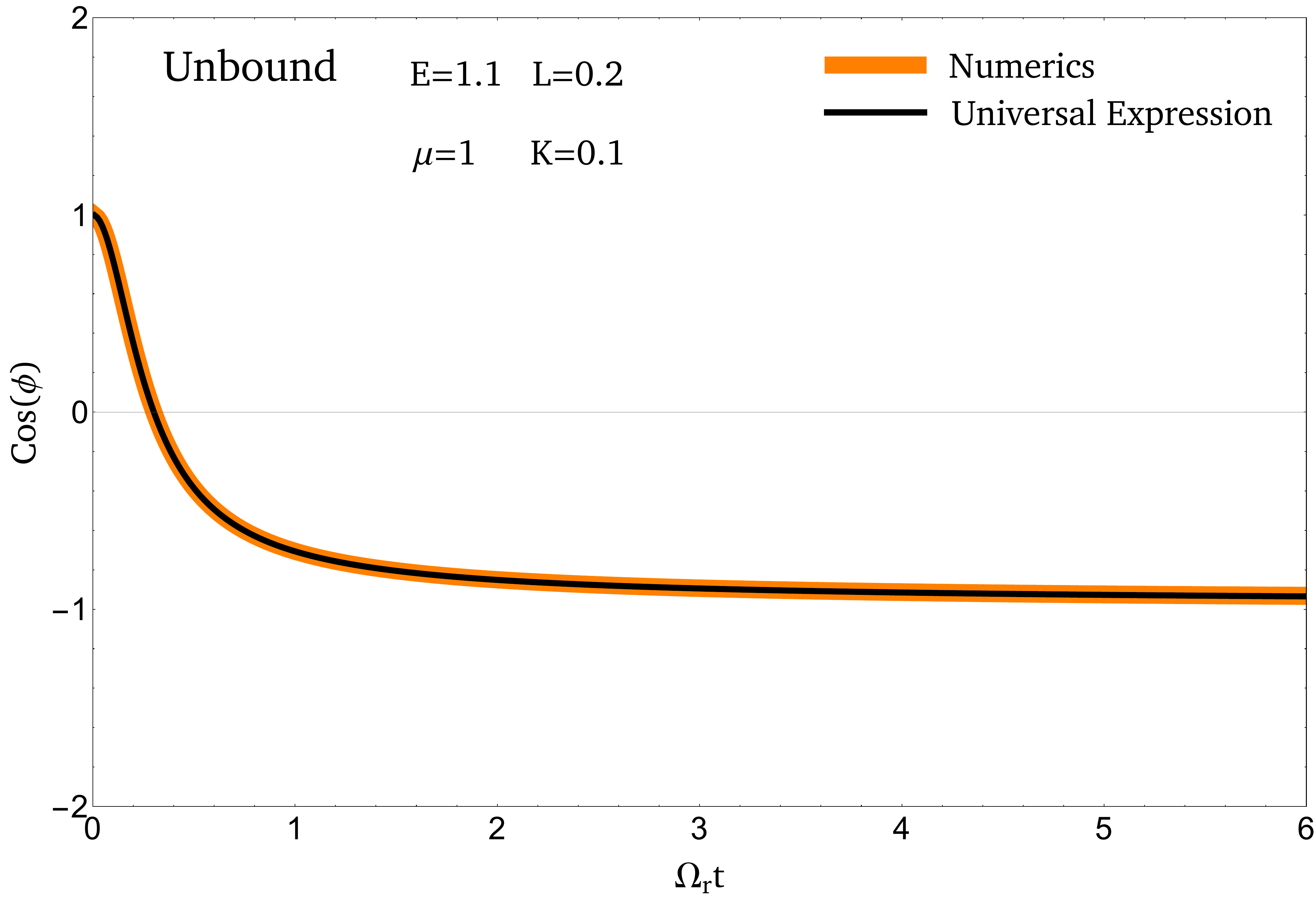}
         \caption{Unbound}
\end{subfigure}
     \hspace{25pt}
     \vspace*{-15pt}
        \hspace{-2pt}  \begin{subfigure}{0.465\textwidth}
         \centering \includegraphics[width=\textwidth]{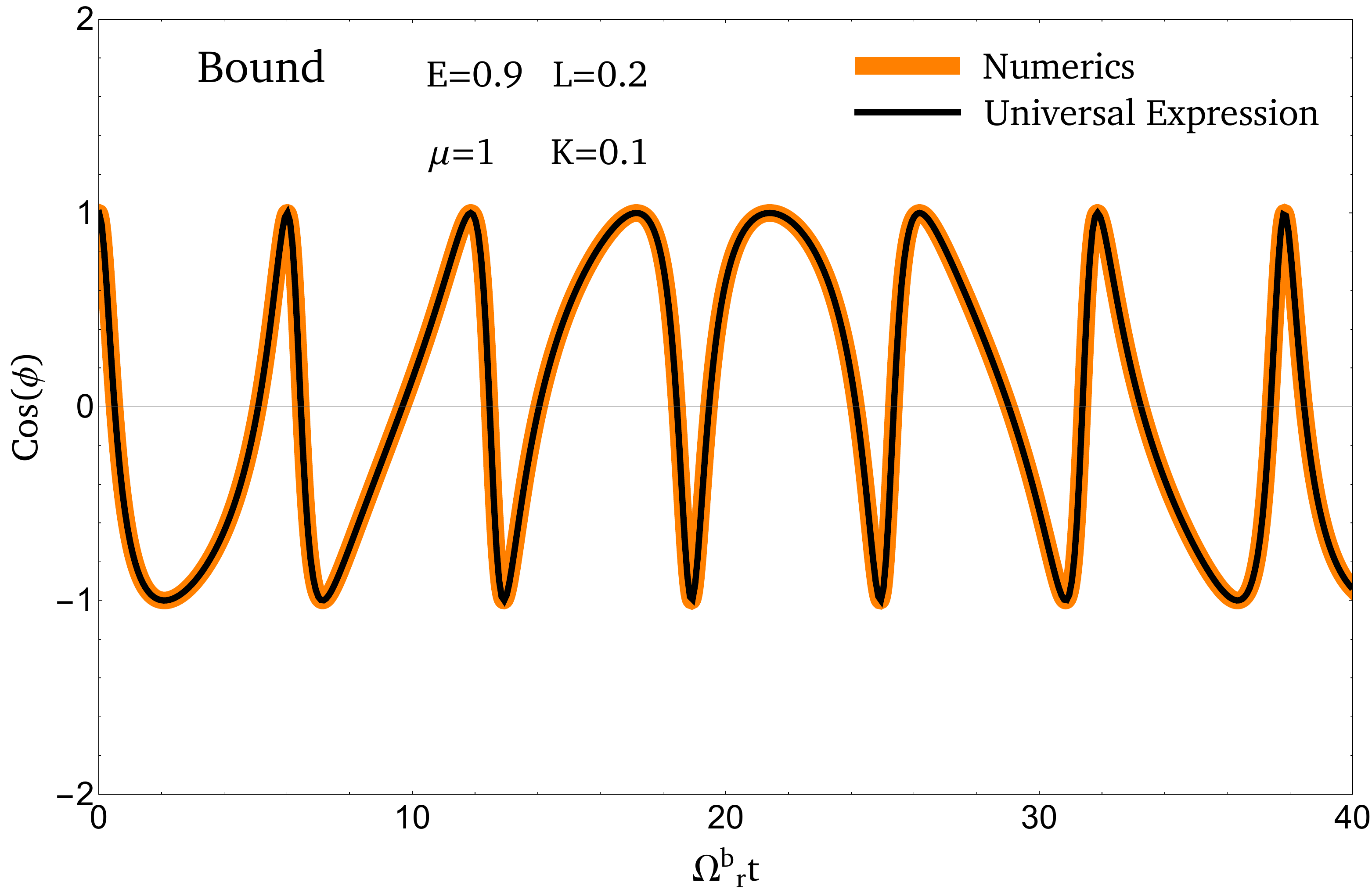}
         \caption{Bound}
     \end{subfigure}
     \vspace*{15pt}
     \caption{$\cos(\varphi(t))$ for relativistic Keplerian motion: comparison of universal expression (inverse Laplace/Fourier transform of Laplace observable) with numerical solutions. The units for the parameters $E,L,\mu,K$ are given in Table~\ref{tab:units}. Note that for bound motion, $\cos(\varphi(t))$ is not periodic, as relativistic Keplerian motion involves \textit{precession}.}\label{fig:Unbound_Bound_phi}
     \end{figure}
\begin{figure}[ht!]
     \centering  
     \vspace{15pt}
     \begin{subfigure}{0.455\textwidth}
         \includegraphics[width=\textwidth]{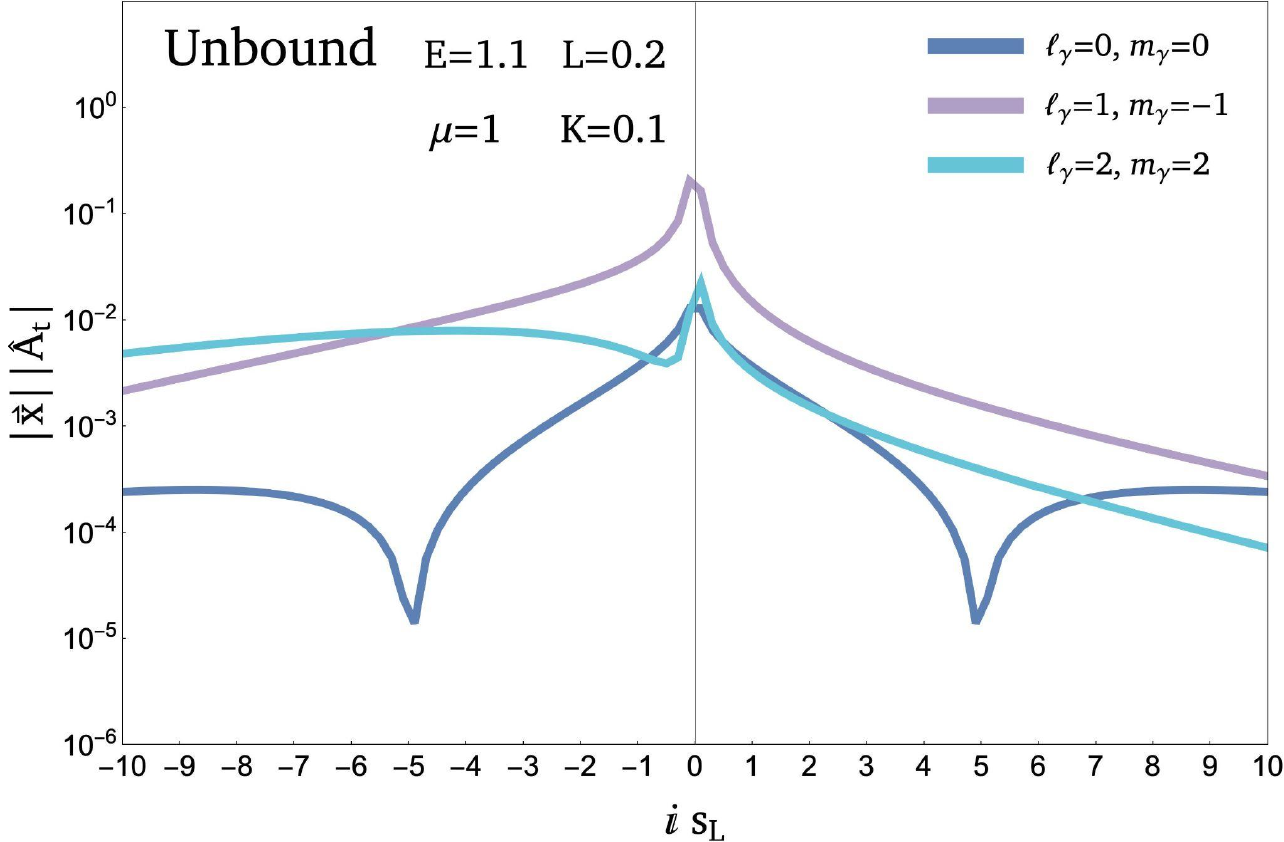}   
         \end{subfigure}
         \hspace{18pt}
    \begin{subfigure}{0.449\textwidth}
         \includegraphics[width=\textwidth]{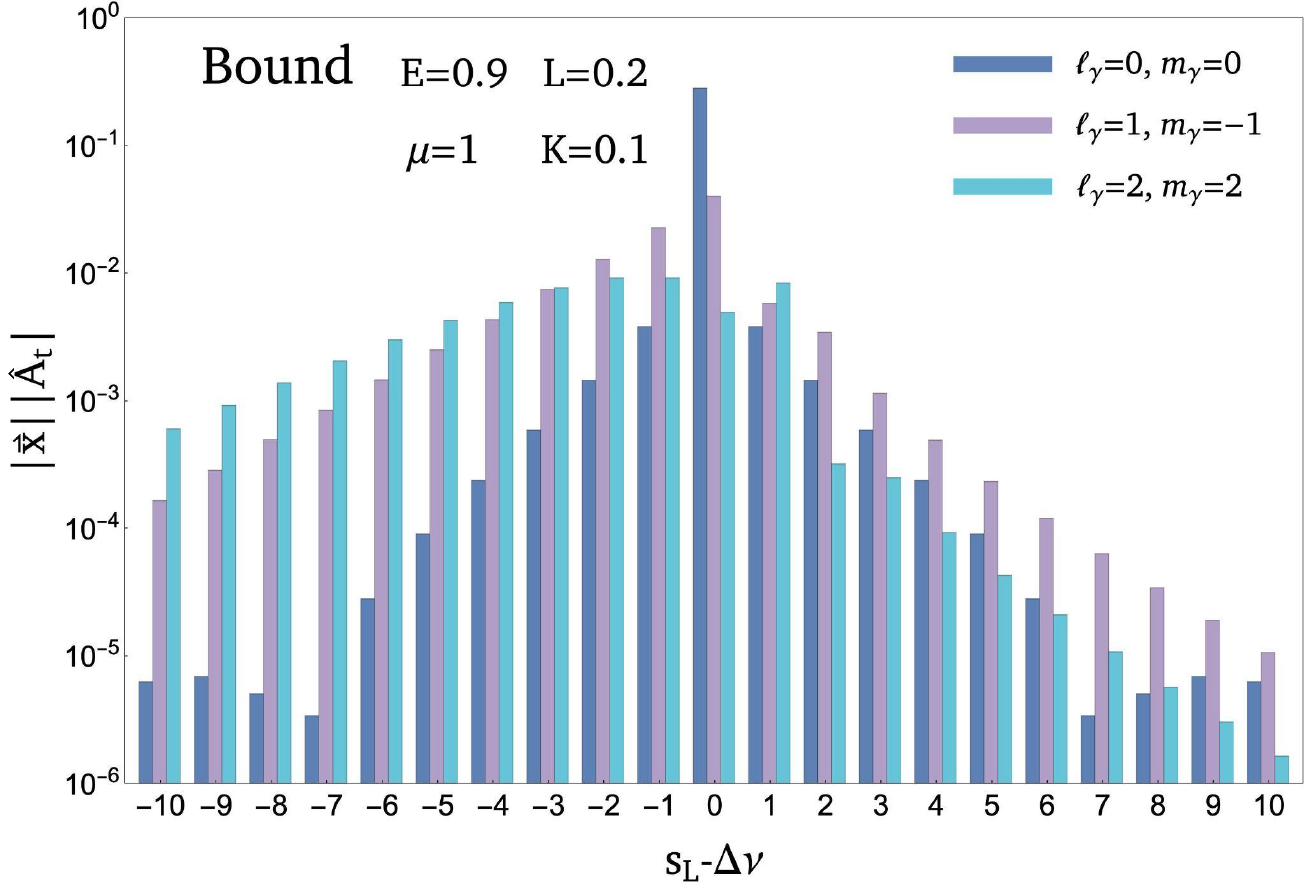}
     \end{subfigure}\\
     
     \hspace{-3pt}
     \begin{subfigure}{0.455\textwidth}
         \includegraphics[width=\textwidth]{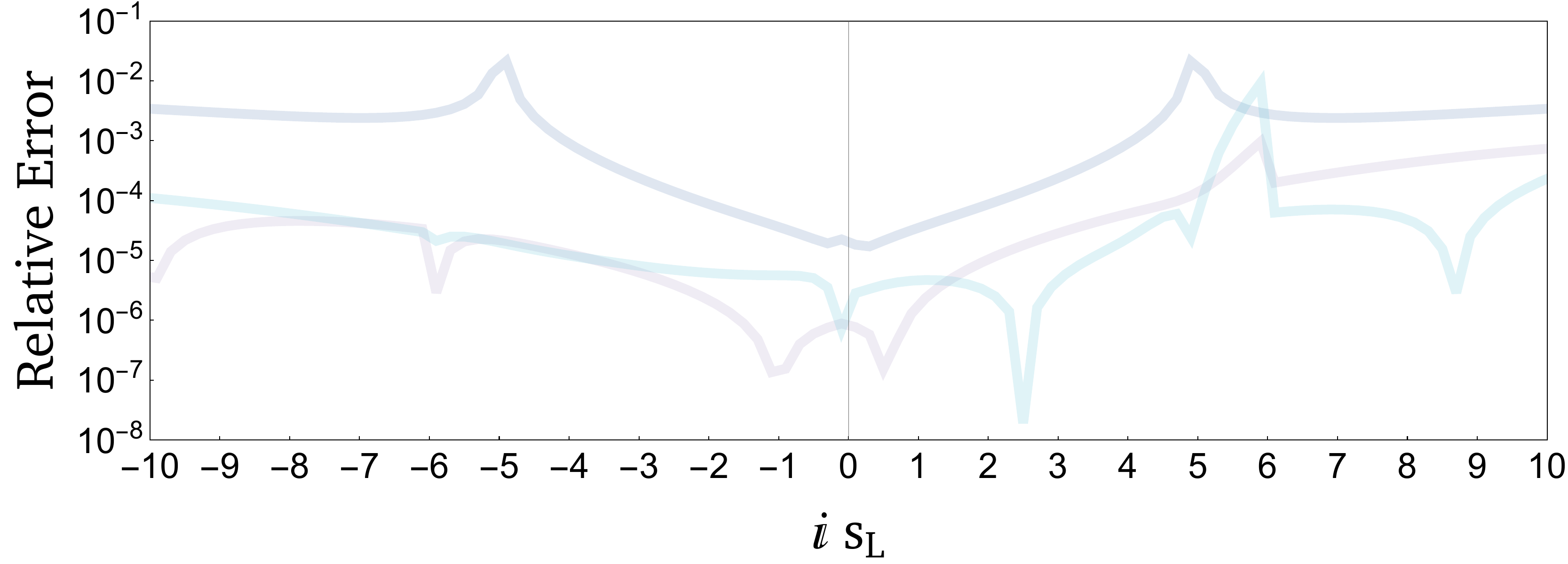}
         \caption{Unbound}
         \label{fig:unboundAt}
     \end{subfigure}
     \hspace{17pt}
     \begin{subfigure}{0.450\textwidth}
         \includegraphics[width=\textwidth]{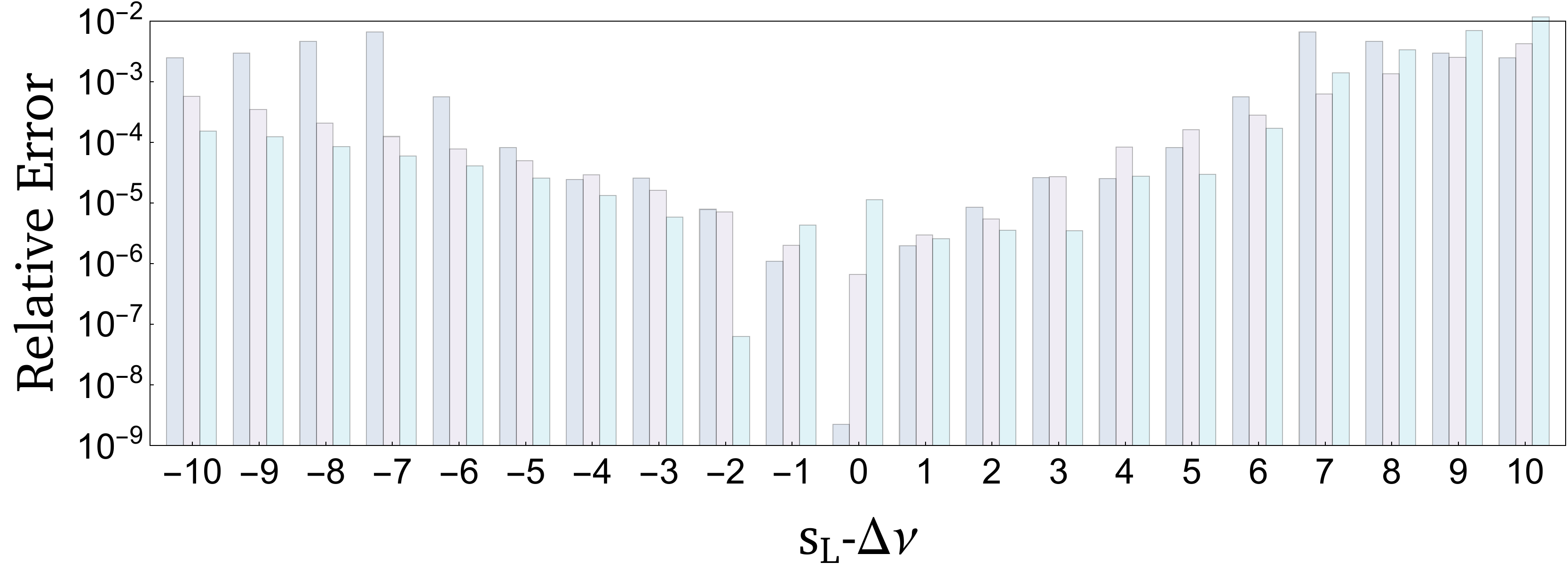}
         \caption{Bound}
         \label{fig:boundAt}
     \end{subfigure}
        \caption{The Laplace/Fourier EM field $A_t$ radiated by a charge in relativistic Keplerian motion. The units for the parameters $E,L,\mu,K$ are given in Table~\ref{tab:units}. Top left: Laplace observable for unbound motion vs $i s_L\in\mathbb{R}$. Top right: Fourier coefficients for bound motion vs $s_L{-}\Delta\nu\in\mathbb{Z}$. Bottom: comparison of our analytic universal expression with numerical solutions. The relative difference between the two is due to the error in the numerical Laplace/Fourier transform of the time-dependent numerical result, and can be systematically minimized with more numerical computation time. The analytic expression involves a truncated infinite sum and can be systematically improved by choosing a higher cutoff if needed.}\label{fig:Unbound_Bound_At}       
\end{figure}
Now consider a time-dependent operator $\mathcal{O}(t)$ of the system. This observable may be the distance of the probe particle from the origin, its azimuthal angle, the field radiated by its conservative trajectory, etc. Taking a two-sided \textit{Laplace transform} of $\mathcal{O}\left(t\right)$, we have
\begin{eqnarray}\label{eq:LaplaceFourier}
\hat{\mathcal{O}}(s_L)&=&\frac{\Omega_r}{2\pi}\,\int_{-\infty}^{\infty}\,dt~\mathcal{O}\left(t\right)\,e^{-s_L \Omega_r t}\,.
\end{eqnarray}
For the ``nice" enough observables we consider, this integral converges in a strip\footnote{If not, we can always choose a suitable $\epsilon$ prescription to guarantee convergence on a strip of width $\mathcal{O}(\epsilon)$.} $\sigma^{min}_{r}\leq {\rm Re}(s_L)\leq \sigma^{max}_{r}$. The corresponding inverse Laplace transform is then given by \textit{Mellin's inversion formula}, also known as the \textit{Bromwich integral} \cite{LEPAGEWilburR1961CVat}:
\begin{eqnarray}\label{eq:LaplaceFourierinv}
\mathcal{O}\left(t\right)&=&-i\,\mathcal{P}\int_{\Gamma_r-i\infty}^{\Gamma_r+i\infty}\,\hat{\mathcal{O}}(s_L)\,e^{s_L\Omega_r t}\,ds_L\,,
\end{eqnarray}
where $\mathcal{P}$ stands for the principal value of the integral, and $\Gamma_{r}$ is a real number so that\footnote{Note that $\mathcal{O}(\alpha_r,\alpha_\varphi)$ is uniquely determined from $\hat{\mathcal{O}}(s_L)$ \textit{and} $\Gamma_{r}$, i.e. the integration contour is required as input for the inverse Laplace transform.} $\sigma^{min}_{r}\leq \Gamma_r\leq \sigma^{max}_{r}$. 
In the bound regime, we can define the real and positive $\Omega^b_r\equiv i \Omega_r$, and so the Laplace transform becomes a Fourier transform,
\begin{eqnarray}\label{eq:LaplaceFourierb}
\widetilde{\mathcal{O}}(s_L)&=&i\hat{\mathcal{O}}(s_L)=\frac{\Omega^b_r}{2\pi}\,\int_{-\infty}^{\infty}\,dt~\mathcal{O}\left(t\right)\,e^{is_L \Omega^b_r t}\,,
\end{eqnarray}
and its inverse is the inverse Fourier transform
\begin{eqnarray}\label{eq:LaplaceFourierbinv}
\mathcal{O}(t)&=&\,\mathcal{P}\int_{-\infty}^{\infty}\,ds_L~\widetilde{\mathcal{O}}\left(s_L\right)\,e^{-is_L \Omega^b_r t}\,,
\end{eqnarray}
Closing the contour in the lower {(upper)} half plane {when $t>0$ ($t<0$)}, we get a double Fourier series from the poles of $\hat{\mathcal{O}}\left(s_L\right)$, namely 
\begin{eqnarray}\label{eq:LaplaceFourierinvrsb}
\mathcal{O} (t)&=&2\pi\,{{\rm{sign}}(t)}\sum\,\underset{s_L}{\rm Res}\,\left[\hat{\mathcal{O}}(s_L)\,e^{-is_L\Omega^b_r t}\right]\,.
\end{eqnarray}
In conclusion, the knowledge of $\hat{\mathcal{O}}(s_L)$ (and a particular integration contour $\Gamma_r$) allows us to reproduce time-dependent $\mathcal{O}\left(t\right)$ in both the unbound and bound regimes, where in the latter case the motion is doubly periodic. In other words, $\hat{\mathcal{O}}(s_L)$ is a ``universal expression" whose inverse {Laplace} (inverse {Fourier}) transform gives time-dependent unbound (bound) motion.

To better illustrate the usefulness of Laplace observables, let us focus on a particular example: $\hat{r}(s_L)$, the Laplace observables  corresponding to the time-dependent radial coordinate of an electron in relativistic Keplerian motion. We have
\begin{eqnarray}\label{eq:Laplacer2}
\hat{r}(s_L)&=&\lim_{\epsilon\rightarrow 0^+}\frac{p}{e^2-1}\frac{e}{{2}\gamma^2_{\infty}}\,\left[{\frac{1}{(s^\epsilon_L)^2 \sin(\pi s^\epsilon_L)}\frac{d}{de}J_{s^\epsilon_L}\left(-e\gamma^2_{\infty} s^\epsilon_L\right)+\left(s_L\to-s_L\right)}\right]\,,
\end{eqnarray}
with $s^\epsilon_L=s_L+\epsilon$. Here $p,e$ are constant functions of $E,\,L$ given in the next section, together with the derivation of \eqref{eq:Laplacer2} and an exposition of the relativistic Kepler problem. {Moreover, $\left(s_L\to-s_L\right)$ means a similar term to the preceding one but with the sign of $s_L$ flipped.} Here we merely use \eqref{eq:Laplacer2} to illustrate its usefulness both in the bound and unbound regimes. In the \textit{unbound regime}, we can directly substitute \eqref{eq:Laplacer2} in \eqref{eq:LaplaceFourierinv} and integrate to get the time domain $r(t)$. The result is depicted in the left panel of Fig.~\ref{fig:Unbound_Bound} and compared to an explicit numerical solution. In the \textit{bound regime}, we pick up the real-axis poles of \eqref{eq:Laplacer2} (ignoring the double pole at $0${, and upholding the initial condition $r\left(0\right)=\frac{p}{1+e}$ instead, see footnote~\ref{ft:poleatzerocav}}). We then have
\begin{eqnarray}\label{eq:rtimeub}
r(t)&=&\frac{p}{1-e^2}\,\left[1+\frac{e^2\gamma^2_\infty}{2}-\frac{2e}{\gamma^2_{\infty}}\,\sum_{s_L=1}^\infty\,\frac{1}{s^2_L}\,\frac{d}{de}J_{s_L}\left(e\gamma^2_{\infty} s_L\right)\,\cos(s_L\Omega^b_rt)\right]\,,
\end{eqnarray}
which is indeed the correct time-dependent motion for the bound case, as can be seen in the right panel of Fig.~\ref{fig:Unbound_Bound}. Our other two analytical results are the azimuthal angle $\varphi(t)$ and the radiated EM field from a relativistic Keplerian electron, which are depicted in Fig.~\ref{fig:Unbound_Bound_phi} and Fig.~\ref{fig:Unbound_Bound_At}, respectively. Their computation using the QSM, which is fully analytical and non-perturbative in the EM coupling $K$, is detailed in Section~\ref{section:QSMKep}, which forms the bulk of this work.

\section{Relativistic Keplerian Motion}\label{section:glance}

The Laplace observable \eqref{eq:Laplacer2} corresponds to the time-dependent radius in \textit{relativistic Keplerian motion}. It is unique in the sense that it can be easily computed from the standard solution to the relativistic Kepler problem. The Laplace observables for other time-dependent observables, such as the azimuthal angle $\varphi(t)$ or the emitted EM field $A_\mu(t)$ are not so easy to compute, and we would have to resort to a more advanced method -- the QSM \cite{Khalaf:2023ozy}. In this section we present an overview of the relativistic Kepler problem, as well as a trivial calculation of the Laplace observable \eqref{eq:Laplacer2}.

Consider the motion of a relativistic electron of mass $\mu$ in a Coulomb potential,
\begin{eqnarray}\label{eq:Abckg}
A_0=\frac{K}{r}~~~~,~~~~\vec{A}=0\,,
\end{eqnarray}
where $K=-Qq/4\pi=-\alpha_{EM}(Qe^{-1}_{EM})(qe^{-1}_{EM})$, i.e. Coulomb's constant multiplied by two electric charges. The relativistic Lagrangian is \cite{Carter:1968ks}
\begin{eqnarray}\label{eq:Lag}
\mathcal{L}=\frac{\mu}{2}\eta_{\mu\nu}\frac{dx^\mu}{d\tau}\frac{dx^\nu}{d\tau}+\frac{dx^\mu}{d\tau} A_\mu\,,
\end{eqnarray}
where $\eta_{\mu\nu}$ is the Minkowski metric in spherical coordinates and a mostly plus signature. A simple Legendre transformation gives the Hamiltonian 
\begin{eqnarray}\label{eq:Ham}
H=\frac{1}{2\mu}\eta^{\mu\nu}(p_\mu-A_\mu)(p_\nu-A_\nu)\,,
\end{eqnarray}
where the conjugate momenta are
\begin{eqnarray}\label{eq:pconj}
p_\mu=\mu\left(-\frac{dt}{d\tau}+\frac{K}{\mu r},\frac{dr}{d\tau},r^2\frac{d\theta}{d\tau},r^2\sin^2(\theta)\frac{d\varphi}{d\tau}\right)\,.
\end{eqnarray}
The problem is energy conserving and spherically symmetric and so the energy $p^{t}=E$ and the angular momentum $\vec{L}$ are conserved. Furthermore, the motion is planar, and so we focus without loss of generality on motion in the XY plane $\theta=0$, with $\vec{L}=L\hat{z}$. The conserved angular momentum is given by
\begin{eqnarray}\label{eq:ELkep}
L=\gamma \mu r^2 \dot{\varphi}\,,
\end{eqnarray}
where the (time-dependent) boost $\gamma(t)$ is given by
\begin{eqnarray}\label{eq:gammaKep}
\gamma=\left(1-\beta^2\right)^{-\frac{1}{2}}~~,~~\beta^2=\dot{r}^2+r^2\dot{\varphi}^2\,.
\end{eqnarray}
 In this section a dot indicates a derivative with respect to \textit{coordinate time} $t$ (as opposed to proper time $\tau$). The relativistic EOM derived from the Hamiltonian \eqref{eq:Ham} are
\begin{eqnarray}\label{eq:dots}
\dot{r}=\frac{\sqrt{U_r(r)}}{\gamma \mu}~~,~~\dot{\varphi}=\frac{L}{\gamma \mu r^2}~~,~~\frac{dt}{d\tau}=\gamma=\frac{E+\frac{K}{r}}{\mu}\,,
\end{eqnarray}
where
\begin{eqnarray}\label{eq:Ur}
U_r(r)=E^2+\frac{2EK}{r}-\frac{\mathcal{N}^2}{r^2}-\mu^2\equiv\frac{(\mu^2-E^2)(r-r_{min})(r_{*}-r)}{r^2}\,,
\end{eqnarray}
and $\mathcal{N}=L\sqrt{1-K^2/L^2}$. Here $U_r(r)$ also serves to define the \textit{radial action} $S^r(r)$ as
\begin{eqnarray}\label{eq:Sr}
S^r(r)=\int_{r_{min}}^r\,\sqrt{U_r(r)}\,dr\,,
\end{eqnarray}
where $r_{min}$ is the turnover radius. When $E^2<\mu^2$, the radial motion is bound between $r_{min}$ and $r_{*}$, the other real positive root of $r^2\,U_r(r)$. When $E^2>\mu^2$, $r_{min}$ is the only real positive root of $r^2\,U_r(r)$, and motion is unbound between $r_{min}$ and $\infty$. We also define $\gamma_{\infty}\equiv E/\mu$, so that $\lim_{r\rightarrow\infty}\gamma=\gamma_{\infty}$.

For future reference, we can directly integrate \eqref{eq:Sr} and get an explicit expression for $S^r(r)$,
\begin{eqnarray}\label{eq:Srexp}
S^r(r)=\frac{2EK}{\sqrt{\mu^2-E^2}}{\rm atan}\left(\sqrt{\frac{r-r_{min}}{r_{*}-r}}\right)+\mathcal{N}\left\{\sqrt{\frac{(r-r_{min})(r_{*}-r)}{r_{min}r_{*}}}-2\,{\rm atan}\left(\frac{r_{*}}{r_{min}}\frac{r-r_{min}}{r_{*}-r}\right)\right\}\,.\nonumber\\
\end{eqnarray}
If we set $r=r_{*}$, we get the radial \textit{action variable} $J_r$, which is the radial equivalent of the angular action variable, $L$. For bound motion, it is given by
\begin{eqnarray}\label{eq:Jr}
J_r=\frac{1}{\pi}S^r(r_{*};E,L)=-\mathcal{N}+\frac{EK}{\sqrt{\mu^2-E^2}}\,.
\end{eqnarray}
It is a constant of motion, and we can express the motion in terms of the action variables $(J_r,L)$ instead of $(E,L)$. For unbound trajectories, $J_r$ is defined by analytic continuation.

In the unbound regime. As we shall see, the expression we derive will have a natural analytical continuation to the bound regime. For unbound motion where $E^2>\mu^2$, the trajectory is hyperbolic and is given by \eqref{eq:rtrel} with $e>1$. The azimuthal angle is then in the range $\varphi_\infty\leq \varphi\leq\varphi_\infty$ where $\varphi_\infty=(L/\mathcal{N})\arccos(-1/e)$.
Consequently, the scattering angle for hyperbolic motion is
\begin{eqnarray}\label{eq:scatchi}
\chi=2\varphi_\infty-\pi=\frac{2L}{\mathcal{N}}\arccos\left(-\frac{1}{e}\right)\,.
\end{eqnarray}
\subsection{Time-Dependent Radius}\label{section:radKep}
To compute the universal expression for the radius, we look for the Laplace-space operator
\begin{eqnarray}\label{eq:Laplacer}
\hat{r}(s_L)&=&\,\frac{\Omega_r}{2\pi}\,\int_{-\infty}^{\infty}\,dt\,r(t)\,e^{-s_L \Omega_r t}\,.
\end{eqnarray}
Changing variables from $\alpha_r$ to $\kappa_r$ via \eqref{eq:kbnrubrel} and from $r$ to $\kappa_r$ via \eqref{eq:rtnrubrel}, we get\footnote{{\label{ft:poleatzerocav}The result is restricted to $s_L\neq 0$ to avoid subtleties at $s_L=0$. This implies neglecting all $\delta(s_L)$ contributions. Since it contributes a constant term to  time-dependent observables, it is easy to compensate for it by upholding initial conditions. }} 
\begin{align}\label{eq:Laplacer2t}
&\hat{r}(s_L)=\frac{1}{2\pi}\frac{p}{e^2-1}{\lim_{\epsilon\to0^+}}\int_{-\infty}^{\infty}\,\left(e\cosh\kappa_r-1\right)\,\,\left(e\gamma^2_{\infty}\cosh \kappa_r-1\right)\,e^{-(s_L{+\epsilon\,\rm{sign}(\kappa_r)}) \left(e\gamma^2_{\infty}\sinh\kappa_r-\kappa_r\right)}\,d\kappa_r\nonumber\\[5pt]
&=\lim_{\epsilon\to0^+}\frac{p}{e^2-1}\frac{e}{{2}\gamma^2_{\infty}}\,\left[{\frac{1}{(s^{\epsilon}_L)^2 \sin(\pi s^{\epsilon}_L)}\frac{d}{de}J_{s^{\epsilon}_L}\left(-e\gamma^2_{\infty} s^{\epsilon}_L\right)+\left(s_L\to-s_L\right)}\right]\,,\nonumber\\
\end{align}
where $s^{\epsilon}_L=s_L+\epsilon$, which leads to \eqref{eq:Laplacer2}. Note that we added an $\epsilon$ prescription to force convergence of the integral. We have already seen in the previous section that the resulting universal expression allows us to fully reproduce $r(t)$ both in the unbound and bound regimes.
\subsection{Pericenter Precession and the Boundary-to-Bound Map}
For bound motion, the radial frequency is given by $\Omega^b_r=i\Omega_r$, while azimuthal frequency is then given by
\begin{eqnarray}\label{eq:omegaphi_b}
\Omega^b_\varphi=\Omega^b_r\,\frac{\varphi(r_{*})-\varphi(r_{min})}{\pi}\,,
\end{eqnarray}
where
\begin{eqnarray}
\varphi(r_{*})-\varphi(r_{min})=\int_{r_{min}}^{r_{*}}\,\frac{\dot{\varphi}}{\dot{r}}\,dr=\pi\frac{L}{\mathcal{N}}\,.
\end{eqnarray}
Consequently, the precession of the pericenter is
\begin{eqnarray}\label{eq:peri}
\delta\varphi\equiv2\varphi(r_{*})-2\varphi(r_{min})-2\pi=2\pi\left(\frac{L}{\mathcal{N}}-1\right)\,.
\end{eqnarray}
One can explicitly check that pericenter precession \eqref{eq:peri} is linked to the analytical continuation of the scattering angle \eqref{eq:scatchi} to the bound case via the boundary-to-bound relation \cite{Kalin:2019rwq,Kalin:2019inp}:
\begin{eqnarray}\label{eq:b2b}
\delta\varphi=\chi(L)+\chi(-L)~~,~~E<\mu\,,
\end{eqnarray}
where we defined $e$ and $\mathcal{N}$ so that they are sensitive to the sign of $L$.
\section{Computing Any Laplace Observable with the QSM}\label{section:QSMKep}
In Section~\ref{section:radKep} we presented a particularly simple scenario in which the Laplace observable \eqref{eq:Laplacer2t} was directly calculable as an integral over the hyperbolic eccentric anomaly $\kappa_r$. For generic variables $\hat{\mathcal{O}}(s_L)$ this is not the case, and so we need a robust method to calculate Laplace observables. Luckily, the QSM provides such a robust algorithm. 
\subsection{The QSM Master Equation for Relativistic Motion}

The QSM \textit{Master Equation} \cite{Khalaf:2023ozy} relates any Laplace observable\footnote{The original QSM was demonstrated for bound states, where the Laplace observable is interpreted as a Fourier coefficient.} to the classical limit of a \textit{quantum matrix element}, taken between eigenstates $\left|j_r,\ell,m\right\rangle$ of the corresponding quantum system. More concretely, for any observable $\mathcal{O}$:
\begin{eqnarray}\label{eq:QSMMaster}
\hat{\mathcal{O}}(s_L) =\sum_{\Delta\ell,\Delta m}\,(-1)^{\Delta \ell}\lim_{\hbar\rightarrow 0}\,\left\langle j'_r,\ell',m'\right|\mathcal{O}\left|j_r,\ell,m\right\rangle\,,
\end{eqnarray}
where on the RHS $\mathcal{O}$ is interpreted as an operator whose matrix element is taken between a bra and a ket. We set $(j'_r,\ell',m')=(j_r,\ell,m)-(\Delta j_r,\Delta \ell,\Delta m)$ and $(j_r,\ell,m)=\hbar^{-1}(J_r,L,L_z)$ before taking the $\hbar\rightarrow 0$ limit. Note that for planar motion in the XY plane we can always set $L_z=L$. Finally, $\Delta j_r=-s_L-f_\varphi\Delta\ell$ where $f_\varphi\equiv\Omega_\varphi/\Omega_r=L/\mathcal{N}$.

Using the master equation \eqref{eq:QSMMaster}, the extraction of Laplace observables becomes a streamlined task. For non-relativistic spherical motion, the algorithm to compute time-dependent observables for any spherically symmetric motion is then
\begin{enumerate}
\item Write down the Hamiltonian $H$ and find a complete \textit{quantum} eigenbasis $\left|j_r,\ell,m\right\rangle$ for it.
\item For any observable $\mathcal{O}$, compute its Laplace representation via \eqref{eq:QSMMaster}.
\item To get the time-domain observable, use \eqref{eq:LaplaceFourierinv} for unbound motion and \eqref{eq:LaplaceFourierinvrsb} for bound motion.
\end{enumerate}
This algorithm was demonstrated in \cite{Khalaf:2023ozy} for bound, non-relativistic Keplerian motion, where it was used to compute $r(t)$ and the emitted EM field $A_\mu(t)$ from a classical electron in Keplerian motion.
In this paper, we focus on relativistic motion and so we need to slightly generalize the above algorithm. Importantly, the proof of the Master Equation \eqref{eq:QSMMaster} relies solely on the WKB approximation, which becomes exact in the $\hbar\rightarrow 0$ limit. In fact, it does not rely on a proper quantization of the system. For this reason, it has a straightforward relativistic generalization, in which the Schr\"{o}dinger equation is replaced by the \textit{Klein-Gordon} (KG) one. We note that when we use the word ``quantum" in a relativistic context we do not really mean a proper quantization, as the latter inevitably leads to quantum field theory. Instead, we will see that the solutions to the KG equation satisfy the same simple WKB relation to the classical Hamilton-Jacobi radial action as in the non-relativistic case, and so the QSM can be immediately extended to this regime. In other words, WKB holds regardless of the fact that the KG equation is a \textit{classical field equation} rather than a quantum equation.

The latter relativistic generalization is valid in any spherically symmetric motion, even in curved space. It was demonstrated explicitly for motion in a Schwarzschild background in 
the last section of \cite{Khalaf:2023ozy}. Here we will demonstrate it in the context of \textit{relativistic Keplerian motion}. For that purpose, we consider the KG equation in a background Coulomb potential. The KG equation is then
\begin{eqnarray}\label{eq:KGPhi}
\left(\eta^{\mu\nu}(\partial_\mu-A_\mu)(\partial_\nu-A_\nu)-\frac{\mu^2}{\hbar^2}\right) \Phi=0\,,
\end{eqnarray}
where $\eta^{\mu\nu}={\rm diag}(-1,1,r^2,r^2\sin^2\theta)$ in $(t,r,\theta,\varphi)$ coordinates. The solution to this KG equation is 
\begin{eqnarray}\label{eq:fullCoulsol}
\Phi=e^{-i\hbar^{-1}Et}\,R_{j_r,\ell}(k_{j_r,\ell}r)\,Y_{\ell}^{{m}}\left(\theta,\varphi\right)\,.
\end{eqnarray}
To present the radial function, we first define $\nu$, the relativistically deformed version of the angular momentum quantum number $\ell$, as
\begin{eqnarray}\label{eq:nu}
\nu=\frac{\mathcal{N}}{\hbar}=\sqrt{\left(\ell+\frac{1}{2}\right)^2-\frac{K^2}{\hbar^2}}-\frac{1}{2}\,.
\end{eqnarray}
The energy $E$, asymptotic velocity $\beta_{\infty}$ and wavenumber $k$ are given in terms of the quantum numbers as
\begin{eqnarray}\label{eq:disp}
E_{j_r,\ell}=\mu\gamma^{j_r,\ell}_\infty=\frac{\mu}{\sqrt{1-(\beta^{j_r,\ell}_{\infty})^2}}~~,~~
\beta^{j_r,\ell}_\infty=-i\frac{K}{\hbar\,(j_r+\nu+1)}~~,~~k_{j_r,\ell}=\frac{E_{j_r,\ell}\beta^{j_r,\ell}_{\infty}}{\hbar}\,.~~~
\end{eqnarray}
Finally, the explicit (regular at $r=0$) solution for the radial Coulomb function is
\begin{eqnarray}\label{eq:Regsol}
&&R_{j_r,\ell}(z)=C_{j_r,\ell}\,(2z)^{\nu}\,e^{-i z}\,\,_{1}\widetilde{F}_{1}\left(-j_r,2\nu+2,2iz\right)\,.
\end{eqnarray}
while 
\begin{eqnarray}\label{eq:Cco}
C_{j_r,\ell}\,\equiv\,\sqrt{\frac{2\Omega_r}{\pi\beta^{j_r,\ell}_{\infty}}}\,k_{j_r,\ell}\,e^{i\frac{\pi}{2}(j_r+\nu+1)}\,\left|\Gamma\left(-j_r\right)\right|\,.
\end{eqnarray}
Note that we normalize this solution with the extra square-root factor, which is crucial for the QSM master equation, with $\Omega_r$ being the relativistic radial Keplerian fundamental frequency given by {
\begin{eqnarray}\label{eq:fundrl2}
\Omega_r=\frac{\mu}{K}\left(\gamma^2_{\infty}-1\right)^{\frac{3}{2}}\,.
\end{eqnarray}}
In the WKB limit, we have
\begin{eqnarray}\label{eq:WKBKG}
&&\lim_{\hbar\rightarrow 0}R_{j_r,\ell}(k_{j_r,\ell}r)=\left(\frac{2E\Omega_r}{\pi  r^2\sqrt{U_r(r)}}\right)^{\frac{1}{2}}\,\sin\left({\frac{S^r(r)}{\hbar}+\frac{\pi}{4}}\right)\,,~~
\end{eqnarray}
where $(j_r,\ell,m)=\hbar^{-1}(J_r,L,L)$, and $E=E(J_r,L)$ as in \eqref{eq:Jr}. Similarly to the non-relativistic case, the function $S^r(r)$ appearing in the WKB wave function is non other than the classical \textit{Hamilton-Jacobi} radial action \eqref{eq:Sr}. Its appearance guarantees that the proof of the QSM master equation presented in \cite{Khalaf:2023ozy} generalizes to the relativistic case. Before concluding this part, we note one other technical yet crucial modification required in the relativistic case. Due to the mismatch between coordinate and proper time, the ``wave function" $\Psi_{j_r,\ell,m}$ corresponding to $\left|j_r,\ell,m\right\rangle$ appearing in \eqref{eq:QSMMaster} is not exactly \eqref{eq:fullCoulsol}, but rather proportional to it via
\begin{eqnarray}\label{eq:Psidef}
\Psi_{j_r,\ell,m}(r,\theta,\varphi)\equiv\sqrt{\frac{\gamma^{j_r,\ell}}{\gamma^{j_r,\ell}_\infty}}\,R_{j_r,\ell}(k_{j_r,\ell}r)\,Y_{\ell}^{{m}}\left(\theta,\varphi\right)\,,
\end{eqnarray}
where $\gamma^{j_r,\ell}=dt/d\tau=\gamma^{j_r,\ell}_\infty(1+\frac{K}{{E_{j_r,\ell}}r})$ in the relativistic Keplerian case. With this definition, the matrix element entering the {master equation} \eqref{eq:QSMMaster} is
\begin{eqnarray}\label{eq:ME}
\left\langle j'_r,\ell',m'\right|\mathcal{O}\left|j_r,\ell,m\right\rangle\equiv \int\,d^3r\,\Psi^*_{j'_r,\ell',m'}(r,\theta,\varphi)\,\mathcal{O}\,\Psi_{j_r,\ell,m}(r,\theta,\varphi)\,.
\end{eqnarray}
\subsection{Complexified Coulomb Wave Function}\label{section:compc}
In addition to the regular solution \eqref{eq:Regsol}, there is also an irregular solution, explicitly given by
\begin{eqnarray}
R^{c}_{j_r,\ell}(z)=\sqrt{\frac{2\Omega_r}{\pi\beta^{j_r,\ell}_{\infty}}}\,\frac{k_{j_r,\ell}}{z}\,e^{-i\sigma(z)}(2iz)^{-j_r}\,U(-j_r,2\nu+2,2iz)\,,
\end{eqnarray}
where
\begin{eqnarray}
\sigma(z)=z+n\log(2z)-\frac{\pi \nu}{2}-{\rm arg}\,\Gamma\left(-j_r\right)\,,
\end{eqnarray}
is the Coulomb phase and $n=i(j_r+\nu+1)$. This solution is defined so that
\begin{eqnarray}
-{\rm Im}\left[R^{c}_{j_r,\ell}(z)\right]=R_{j_r,\ell}(z)\,.
\end{eqnarray}
Furthermore, it has the WKB limit
\begin{eqnarray}\label{eq:WKBKGc}
&&\lim_{\hbar\rightarrow 0}R^c_{j_r,\ell}(k_{j_r,\ell}r)\equiv R^{c, WKB}_{j_r,\ell}(k_{j_r,\ell}r)=\left(\frac{2E\Omega_r}{\pi  r^2\sqrt{U_r(r)}}\right)^{\frac{1}{2}}\,\exp\left(-{i\frac{S^r(r)}{\hbar}-\frac{i\pi}{4}}\right)\,.\nonumber\\
\end{eqnarray}
While we will not make use of this complexified solution in our computations for relativistic Keplerian motion, we will use it in Section~\ref{section:Schwarz} as a building block for the semiclassical wave function in Schwarzschild.
\subsection{Azimuthal Motion and Scattering Angle}

To demonstrate the power of the QSM, let us now calculate the Laplace-Fourier coefficients of the classical observable
\begin{eqnarray}
\mathcal{O}^{\varphi}\equiv e^{i\varphi}\,,
\end{eqnarray}
for relativistic Keplerian motion, which includes precession. For bound motion, this results in the precession of $\varphi$ over each radial period. The {m}aster {e}quation \eqref{eq:QSMMaster} applied to this operator gives
\begin{eqnarray}\label{eq:QSMMasterphi}
\hat{\mathcal{O}}^{\varphi}(s_L) &=&\sum_{\Delta\ell,\Delta m}\lim_{\hbar\rightarrow 0}\,(-1)^{\Delta \ell}\,\left\langle j'_r,\ell',m'\right|e^{i\varphi}\left|j_r,\ell,m\right\rangle=\nonumber\\[5pt]
&&-\sqrt{\frac{8\pi}{3}}\sum_{\Delta\ell,\Delta m}\lim_{\hbar\rightarrow 0}\,(-1)^{\Delta \ell}\,\left\langle \ell',m'\right|Y_{1}^{1}\left(\frac{\pi}{2},\phi\right)\left|\ell,m\right\rangle\times \left\langle j'_r,\ell',m'|j_r,\ell,m\right\rangle\,.\nonumber\\
\end{eqnarray}
where $(j_r,\ell,m)=\hbar^{-1}(J_r,L,L)$ and $(j'_r,\ell',m')=(j_r,\ell,m)-(\Delta j_r,{\Delta \ell},\Delta m)$ and $\Delta j_r=-s_L-\Delta\nu$. We also define the quantum number $\nu'=\sqrt{(\ell'+\frac{1}{2})^2-\frac{K^2}{\hbar^2}}-\frac{1}{2}$ so that
\begin{eqnarray}
\Delta \nu=\nu-\nu'=f_\varphi\Delta\ell+\mathcal{O}(\hbar)\,,
\end{eqnarray}
where $f_\varphi=L/\mathcal{N}$. In this case, $\Delta\nu=-L/\mathcal{N}+\mathcal{O}(\hbar)$ since $\Delta\ell=-1$. 
Using the angular matrix element in Appendix~\ref{App:classspherelem}\footnote{{While we actually compute
the classical limit of the matrix elements of $Y_{\ell}^{m} (\theta,\varphi)$ in the Appendix, the result is the same for $Y_{\ell}^{m} (\frac{\pi}{2},\varphi)$ since the motion is constrained to the XY plane.}}, we have 
\begin{eqnarray}\label{eq:QSMMasterphi4}
\hat{\mathcal{O}}^{\varphi}(s) &=&\lim_{\hbar\rightarrow 0}\,\left\langle j'_r,\ell',m'|j_r,\ell,m\right\rangle\,,
\end{eqnarray}
where $\Delta\ell=\Delta m=-1$.
By definition, the radial matrix element equals
\begin{eqnarray}\label{eq:QSMMasterphi2}
\lim_{\hbar\rightarrow 0}\left\langle j'_r,\ell',m'|j_r,\ell,m\right\rangle&=&\lim_{\hbar\rightarrow 0}\,\int_0^{\infty}\,\frac{\gamma^{j_r,\ell}}{\gamma^{j_r,\ell}_{\infty}}\,r^2\,R_{j'_r,\ell'}(k_{j'_r,\ell'}\,r)\,R_{j_r,\ell}(k_{j_r,\ell}\,r)\,dr\,,
\end{eqnarray}

Unpacking the matrix element in \eqref{eq:QSMMasterphi2}, we have
\begin{eqnarray}\label{eq:QSMMasterphi3}
&&\hat{\mathcal{O}}^{\varphi}(s_L) =\lim_{\hbar\rightarrow 0}\,\left\{I_0+\frac{K}{E}\,I_{-1}\right\}\,,\nonumber\\[5pt]
&&I_j\equiv \int_0^{\infty}\,r^{2+j}\,R^*_{j'_r,\ell'}(k_{j'_r,\ell'}\,r)\,R_{j_r,\ell}(k_{j_r,\ell}\,r)\,dr\,.
\end{eqnarray}
where again $J_r,\,L$ and $E$ are related by \eqref{eq:Jr}. $I_j$ itself can be expressed using the \textit{Gordon integral} \cite{Gordon1929,Matsumoto1991},
\begin{equation} \label{eq:IntIdenKummer}
\int_{0}^{\infty}e^{-s r}r^{\rho-1}\,_{1}\widetilde{F}_{1}\left(a;b;pr\right)\,_{1}\widetilde{F}_{1}\left(c;d;qr\right)\,dr =s^{-\rho}\,\frac{\Gamma\left(\rho\right)}{\Gamma\left(b\right)\Gamma\left(d\right)}F_{2}\left(\rho,a,c,b,d;\frac{p}{s},\frac{q}{s}\right)\,,
\end{equation}
where $F_2$ is the second Appell Function \cite{bateman1953higher}. Consequently,
\begin{eqnarray} \label{eq:I0tt}
I_j=&&C_{j'_r,\nu'}C_{j_r,\ell}(2k_{j_r,\ell})^{\nu}(2k_{j'_r,\ell'})^{\nu'}\frac{\Gamma(\nu+\nu'+j+3)}{\Gamma(2\nu+2)\Gamma(2\nu'+2)}\left(ik_{j_r,\ell}+ik_{j'_r,\ell'}\right)^{-\nu-\nu'-j-3}\,\times\nonumber\\[5pt]
&&F_2\left(\nu+\nu'+j+3,-j_r,-j'_r,2\nu+2,2\nu'+2;\frac{2k_{j_r,\ell}}{k_{j_r,\ell}+k_{j'_r,\ell'}},\frac{2k_{j'_r,\ell'}}{k_{j_r,\ell}+k_{j'_r,\ell'}}\right)\,.
\end{eqnarray}
Using the classical limit from Appendix~\ref{AjDeltal} in \eqref{eq:QSMMasterphi3}, we get
\begin{eqnarray}\label{eq:QSMMasterphi5}
&&\hat{\mathcal{O}}^{\varphi}(s_L) =\mathcal{A}_0^{\Delta \ell}(\Delta j_r)+\frac{K}{E}\,\mathcal{A}_{-1}^{\Delta \ell}(\Delta j_r)\,,
\end{eqnarray}
where $\Delta\ell=-1$ and $\Delta j_r=-s_L-\Delta\nu=-s_L+L/\mathcal{N}$. The function $\mathcal{A}_j^{\Delta \ell}(\Delta j_r)$ is given in \eqref{eq:Aeq}, and it has simple poles at integer values of $\Delta j_r{+2\Delta\nu}${, and a pole at $\Delta j_r=-\Delta\nu$}\footnote{{It might seem like there are also poles at non-zero integer $\Delta j_r +\Delta \nu$, but the expression is regular at these points.}}. We ignore the pole at $\Delta j_r=-\Delta\nu$, and uphold the initial condition $\mathcal{O}^{\varphi}\left(0\right)=1$ instead, see footnote~\ref{ft:poleatzerocav}. Substituting the Laplace operator \eqref{eq:QSMMasterphi5} into the inverse transforms \eqref{eq:LaplaceFourierinv} (with $\Gamma_r={0}$) and \eqref{eq:LaplaceFourierinvrsb}, we get the exact time dependence of the azimuthal angle, as shown in Fig.~\ref{fig:Unbound_Bound_phi}, and compare it to explicit numerical solutions. {It is necessary to employ the $\epsilon$ prescription by shifting $s_L\to s^\epsilon_L$ when picking up the poles in the bound case.}

Using the QSM, we can straightforwardly compute the scattering angle $\chi$ for unbound motion. This scattering angle is related to the asymptotic value of $\varphi$ via
\begin{equation}
2i\sin{\frac{\pi-\chi}{2}}=e^{i \varphi(\infty)}-e^{i \varphi(-\infty)}\,.
\end{equation}
Using the final value theorem for two sided Laplace transform \cite{LEPAGEWilburR1961CVat}, one obtains
\begin{equation}
e^{i \varphi(\infty)}-e^{i \varphi(-\infty)}=2\pi\,\underset{s_L\rightarrow {0}}{\rm Res}\left[\hat{\mathcal{O}}^\varphi(s_L)\right]\,.
\end{equation}
Taking the residue of the explicit expression \eqref{eq:QSMMasterphi5}, we reproduce the correct classical result \eqref{eq:scatchi}. Similarly, in the bound case, the Fourier series \eqref{eq:LaplaceFourierinvrsb} for $e^{i\varphi}$ allows us to directly calculate the pericenter precession, 
\begin{eqnarray}\label{eq:LaplaceFourierinvrsbphi}
e^{i\Delta\varphi}=e^{i\left[\varphi(t=T_r)-\varphi(t=0)\right]}&=&2\pi\,\sum_{{k}\in\mathbb{Z}}\,\underset{s_L\rightarrow \Delta\nu{+k}}{\rm Res}\,\left[\hat{\mathcal{O}}^{\varphi}(s_L)\,e^{{-}2\pi i s_L}\right]=\,e^{2\pi i (L/\mathcal{N}-1)}\,,\nonumber\\
\end{eqnarray}
which reproduces the classical result \eqref{eq:peri}, as well as the boundary-to-bound map \eqref{eq:b2b}.

\subsection{All-Order Electromagnetic Waveform from Keplerian Orbit}
As a final demonstration of the QSM, we consider the case of a relativistic classical electron undergoing Keplerian motion. As the electron is accelerating, it is radiating an EM field $A_\mu$. We are interested in the exact computation of this radiated field, assuming that the source electron moves on an exact (relativistic) Keplerian orbit. In \cite{Khalaf:2023ozy}, a similar computation was performed for \textit{non-relativistic}, \textit{bound} motion. Here we generalize the latter to fully relativistic motion, and derive a Laplace operator that allows us to compute the radiated $A_\mu$ both for bound and unbound motion. The added value of this computation is threefold:
\begin{itemize}
\item A generalization to relativistic motion{.}
\item A way to benchmark the relativistic \textit{unbound} results vs PL calculations, while the Laplace operator is directly applicable both to bound and unbound motion{.}
\item The relativistic Keplerian calculation will be the `atomic' unit in the computation of the Schwarzschild case. 
\end{itemize}
We now present our exact computation of the radiated $A_\mu$ from a relativistic Keplerian orbit. 
\subsubsection{Current and Green's Function}
The EM field generated by a Keplerian electron is given by \cite{Jackson1998}
\begin{equation}\label{eq:generalAmu}
A^{\mu}(t,\vec{x})=\int\,d^{4}x'\,G_{{\rm ret}}^{\mu\nu}(t,\vec{x};t',\vec{x}')\,J_{\nu}(t',\vec{x}')\,\,,
\end{equation}
where $G_{{\rm ret}}^{\mu\nu}(t,\vec{x};t',\vec{x}')$ is the retarded Green's function of the EM field, and $J_{\nu}(t',\vec{x}')$ is the 4-current density from the motion of the electron. Explicitly, $J_\mu$ is given by
\begin{equation}\label{eq:current}
J^{\mu}\left(t',\vec{x}'\right)=\frac{q}{\mu\gamma(t')}\,p^{\mu}(t')\,\delta^{\left(3\right)}\left(\vec{x}'-\vec{r}\left(t'\right)\right)\,,
\end{equation}
where $p^\mu(t')$ and $\vec{r}(t')$ are the 4-momentum and 3-position at time $t'$ of the electron. Furthermore, the boost $\gamma(t')$ is given by \eqref{eq:dots}.
On the other hand, the retarded Green's function for the EM field is famously \cite{Jackson1998}
\begin{equation}
G_{{\rm ret}}^{\mu\nu}(t,\vec{x};t',\vec{x}')=\eta^{\mu\nu}\frac{\Theta(t-t')}{4\pi R}\,\delta(t-t'-R)\,,
\end{equation}
where $R=|\vec{x}-\vec{x}'|$, and $\eta^{\mu\nu}$ is the Minkowski metric in mostly-plus signature. For our purposes, it is convenient to use the Fourier representation of the delta function
\begin{equation}
G_{{\rm ret}}^{\mu\nu}(t,\vec{x};t',\vec{x}')=\eta^{\mu\nu}\frac{\Theta(t-t')}{2\pi}\,\int_{-\infty}^{\infty}\,d\omega'\,e^{-i\omega'(t-t')}\,\frac{e^{i\omega' R}}{4\pi R}\,.
\end{equation}
We now expand $e^{i\omega' R}/(4 \pi R)$ in multipoles and get
\begin{eqnarray}\label{eq:multi}
&&G_{{\rm ret}}^{\mu\nu}(t,\vec{x};t',\vec{x}')=\eta^{\mu\nu}\frac{\Theta(t-t')}{2\pi}\,\int_{-\infty}^{\infty}\,d\omega'\,e^{-i\omega'(t-t')}\,\times~~~~~~~~\nonumber\\[5pt]
&&~~~~~~~~~~~~~~~~~\left\{ i\omega'\,\sum_{\ell_{\gamma}=0}^{\infty}\,j_{\ell_{\gamma}}(\omega'\, r_{<})\,h_{\ell_{\gamma}}^{(1)}\left(\omega'\, r_{>}\right)\,\sum_{m_{\gamma}=-\ell_{\gamma}}^{\ell_{\gamma}}\,Y_{\ell_{\gamma}}^{m_{\gamma}}(\theta',\varphi')Y_{\ell_{\gamma}}^{m_{\gamma}*}(\theta,\varphi)\right\} \,,~~~
\end{eqnarray}
where $\Theta$ is the Heaviside function, $Y_{\ell_{\gamma}}^{m_{\gamma}}$
is a spherical harmonic, $j_{\ell_{\gamma}}$ is a spherical Bessel function,
and $h_{\ell_{\gamma}}^{(1)}$ is a spherical Hankel function of the first kind. Finally, $\{r_{<},r_{>}\}=\{{\rm{min}}(|\vec{x}|,|\vec{x}'|),{\rm{max}}(|\vec{x}|,|\vec{x}'|)\}$.
From here on, we will focus on the case where $|\vec{x}|$ is larger than $|\vec{x}'|$, so that $r_{>}=|\vec{x}|,\,r_{<}=|\vec{x}'|$. Furthermore, we are interested in the radiation region of large $|\vec{x}|$, where $h_{\ell_\gamma}(\omega' |\vec{x}|)\rightarrow (-i)^{\ell_\gamma+1}\,e^{i\omega' |\vec{x}|}/(\omega' |\vec{x}|)$. Substituting the current \eqref{eq:current} and the multipole-expanded Green's function \eqref{eq:multi} in the general expression \eqref{eq:generalAmu}, we get
\begin{eqnarray}\label{eq:eq:AmuM}
&&A^{\mu}(t,\vec{x})=\frac{q}{2\pi\mu|\vec{x}|}\int_{-\infty}^{\infty}\,d\omega'\,e^{-i\omega'u}\,\sum_{\ell_{\gamma}=0}^{\infty}\,\sum_{m_{\gamma}=-\ell_{\gamma}}^{\ell_{\gamma}}\,(-i)^{\ell_\gamma}\,Y_{\ell_{\gamma}}^{m_{\gamma}*}(\theta,\varphi)\times\nonumber\\[5pt]
&&\int_{-\infty}^t\,dt'\,e^{i\omega't'}\,\mathcal{M}_{\ell_{\gamma},m_{\gamma}}^{\mu}(\omega',t')\,,
\end{eqnarray}
where $u\equiv t-|\vec{x}|$ is the retarded coordinate of the observation point. Here, the only part of the expression containing the details of the source (the current from the Keplerian electron) are encapsulated in the multipole factor $\mathcal{M}_{\ell_{\gamma},m_{\gamma}}^{\mu}(\omega',t')$. It is explicitly given by
\begin{equation} \label{Amu1}
\mathcal{M}_{\ell_{\gamma},m_{\gamma}}^{\mu}(\omega',t')\equiv j_{\ell_\gamma}\left[\omega' r(t')\right]\,Y_{\ell_{\gamma}}^{m_{\gamma}}\left[\hat{r}(t')\right]\,\frac{\mu p^{\mu}(t')}{E+\frac{K}{r\left(t'\right)}}\,,
\end{equation}
where again $r(t'),p^{\mu}(t')$ refer to the relativistic Keplerian motion of the source, and we used the explicit expression $\gamma(t')=\frac{E+\frac{K}{r(t')}}{\mu}$. 
\iffalse
\begin{align}
    A^{\mu}(t,\vec{x}) = \frac{1}{4\pi |\vec{x}|} \int^\infty_{-\infty} d\omega' \widetilde{A}(\omega',\vec{x}) e^{-i\omega'u},
\end{align}
where $u=t-|\vec{x}|$ is the retarded time and the frequency-domain $\tilde{A}(\omega',\vec{x})$ reads
\begin{eqnarray}\label{eq:AmuM}
\widetilde{A}^{\mu}(\omega,\vec{x})=\frac{{2} q}{\mu}\,\sum_{\ell_{\gamma}=0}^{\infty}\,\sum_{m_{\gamma}=-\ell_{\gamma}}^{\ell_{\gamma}}\left(-i\right)^{\ell_{\gamma}}\,Y_{\ell_{\gamma}}^{m_{\gamma}*}(\theta,\varphi)\,\int_{-\infty}^{t}\,dt'\,e^{i\omega't'}\,\mathcal{M}_{\ell_{\gamma},m_{\gamma}}^{\mu}(\omega',t')\,.
\end{eqnarray}

\begin{eqnarray}\label{eq:AmuM}
&&A^{\mu}(t,\vec{x})=\nonumber\\[5pt]
&&\frac{q}{\mu r}\,\sum_{\ell_{\gamma}=0}^{\infty}\,\sum_{m_{\gamma}=-\ell_{\gamma}}^{\ell_{\gamma}}\left(-i\right)^{\ell_{\gamma}}\,Y_{\ell_{\gamma}}^{m_{\gamma}*}(\theta,\varphi)\,\int_{-\infty}^{\infty}\,dt'\,\frac{\Theta(t-t')}{2\pi}\,\int_{-\infty}^{\infty}\,d\omega'\,e^{-i\omega'(t-t')}\,e^{i \omega' |\vec{x}|}\,\mathcal{M}_{\ell_{\gamma},m_{\gamma}}^{\mu}(\omega',t')\,.\nonumber\\
\end{eqnarray}
\fi
\subsubsection{Laplace Transform of $A^\mu$}
In keeping with our general algorithm, we now present $A_\mu(t,\vec{x})$ as an inverse Laplace transform \eqref{eq:LaplaceFourierinv} of its Laplace observable $\hat{A}^\mu(s,\vec{x})$, namely 
\begin{equation}\label{eq:ALap}
A^{\mu}(t,\vec{x})=-i\,\mathcal{P}\int_{-i\infty}^{i\infty}\,\hat{A}^\mu(s_L,\vec{x})\,\,e^{s_L \Omega_r t}\,ds_L\,.
\end{equation}
The expression \eqref{eq:eq:AmuM} allows us to relate the Laplace observable for $A^\mu$ with the one for the multipole factor 
$\mathcal{M}_{\ell_{\gamma},m_{\gamma}}^{\mu}(\omega',t')$. The latter is defined as
\begin{equation}\label{eq:MLap}
\mathcal{M}_{\ell_{\gamma},m_{\gamma}}^{\mu}(\omega',t')=-i\mathcal{P}\int_{-i\infty}^{i\infty}\,\hat{\mathcal{M}}_{\ell_{\gamma},m_{\gamma}}^{\mu}(\omega',s_L)\,e^{s_L\Omega_r t'}\,ds_L\,.
\end{equation}
To see the relation between the two, we substitute the definition \eqref{eq:MLap} in \eqref{eq:eq:AmuM}, and get  
\begin{eqnarray}\label{eq:AmuM2}
A^{\mu}(t,\vec{x})=-i\,&&\mathcal{P}\int_{-i\infty}^{i\infty}\,ds_L\,\int_{-\infty}^{\infty}\,d\omega'\,e^{-i\omega'(t-|\vec{x}|)}\,\left\{\frac{q}{\mu |\vec{x}|}\,\sum_{\ell_{\gamma}=0}^{\infty}\sum_{m_{\gamma}=-\ell_{\gamma}}^{\ell_{\gamma}}\left(-i\right)^{\ell_{\gamma}}\,Y_{\ell_{\gamma}}^{m_{\gamma}*}(\theta,\varphi)\,\right.\nonumber\\
&&\left.\hat{\mathcal{M}}_{\ell_{\gamma},m_{\gamma}}^{\mu}(\omega',s_L)\int\,dt'\,\frac{\Theta(t-t')}{2\pi}\,e^{\left(i \omega'+s_L\Omega_r\right) t'}\right\}\,.
\end{eqnarray}
We can now carry out the $t'$ and $\omega'$ integrals directly to obtain the form \eqref{eq:ALap} with
\begin{equation}\label{eq:Ahat}
\hat{A}^{\mu}\left(s_L,\vec{x}\right)=\frac{q}{\mu |\vec{x}|}\,\sum_{ \ell_{\gamma}=0}^{\infty}\,\sum_{m_{\gamma}=-\ell_{\gamma}}^{\ell_{\gamma}}\left(-i\right)^{\ell_{\gamma}}\,Y_{\ell_{\gamma}}^{m_{\gamma}*}(\theta,\varphi)\,\,e^{-s_L\Omega_r\,|\vec{x}|}\,\hat{\mathcal{M}}_{\ell_{\gamma},m_{\gamma}}^{\mu}(\omega,s_L)\,\,,
\end{equation}
where $\omega=is_L
\Omega_r$. 
\subsubsection{QSM Result}
Employing the master equation~\eqref{eq:QSMMaster} for $\hat{\mathcal{M}}_{\ell_{\gamma},m_{\gamma}}^{\mu}$, we get 
\begin{align}\label{eq:MQSM}
&\hat{\mathcal{M}}_{\ell_{\gamma},m_{\gamma}}^{\mu}(\omega,s_L)=\sum_{\Delta \ell,\Delta m}\,(-1)^{\Delta\ell}\lim_{\hbar\rightarrow0}\,\left\langle j_{r}',\ell',m'\right|\,j_{\ell}(\omega r)\,Y_{\ell_{\gamma}}^{m_{\gamma}}(\hat{r})\,\frac{\mu\,p^{\mu}}{E+\frac{K}{r}}\,\left|j_{r},l,l\right\rangle\,,
\end{align}
% \nonumber\\ &\hspace{5cm}\equiv(-1)^{\Delta\ell}\mathcal{M}_{\Delta \ell,\Delta m,\ell_{\gamma},m_{\gamma}}^{\mu}\left(s,J_{r},L\right)
where $\left(j_{r}',\ell',m'\right)=\left(j_{r},\ell,m\right)-\left(\Delta j_r,\Delta \ell,\Delta m\right)$, and $\Delta j_r=-s_L-\Delta\nu$. 
% Hence, we obtain
% Using a similar (but slightly generalized to relativistic+unbound) derivation as the first QSM paper, show that $\hat{A}_{\mu}(\omega)\equiv\int A_{\mu}\left(t\right)e^{i\omega (t-r)}\,\,dt$ is given by (in the asymptotic "radiation" region of large $r$)  
% \begin{equation} \label{GenAmu}
% \hat{A}^{\mu}\left(\omega\right)=\frac{2\pi q}{\Omega_r\mu r}\,\sum_{\ell_{\gamma}=0}^{\infty}\sum_{m_{\gamma}=-\ell_{\gamma}}^{\ell_{\gamma}}(-i)^{\ell_{\gamma}}\,Y_{\ell_{\gamma}}^{m_{\gamma}*}\left(\theta,\varphi\right)\,\sum_{\Delta \ell}\,(-1)^{\Delta\ell}\mathcal{M}_{\Delta \ell,\ell_{\gamma},m_{\gamma}}^{\mu}\left(\omega,J_r,L\right)\,,
% \end{equation}
% where
% \begin{equation}
% \,\mathcal{M}_{\Delta \ell,\ell_{\gamma},m_{\gamma}}^{\mu}\left(\omega,J_r,L\right)=\,\lim_{\hbar\rightarrow0}\,\sum_{\Delta m}\left\langle j_r',\ell',m'\right|\,j_{\ell}(\omega r)\,Y_{\ell_{\gamma}}^{m_{\gamma}}(\hat{r})\,\frac{\mu\, p^{\mu}}{E+\frac{K}{r}}\,\left|j_r,l,l\right\rangle \,,
% \end{equation}
% and $\left(j_r',\ell',m'\right)=\left(j_r,\ell,m\right)-\left(s,\Delta \ell,\Delta m\right)$ and $s=-\Delta \nu + i\frac{\omega}{\Upsilon_r}$. 
Using angular momentum identities, we can unpack the latter expression as
\begin{eqnarray} \label{eq:CurlyMt}
\hat{\mathcal{M}}_{\ell_{\gamma},m_{\gamma}}^{t}=\sum_{\Delta \ell,\Delta m}\,
(-1)^{\Delta\ell}\mu\lim_{\hbar\rightarrow0}\,\left\langle j_r',\ell'\right|j_{\ell_{\gamma}}\left(\omega r\right)\,\left|j_r, l\right\rangle \,\lim_{\hbar\rightarrow0}\,\left\langle \ell',m'\right|\,Y_{\ell_{\gamma}}^{m_{\gamma}}(\theta,\varphi)\,\left|l,l\right\rangle 
\end{eqnarray}
for the temporal component and
\begin{eqnarray}
&&i\mathcal{\vec{\hat{M}}}_{\ell_{\gamma},m_{\gamma}}=\sum_{\Delta \ell,\Delta m}\,(-1)^{\Delta \ell}\,\times\nonumber\\[5pt]
&&\left\{\lim_{\hbar\rightarrow0}\left\langle j_r',\ell'\right|\,\frac{\mu j_{\ell_{\gamma}}\left(\omega r\right)}{r\left(E+\frac{K}{r}\right)}\left|j_r,l\right\rangle ~\,\lim_{\hbar\rightarrow0}\,\hbar\sum_{q=-1}^{1}\,\left\langle \ell',m'\right|Y_{\ell_{\gamma}}^{m_{\gamma}}(\theta,\varphi)(\overrightarrow{\nabla}_{\Omega})_{q}\left|l,l\right\rangle \,\vec{\varepsilon}_{q}+\,\right.\nonumber\\[5pt]
&&\left.~~\lim_{\hbar\rightarrow0}\hbar\left\langle j_r',\ell'\right|\,\frac{\mu j_{\ell_{\gamma}}\left(\omega r\right)}{E+\frac{K}{r}}\,\partial_{r}\left|j_r,l\right\rangle ~\,\lim_{\hbar\rightarrow0}\sum_{q=-1}^{1}\,\left\langle \ell',m'\right|Y_{\ell_{\gamma}}^{m_{\gamma}}(\theta,\varphi)(\hat{r})_{q}\left|l,l\right\rangle \,\vec{\varepsilon}_{q}\right\}\,\,,
\end{eqnarray}
for the spatial components. Here $\vec{\varepsilon}_{0}=\hat{z},\,\vec{\varepsilon}_{\pm}=\tfrac{1}{\sqrt{2}}\left(i\hat{x}\pm\hat{y}\right)$
and $(\vec{v})_{q}=\vec{v}\cdot\vec{\varepsilon}_{q}^{\,*}$. The classical limit of the spherical matrix elements is given in Appendix.~\ref{App:classspherelem}, and they involve the selection rule on $\Delta m$.
% The classical limit of the spherical elements was calculated in the first QSM paper, 
% \begin{equation} \label{eq:classYlmelem}
% \lim_{\hbar\rightarrow0}\left\langle \ell',m'\right|Y_{\ell_{\gamma}}^{m_{\gamma}}\left(\hat{r}\right)\left|l,l\right\rangle =\delta_{\ell',m'}\delta_{-\Delta \ell,m_{\gamma}}\,f_{\ell_{\gamma},m_{\gamma}}\,,
% \end{equation}
% \begin{equation}
% \lim_{\hbar\rightarrow0}\left\langle \ell',m'\right|Y_{\ell_{\gamma}}^{m_{\gamma}}\left(\hat{r}\right)\left(\hat{r}\right)_{q}\left|l,l\right\rangle =\delta_{\ell',m'}\delta_{-\Delta \ell,m_{\gamma}+q}\,\frac{1}{\sqrt{2}}\left(\delta_{q,1}-\delta_{q,-1}\right)\,f_{\ell_{\gamma},m_{\gamma}}\,,
% \end{equation}
% \begin{equation}
% \lim_{\hbar\rightarrow0}\,\hbar\left\langle \ell',m'\right|Y_{\ell_{\gamma}}^{m_{\gamma}}\left(\hat{r}\right)\left(\vec{\nabla}_{\Omega}\right)_{q}\left|l,l\right\rangle =-\frac{L}{\sqrt{2}}\delta_{\ell',m'}\delta_{-\Delta \ell,m_{\gamma}+q}\,\left(\delta_{q,1}+\delta_{q,-1}\right)\,f_{\ell_{\gamma},m_{\gamma}}\,,
% \end{equation}
% where
% \begin{equation}\label{flm}
% f_{\ell_{\gamma},m_{\gamma}}\equiv Y_{\ell_\gamma}^{-m_\gamma}\left(\frac{\pi}{2},0\right)=\frac{\cos\left[\tfrac{\pi(\ell_{\gamma}-m_{\gamma})}{2}\right]}{2\pi}\sqrt{\frac{\left(2\ell_{\gamma}+1\right)\Gamma\left(\frac{\ell_{\gamma}+m_{\gamma}+1}{2}\right)\Gamma\left(\frac{\ell_{\gamma}-m_{\gamma}+1}{2}\right)}{\Gamma\left(\frac{\ell_{\gamma}+m_{\gamma}}{2}+1\right)\Gamma\left(\frac{\ell_{\gamma}-m_{\gamma}}{2}+1\right)}}\,\,.
% \end{equation}
Using the power series of the spherical Bessel function, the classical limit of the radial matrix elements is obtained by
\begin{eqnarray} \label{eq:besseljmatelem}
\lim_{\hbar\rightarrow0}&&\left\langle j_r',\ell'\right|j_{\ell_{\gamma}}\left(\omega r\right)\left|j_r, l\right\rangle = 2^{\ell_{\gamma}}\sum_{\kappa=0}^{\infty}\frac{\left(-1\right)^{\kappa}\left(\kappa+\ell_{\gamma}\right)!\,\omega^{2\kappa+\ell_{\gamma}}}{\kappa!\left(2\kappa+2 \ell_{\gamma}+1\right)!}\left[\mathcal{A}_{2\kappa+\ell_{\gamma}}^{\Delta \ell}\left(\Delta j_r\right)+\frac{K}{E}\mathcal{A}_{2\kappa+\ell_{\gamma}-1}^{\Delta \ell}\left(\Delta j_r\right)\right]\,,\nonumber\\
% \lim_{\hbar\rightarrow0}\,C_{j_r',\ell'}C_{j_r,l }\left(I_{2\kappa+2 \ell_{\gamma}}+\frac{K}{E} I_{2\kappa+2 \ell_{\gamma}-1}\right)\,, 
\end{eqnarray}
as well as
\begin{equation}
\lim_{\hbar\rightarrow0}\left\langle j_r',\ell'\right|\,\frac{\mu j_{\ell_{\gamma}}\left(\omega r\right)}{r\left(E+\frac{K}{r}\right)}\left|j_r,l\right\rangle = \frac{\mu}{E} 2^{\ell_{\gamma}}\sum_{\kappa=0}^{\infty}\frac{\left(-1\right)^{\kappa}\left(\kappa+\ell_{\gamma}\right)!\,\omega^{2\kappa+\ell_{\gamma}}}{\kappa!\left(2\kappa+2 \ell_{\gamma}+1\right)!}\,\mathcal{A}_{2\kappa+\ell_{\gamma}-1}^{\Delta \ell}\left(\Delta j_r\right)\,,
% \,\lim_{\hbar\rightarrow0}\,C_{j_r',\ell'}C_{j_r,l }\,I_{2\kappa+2 \ell_{\gamma}-1}\,,
\end{equation}
\begin{equation}
\lim_{\hbar\rightarrow0}\hbar\left\langle j_r',\ell'\right|\,\frac{\mu j_{\ell_{\gamma}}\left(\omega r\right)}{E+\frac{K}{r}}\,\partial_{r}\left|j_r,l\right\rangle = \frac{\mu}{E} 2^{\ell_{\gamma}}\sum_{\kappa=0}^{\infty}\frac{\left(-1\right)^{\kappa}\left(\kappa+\ell_{\gamma}\right)!\,\omega^{2\kappa+\ell_{\gamma}}}{\kappa!\left(2\kappa+2 \ell_{\gamma}+1\right)!}\,\mathcal{B}_{2\kappa+\ell_{\gamma}}^{\Delta \ell}\left(\Delta j_r\right)\,,
\end{equation}
where
\begin{eqnarray}\label{eq:AB}
\mathcal{A}_j^{\Delta \ell}\left(\Delta j_r\right)&\equiv&\lim_{\hbar\rightarrow0}\,I_{j}=\lim_{\hbar\rightarrow0}\left\langle j_r-\Delta j_r,\ell-\Delta\ell\right|\frac{r^j}{1+\frac{K}{Er}}\left|j_r,l\right\rangle\,,\nonumber\\[5pt]
\mathcal{B}_{j}^{\Delta \ell}\left(\Delta j_r\right)&\equiv&\lim_{\hbar\rightarrow0}\hbar\left\langle j_r-\Delta j_r,\ell-\Delta\ell\right|\,\frac{r^j}{1+\frac{K}{Er}}\,\partial_{r}\left|j_r,l\right\rangle
\end{eqnarray}
are calculated in \eqref{eq:Aeq} and \eqref{eq:Beq}, respectively. With these explicit expressions, we can evaluate $\hat{\mathcal{M}}_{\ell_{\gamma},m_{\gamma}}^{\mu}(\omega,s_L)$ in \eqref{eq:MQSM} exactly. Using
\eqref{eq:Ahat} and \eqref{eq:ALap}, we can compute the emitted $A_\mu$ from a relativistic Keplerian electron, non-perturbatively in $K$. As an example, in Fig.~\ref{fig:Unbound_Bound_At} we plot the analytical result for $\hat{A}^\mu$ for large $\left|\vec{x}\right|$, both in the bound and in the unbound cases. Comparing with an explicit numerical solution, we get an exact match in both cases. 
%===========================
\section{Perturbative Check of Emitted $A^\mu$}\label{section:pert} Though our analytical result for $A^t$ is verified by comparison to numerics in Fig.~\ref{fig:Unbound_Bound_At}, it is nevertheless instructive to cross-check our unbound result with an analytic PL calculation in the unbound regime. In this section, this cross-check is done at the 1PL order. To do this, we have to expand the non-perturbative QSM result to first order in $K$ and resum the result. This resummation is done analytically in the non-relativistic limit, and numerically in the relativistic one. Let us first derive the leading PL result, which follows closely to the gravitational analog~\cite{Goldberger:2016iau,Jakobsen:2021smu,DeAngelis:2023lvf}. We will adopt the worldline approach~\cite{Goldberger:2016iau}, but the integrand can be obtained from the amplitudes-based approach as well~\cite{Kosower:2018adc}.

\subsection{1PL Result}\label{section:1pll}
We start with the EOM in the Lorentz gauge,
\begin{align}\label{eq:CSEOM}
    \Box A^\mu(x) =& -J^\mu(x), \\
    J^\mu(x) =& \sum_{a=0,1} q_a \int {d\tau}_a\, v_a^\mu(\tau) \,\delta(x-x_a(\tau)), \\
    \frac{d^2 x^\mu_a}{d\tau^2} =& \frac{q_a}{m_a} \,F^\mu{}_\nu\left(x_a\right) \,v^\nu_a\,,
\end{align}
where $a=0,1$ labels the particle, $q_a,m_a,x_a,\tau_a$ are the charge, mass, position, and proper time of particle $a$. The proper velocity $v^\mu_a=dx_a^\mu/d\tau$. The particle $0$ is the one that induces the static Coulomb background,~\eqref{eq:Abckg}. The first two are equivalent to~\eqref{eq:generalAmu} and \eqref{eq:current}, where the latter is matched by integrating out the proper time. We consider a two-particle system but will take the probe limit to match the calculation in earlier sections.

For the classical scenario, we are interested in the asymptotic value of the retarded gauge field at large $|\vec{x}|$ and finite retarded time $u=t-|\vec{x}|$.
Since the current is relatively localized in this case, we have
\begin{align}
    t-t'-R = u+\hat{x}\cdot \vec{x}'-t' +\order{|\vec{x}|^{-1}}.
\end{align}
Combining the above expansion and the current in momentum space with~\eqref{eq:generalAmu} yields
\begin{align}
    A^\mu(x) =& \int \dfk \int d^4x' \frac{\Theta(t-t'-R)}{4\pi R} \delta(t-t'-R) J^\mu(k) e^{ik\cdot x'} \nn \\
    =& \frac{1}{4\pi |\vec{x}|}\int \frac{d\omega}{2\pi} \int \dtk \int d^3\vec{x}' e^{-i\omega u} e^{-i(\omega\hat{x}-\vec{k})\cdot \vec{x}'} J^\mu(k) +\order{|\vec{x}|^{-2}} \nn \\
    =& \frac{1}{4\pi |\vec{x}|}\int^\infty_{-\infty} d\omega e^{-i\omega u} \frac{J^\mu (k^*)}{2\pi} +\order{|\vec{x}|^{-2}}
    \label{eq:asym_Amu}
\end{align}
where $\hat{x}$ is the unit vector pointing in the direction of $\vec{x}$ and $k^*=\omega(1,\hat{x})$. We find the well-known result that the asymptotic value of the gauge field is given by the on-shell part of the current,
\begin{align}
\label{eq:Atilde}
    \widetilde{A}^\mu (\omega,\vec{x}) = \frac{J^\mu (k^*)}{8\pi^2 |\vec{x}|}.
\end{align}

To derive the background gauge field, we first consider the gauge field induced by the particle $0$ that is at rest, $x_0^\mu = v_0 t$ with $v_0 =(1,0,0,0)$, with charge $q_0=Q$. The induced gauge field reads\footnote{We only include the gauge field from particle 0, but not particle 1 (the probe particle). The gauge field from the probe particle is independent from $K$ but only has support at $\omega=0$.}
\begin{align}
    A^\mu (k) =& \frac{Q v_0^\mu}{k^2}  \hat{\delta}(k\cdot v_0), \\
    A^\mu (t,\vec{x}) =& \frac{Q v_0^\mu}{4\pi |\vec{x}|}
\end{align}
where $\hat{\delta}(x)=2\pi\,\delta(x)$, and $1/k^2=1/(\vec{k}^2-(\omega+i0)^2)$ is the retarded Green's function in momentum space. In the second line we derive the position space result.
%Matching it to the result in.~\eqref{eq:Abckg} leads to $K=-Q/4\pi$. 
We consider the probe limit where the particle $0$ is fixed at the origin in the presence of particle 1. From now on, we only consider particle 1, so we drop the particle label for simplicity. $K$ is set to be positive for an attractive potential. The mass of particle $1$ is $m_1=\mu$ to match the QSM notation.

The PL perturbation expands the trajectory around a free-particle one,
\begin{align}\label{eq:traj}
    x^\mu (\tau) &= b^\mu + v^\mu \tau +\delta x^{\mu}_a(\tau_a) \,,
\end{align}
where $v$ is the proper velocity of the free particle with $v^2=1$. The correction to the trajectory follows from EOM
\begin{align}
    \frac{d^2 \delta x^\mu}{d\tau^2} =&
    \frac{iq}{\mu} \int \dfk \, e^{ik\cdot x}\, \left(k^\mu A_\nu(k)- k_\nu A^\mu(k) \right) v^\nu \\
    =& \frac{iQ q}{\mu} \int \dfk e^{ik\cdot b} e^{ik\cdot v \tau} \hat{\delta}(k\cdot v_0)
    \frac{1}{k^2} \left((v\cdot v_0)k^\mu-(k\cdot v) v^\mu_0 \right)
\end{align}
Integrating the above yields
\begin{align}
    \delta v^\mu
    =& \frac{iQq}{\mu} \int \dfk e^{ik\cdot b} e^{ik\cdot v \tau} \hat{\delta}(k\cdot v_0)
    \frac{-i}{k^2 (k\cdot v-i0)} \left((v\cdot v_0)k^\mu-(k\cdot v) v^\mu_0 \right) \\
    \delta x^\mu 
    =& \frac{iQ q}{\mu} \int \dfk e^{ik\cdot b} e^{ik\cdot v \tau} \hat{\delta}(k\cdot v_0)
    \frac{-1}{k^2 (k\cdot v-i0)^2} \left((v\cdot v_0)k^\mu-(k\cdot v) v^\mu_0 \right),
\end{align}
where $\delta v^\mu=\frac{d\delta x^\mu}{d\tau} $.
We use the prescription $k\cdot v \rightarrow k\cdot v-i0$ to enforce the initial condition that $\delta x^\mu=0$ at $\tau= -\infty$.

\begin{figure}[t]
    \centering \includegraphics[trim=1cm 26cm 51cm 0cm, clip, width=0.3\textwidth]{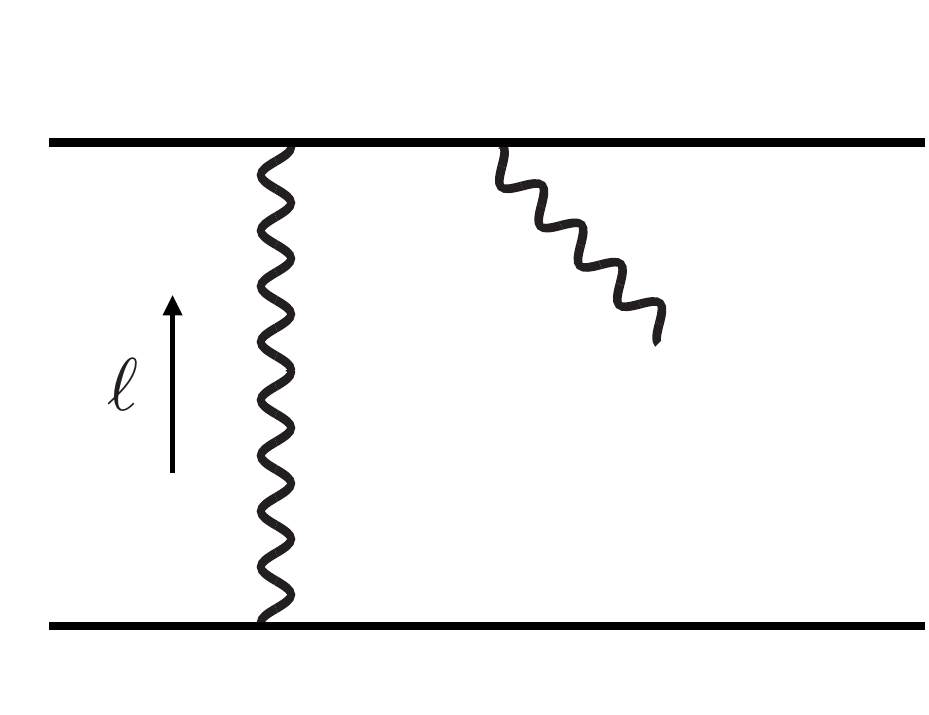}
     \caption{The diagram for the radiation at 1PL order. The wiggle lines are gauge fields while the two solid lines are the two charged particles. We consider the probe limit where the bottom solid line is much heavier than the top solid one. The radiation in the probe limit is given by the emission from the light particle.}\label{fig:PL}
     \vspace*{15pt}
\end{figure}

We are interested in the asymptotic gauge field induced by particle 1, or the on-shell part of the current via~\eqref{eq:asym_Amu}. The full current from the trajectory in~\eqref{eq:traj} reads
\begin{align}
    J^\mu(k) =& q\int d\tau (v^\mu + \delta v^\mu) e^{-ik\cdot (b+v\tau+\delta x)}
\end{align}
At the zeroth order, the free-particle trajectory gives a current proportional to $\delta(\omega)$, so the gauge field is static. The leading non-static gauge field arises from the 1PL perturbation
\begin{align}
    \delta J^\mu(k) 
    =& q\int d\tau e^{-ik\cdot (b+v\tau)} \left(\delta v^\mu - iv^\mu (k\cdot \delta x) \right) \\
    =& \frac{Q q^2}{\mu} \int \frac{d^4\ell}{(2\pi)^4} e^{-i(k-\ell)\cdot b}
    \hat{\delta}((k-\ell)\cdot v) \hat{\delta}(\ell\cdot v_0)
    \frac{1}{\ell^2 (\ell\cdot v-i0)^2} \nn \\
    &\qquad \times \left[
    (\inn{\ell}{v})((\inn{v}{v_0}) \ell^\mu -(\inn{\ell}{v})v_0^\mu)
    -v^\mu((\inn{v}{v_0}) (\inn{k}{\ell}) -(\inn{\ell}{v}) (\inn{k}{v_0}))
    \right].
    \label{eq:1PL_integrand}
\end{align}
As a non-trivial check, the current obeys $k_\mu \delta J^\mu(k)=0$ for any $k$ as a consequence of current conservation. This current arises from the diagram in Fig.~\ref{fig:PL}.

To integrate the result in \eqref{eq:1PL_integrand}, we notice that under the delta functions, 
the result only depends on one integral
\begin{align}
    \delta J^\mu(k) =&
    \frac{Q q^2}{\mu} \frac{e^{-ik\cdot b}}{(\inn{k}{v})^2}
    \left[
    (\inn{k}{v})((\inn{v}{v_0}) \eta^{\mu\nu} -v_0^\mu v^\nu)
    -((\inn{v}{v_0}) v^\mu k^\nu -(\inn{k}{v_0}) v^\mu v^\nu)
    \right] I_\nu(k) \nn \\
    I^\nu(k) =& \int \frac{d^4\ell}{(2\pi)^4}
    \hat{\delta}((k-\ell)\cdot v) \hat{\delta}(\ell\cdot v_0)
    \frac{\ell^\nu}{\ell^2}\, e^{i\ell\cdot b}.
\end{align}
This integral has been computed in Appendix C of \cite{Cristofoli:2021vyo},
\begin{align}
    I^\nu(k) =& \frac{k \cdot v}{2\pi\gamma^3\beta^3}\left[\left(v^\nu-\gamma v_0^\nu\right) K_0\left[z(k)\right]-i\,\textrm{sgn}(\omega)\gamma\beta\,\hat{b}^\nu \,K_1\left[z(k)\right]\right],
\end{align}
where $z(k) = b|k\cdot v|$.
The asymptotic gauge field then follows from \eqref{eq:Atilde}
\begin{align} \label{eq:CHSAmu}
    \widetilde{A}^{\mu}(\omega, \vec{x})
    =& \frac{qK}{4\pi^2 |\vec{x}| \mu (\gamma \beta)^3} \frac{e^{-ik_*\cdot b}}{k_*\cdot v} \\
    &\,\, \times \left[((\inn{k_*}{v_0})v^\mu-(\inn{k_*}{v})v_0^\mu)K_0(z(k_*))
    -i\,\textrm{sgn}(\omega)\gamma^2 \beta ((\inn{k_*}{v})\hat{b}^\mu-(\inn{k_*}{\hat{b}})v^\mu) K_1(z(k_*))
    \right],\nn
\end{align}
where we use $K=-Qq/4\pi$.
The sign function ensures that $\widetilde{A}^{\mu}(-\omega, \hat{x})=-\widetilde{A}^{\mu}(\omega, \hat{x})^*$ such that ${A}^{\mu}(t,\hat{x})$ is real.
One can check that the transverse condition $k_*\cdot \widetilde{A}^{\mu}(\omega, \hat{x})=0$ is indeed satisfied. For {later analytic comparison with the QSM at 1PL, we adapt the notation $\beta\to\beta_\infty$ and $\gamma\to\gamma_\infty$}, and consider the frame\footnote{The initial velocity in the setup of QSM is not exactly along the $y$-direction, but the difference is of higher order in PL expansion.}
\begin{align}
	v^\mu = (\gamma_{\infty},  0, \gamma_{\infty} \beta_{\infty} ,0), \quad
	b^\mu = b \hat{b}^\mu = b(0,1,0,0), \quad b=\frac{L}{\mu\beta \gamma_{\infty}},
\end{align}
and therefore $z(k_*)=\frac{\omega L}{\gamma_{\infty}\beta_{\infty}^2 \mu}(1-\sin\theta\sin\phi)$. In this frame, the explicit expression of the time component reads
\begin{align}\label{eq:CSExp}
\widetilde{A}^t(\omega, \vec{x})
    &=\frac{qK\,e^{-ik_*\cdot b}}{4\pi^2 |\vec{x}|\mu \gamma_{\infty}^3\beta_{\infty}^3\,(\inn{k_*}{v})}
    \,\left\{\left(\,-\gamma_{\infty}\omega
    -(\inn{k_*}{v}) \right)K_0\left[z(k_*)\right]+i\,\gamma_{\infty}^3\beta_{\infty}\, (k_*\cdot \hat{b})\,K_1\left[z(k_*)\right]\right\}\, \nonumber\\
    &=\frac{qK\,e^{-ik_*\cdot b} \sin\theta}{4\pi^2 |\vec{x}|\mu \beta_{\infty}^2\,(1-\beta_{\infty} \sin\theta \sin\phi)}
    \,\left( \sin\phi K_0\left[z(k_*)\right]-i\cos\phi\,K_1\left[z(k_*)\right]\right)+\dots \,,
\end{align}
where in the second line, we keep the first two terms in the non-relativistic expansion.
    
\subsection{Comparison with the QSM}
\begin{figure}[ht!]
         \centering \includegraphics[width=0.7\textwidth]{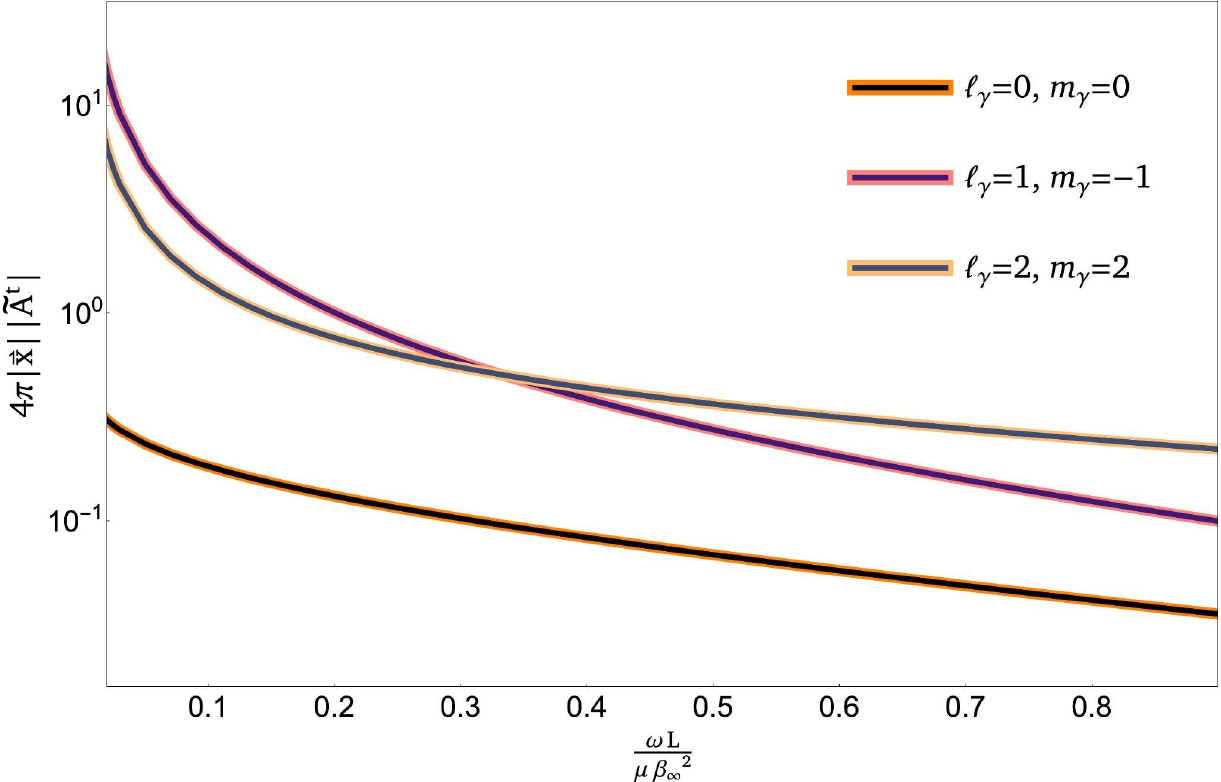}
    \caption{Comparison between relativistic 1PL and the 1PL-expanded QSM results for different partial waves. The thinner lines are the 1PL result, and they coincide with the QSM result, which is expanded to 1PL and resummed numerically.}\label{fig:1PLcmp}
\end{figure}

The expressions \eqref{eq:ALap}-\eqref{eq:Ahat} for the four-potential $A^\mu$ and its Laplace/Fourier transform $\hat{A}^{\mu}(s_L)$ obtained through the QSM, are non-perturbative in the coupling $K$, and therefore should coincide with \eqref{eq:CHSAmu} when expanded to first order in $K$, while leaving the energy and angular momentum fixed. In order to show this, note that~\eqref{eq:CHSAmu} is the Fourier transform of $A^\mu$ with respect to $u=t-|\vec{x}|$, while \eqref{eq:Ahat} is the Laplace transform of $A^\mu$ with respect to $t$. Extracting the Fourier transform with respect to $u$ from \eqref{eq:Ahat}, we have 
\begin{align}\label{eq:AhatFour}
\widetilde{A}^{\mu}\left(\omega,\vec{x}\right)&=e^{-i\omega\, |\vec{x}|}\Omega_r^{-1}\hat{A}^\mu(-i \frac{\omega}{\Omega_r},\vec{x})=\frac{q}{\mu \Omega_r |\vec{x}|}\,\sum_{ \ell_{\gamma}=0}^{\infty}\,\sum_{m_{\gamma}=-\ell_{\gamma}}^{\ell_{\gamma}}\left(-i\right)^{\ell_{\gamma}}\,Y_{\ell_{\gamma}}^{m_{\gamma}*}(\theta,\varphi)\,\hat{\mathcal{M}}_{\ell_{\gamma},m_{\gamma}}^{\mu}(\omega,s_L)\,.
\end{align}
Indeed, the expansion of \eqref{eq:AhatFour} to first order in $K$ coincides with \eqref{eq:CHSAmu}.  However, it is highly nontrivial to show this analytically, since it is necessary to resum the expanded \eqref{eq:AhatFour} in order to compare with \eqref{eq:CHSAmu} . To simplify matters, we show how this occurs for $\widetilde{A}^t$ in the non-relativistic case, resumming into the non-relativistic case of \eqref{eq:CSExp}. In the more general case, we can resum the 1PL expansion of the QSM result \textit{numerically}. The result is shown Fig.~\ref{fig:1PLcmp}, which shows a perfect match between the 1PL expanded \eqref{eq:AhatFour} and the 1PL result \eqref{eq:CSExp}. 

We now provide a few more details about the resummation of the 1PL-expanded QSM result, in the non relativistic limit.
Considering the $t$-component of~\eqref{eq:AhatFour}, and using~\eqref{eq:CurlyMt} and \eqref{eq:classYlmelem}, the Fourier transform of the scalar potential can be written as
\begin{equation} \label{eq:Atnonrel}
\widetilde{A}^{t}\left(\omega,\vec{x}\right)=\frac{q}{\Omega_r |\vec{x}|}\,\sum_{\ell_{\gamma}=0}^{\infty}\sum_{m_{\gamma}=-\ell_{\gamma}}^{\ell_{\gamma}}(-1)^{m_\gamma}(-i)^{\ell_{\gamma}}\,Y_{\ell_{\gamma}}^{m_{\gamma}}\left(\theta,\varphi\right)\, f_{\ell_{\gamma},m_{\gamma}}\lim_{\hbar\rightarrow0}\,\left\langle j_r',\ell+m_\gamma\right|j_{\ell_{\gamma}}\left(\omega r\right)\,\left|j_r, \ell\right\rangle\,,
\end{equation}
where $f_{\ell_{\gamma},m_{\gamma}}$ is defined in~\eqref{flm}, and $j'_r=j-\Delta j_r$ with $\Delta j_r=-\Delta \nu-i\frac{\omega}{\Omega_r}$\footnote{Note that the relation between $\Delta j_r$ and $\omega$ differs by the sign of $\omega$ (or $s_L$) compared with~\eqref{eq:MQSM}. This is equivalent since the radial matrix element is real, and so it is invariant under the simultaneous sign flips of $\Delta j_r$ and $\Delta \ell$.}. In the non-relativistic limit, $\Delta\nu=\Delta\ell$, i.e. an integer. Note that the non-relativistic limit of the matrix element above is obtained by setting $\frac{dt}{d\tau}=1$ in the states, which practically amounts to dropping the second term in~\eqref{eq:besseljmatelem}, yielding
\begin{equation} \label{eq:jlmatelemNR}
\lim_{\hbar\rightarrow0}\,\left\langle j_r',\ell+m_{\gamma}\right|j_{\ell_{\gamma}}\left(\omega r\right)\,\left|j_r, \ell\right\rangle = 2^{\ell_{\gamma}}\sum_{\kappa=0}^{\infty}\frac{\left(-1\right)^{\kappa}\left(\kappa+\ell_{\gamma}\right)!\,\omega^{2\kappa+\ell_{\gamma}}}{\kappa!\left(2\kappa+2 \ell_{\gamma}+1\right)!} \,
\mathcal{A}_{2\kappa+\ell_{\gamma}}^{\Delta \ell=-m_\gamma}\left(m_\gamma-i\frac{\omega}{\Omega_r}\right)\,.
\end{equation}
Further, as can be seen from~\eqref{eq:jlmatelemNR}, the required cases of $\mathcal{A}_{j}^{\Delta \ell}$ for $\widetilde{A}^t$ in the non-relativistic limit satisfy $j+\Delta\ell\ge0$, and so it simplifies greatly, yielding
\begin{align} \label{eq:AjDeltalnonrel}
&\mathcal{A}_{j}^{\Delta \ell}\left(m_\gamma-i\frac{\omega}{\Omega_r}\right) = \left(\frac{p\,e}{e^{2}-1}\right)^{j}\left(-1\right)^{j+\Delta \ell+1}2^{-j-2}\frac{e}{\sinh\left(\frac{\pi\omega}{\Omega_r}\right)}\sum_{s=0}^{j-\Delta \ell+1}\sum_{k=0}^{j+\Delta \ell+1}\left(\begin{array}{c}
j-\Delta \ell+1\\
s
\end{array}\right)\nonumber\\
&\quad
\times\left(\begin{array}{c}
j+\Delta \ell+1\\
k
\end{array}\right){\rm{Re}}\left[i\left(\frac{1-i\sqrt{e^{2}-1}}{e}\right)^{s-k+\Delta \ell}J_{-j-1+k+s+i\frac{\omega}{\Omega_r}}\left(-i\frac{\omega\,e}{\Omega_r}\right)\right]\,.
\end{align}
where the non-relativistic relations for $e,\,p$ and $\Omega_r$ are $e=\sqrt{1+\frac{\beta_\infty^2 L^{2}}{K^{2}}}$, $p=\frac{L^{2}}{\mu K}$, and $\Omega_r=\frac{\mu \beta_\infty^3}{K}$.

To obtain the $\mathcal{O}\left(K\right)$ term in the expansion of $\widetilde{A}^{t}$, it is sufficient to calculate the $\mathcal{O}\left(K^0\right)$ term of~\eqref{eq:AjDeltalnonrel}. The calculation is carried out in Appendix~\ref{AppendixB}, and results in
\begin{align}\label{eq:matelemnonrel0K}
&\mathcal{A}_{j}^{\Delta \ell}\left(m_\gamma-i\frac{\omega}{\Omega_r}\right) =\left(\frac{L}{2\mu\beta_\infty}\right)^{j}\frac{\omega L}{2\pi \mu\beta_\infty^2}\cos\left(\frac{\pi}{2}\left(j-\Delta \ell\right)\right)\sum_{k=0}^{j+\Delta \ell+1}\sum_{s=0}^{j-\Delta \ell+1}\left(-1\right)^{k+j+1}\left(\protect\begin{array}{c}
j-\Delta \ell+1\protect\\
s
\protect\end{array}\right)\nonumber\\
&\quad
\times\left(\protect\begin{array}{c}
j+\Delta \ell+1\protect\\
k
\protect\end{array}\right)\left\{ \left.\frac{d}{d\nu}K_{\nu}\left(\frac{\omega L}{\mu\beta_\infty^2}\right)\right|_{\nu=k+s-j-1}-\frac{\mu\beta_\infty^2}{\omega L}\left(k-s-\Delta \ell\right)K_{k+s-j-1}\left(\frac{\omega L}{\mu\beta_\infty^2}\right)\right\}\,.
\end{align}
Using~\eqref{eq:Atnonrel},~\eqref{eq:jlmatelemNR} and~\eqref{eq:matelemnonrel0K}, the $\mathcal{O}\left(K\right)$ term of $\widetilde{A}^{t}$ is obtained. Resumming the obtained expression is non-trivial and technical, and is deferred to Appendix~\ref{AppendixB}. The resulting expression is 
\begin{align}
&\widetilde{A}^{t}\left(\omega,\vec{x}\right)
=\frac{q K}{ 4\pi^2\mu\beta_\infty^2|\vec{x}|}\frac{\sin\theta\exp\left(-i\frac{\omega L}{\mu\beta_\infty}\sin\theta\cos\phi\right)}{1-\beta_\infty\sin\theta\sin\phi}\Big\{\sin\phi\,K_{0}\left(\frac{\omega L}{\mu\beta_\infty^2}\left(1-\beta_\infty\sin\theta\sin\phi\right)\right)\nonumber\\
&\hspace{4cm}-i\cos\phi\, K_{1}\left(\frac{\omega L}{\mu\beta_\infty^2}\left(1-\beta_\infty\sin\theta\sin\phi\right)\right)\Big\}\,,
\end{align}
which is exactly the non-relativistic limit of the 1PL result~\eqref{eq:CSExp}.
\section{Exact Semiclassical Wave Function in Schwarzschild}\label{section:Schwarz}
\begin{figure}[ht!]
         \centering \includegraphics[width=0.7\textwidth]{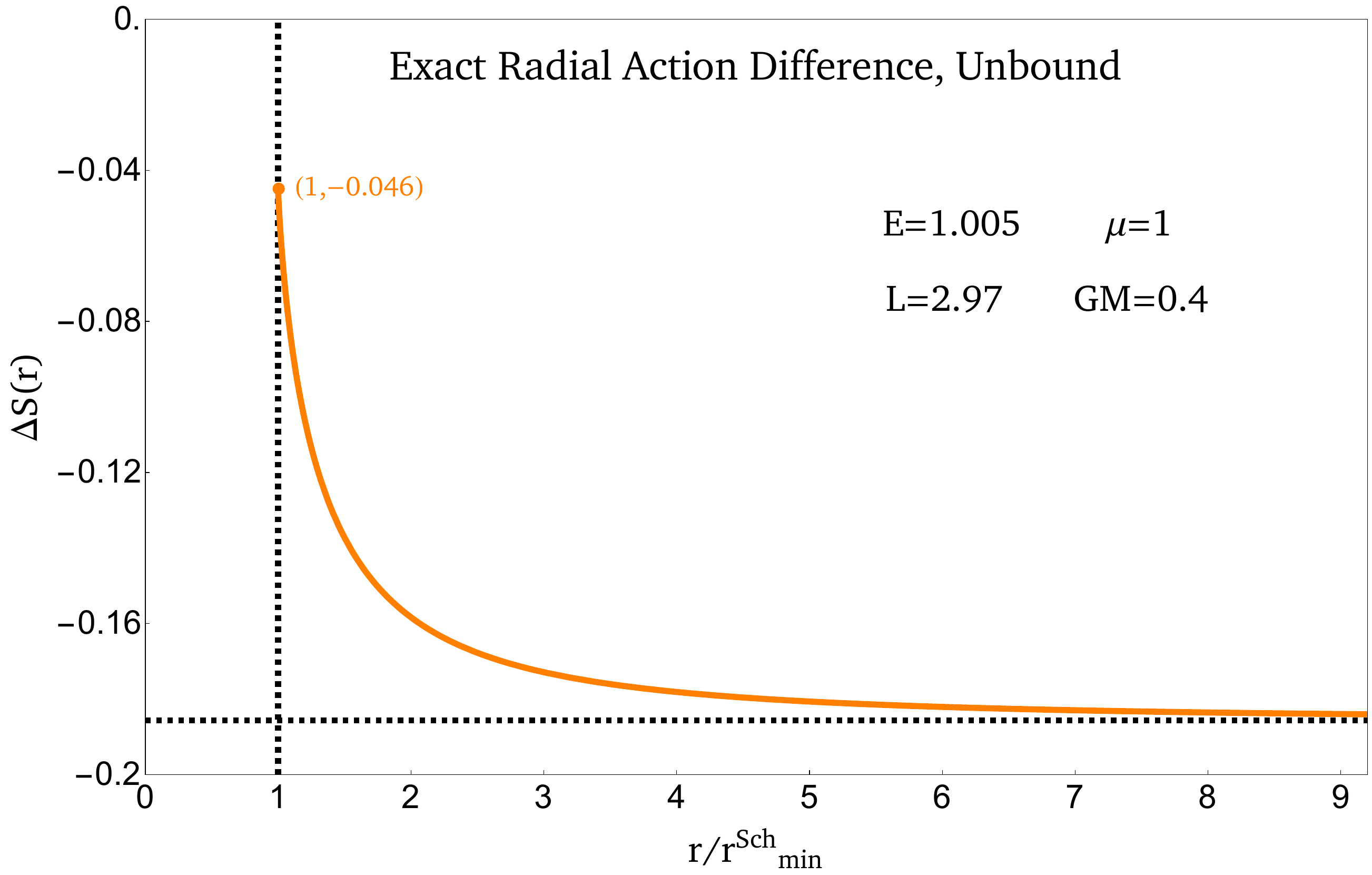}
    \caption{$\Delta S(r)$, the difference between $S^{Sch}_r(r)$ given in \eqref{eq:appeqSsch} and $S^{Coul}_r(r)$ given in \eqref{eq:Srexp}. The result is regular in all of the classically allowed region. Taylor expanding this result leads to an expansion of Schwarzschild Laplace observables in terms of Coulomb/Kepler ones.}\label{fig:DeltaS}
\end{figure}

All of our examples of Laplace observables so far were in the context of relativistic Keplerian motion. Naturally, one wonders about the applicability of our method to Schwarzschild.
In this section, we derive the radial wave function in Schwarzschild in the semi-classical limit. Using this wave function, we can compute any Laplace observable for Schwarzschild at any desired PM/PN order. We leave this explicit computation for upcoming work \cite{Khalaf:2025}. The KG equation in a Schwarzschild background is given by
\begin{eqnarray}\label{eq:KGPhiSch}
\left(g^{\mu\nu}D_\mu D_\nu-\frac{\mu^2}{\hbar^2}\right) \Phi^{Sch}_{j_r,\ell,m}=0\,,
\end{eqnarray}
where $g^{\mu\nu}={\rm diag}\left(-\Delta/r^2,r^2/\Delta,r^2,r^2\sin^2\theta\right)$ and $\Delta=r\,(r-2GM)$, and $D_\mu$ is the curved-space covariant derivative. The corresponding QSM ``wave functions" are then
\begin{eqnarray}\label{eq:PsidefSch}
\Psi^{Sch}_{j_r,\ell,m}\equiv\sqrt{\frac{r^2}{\Delta}}\,R^{Sch}_{j_r,\ell}(k_{j_r,\ell}r)\,Y_{\ell}^{m}\left(\theta,\varphi\right)\,,
\end{eqnarray}
where the prefactor comes from 
 $\gamma^{j_r,\ell}/\gamma^{j_r,\ell}_\infty$ where $\gamma^{j_r,\ell}=dt/d\tau=\gamma^{j_r,\ell}_\infty\,r^2/\Delta$. 
In the WKB limit, the radial wave function $R^{Sch}_{j_r,\ell}(k_{j_r,\ell}r)$ has the form
\begin{eqnarray}\label{eq:WKBKGSch}
\lim_{\hbar\rightarrow 0}R^{Sch}_{j_r,\ell}(k_{j_r,\ell}r)&=& -{\rm Im}\left[R^{Sch,WKB}_{j_r,\ell}(k_{j_r,\ell}r)\right]\nonumber\\[5pt]
R^{Sch,WKB}_{j_r,\ell}(k_{j_r,\ell}r)&=&\left(\frac{2E\Omega^{Sch}_r}{\pi  \Delta\sqrt{U^{Sch}_r(r)}}\right)^{\frac{1}{2}}\,\exp\left(-{\frac{iS^r_{Sch}(r)}{\hbar}-\frac{i\pi}{4}}\right)\,\,,\nonumber\\
\end{eqnarray}
where $(j_r,\ell,m)=\hbar^{-1}(J_r,L,L)$, $\Omega^{Sch}_r$ is the classical radial frequency in Schwarzschild, and $E=E(J_r,L)$ is a known canonical transformation. The functions $S^{Sch}_r(r)$ and $U^{Sch}_r(r)$ are explicitly given by
\begin{eqnarray}\label{eq:Usch}
S^{Sch}_r(r)&=&\int_{r_{min}}^r\,\sqrt{U^{Sch}_r(r')}~dr'\,,\nonumber\\[5pt]
U^{Sch}_r(r)&=&\left(\frac{r^2}{\Delta}\right)^2\,\left[E^2-\frac{\Delta}{r^2}\left(\frac{L^2}{r^2}+\mu^2\right)\right]\equiv\frac{(\mu^2-E^2)}{r-2GM}(r-r_{b})(r-r_{min})(r_{*}-r)\,,\nonumber\\
\end{eqnarray}
where $r_{b}<r_{min}$. In fact, we computed $S^{Sch}_r(r)$ exactly, and it is given in Appendix~\ref{sec:radialSch}. To be able to use the QSM, we are looking for a function $R^{Sch}_{j_r,\ell}(k_{j_r,\ell}r)$ that reduces to $R^{Sch,WKB}_{j_r,\ell}(k_{j_r,\ell}r)$ in the WKB limit. Luckily, we can construct it in a sleek way, using the complexified Coulomb wave function of Section~\ref{section:compc}. We define
\begin{eqnarray}\label{eq:Schslick}
R^{Sch}_{j_r,\ell}(k_{j_r,\ell}r)=-{\rm Im}\left[\mathcal{F}_{Sch}\,R^c_{j_r,\ell}(k_{j_r,\ell}r)\right]\,,
\end{eqnarray}
where
\begin{eqnarray}\label{eq:Fsch}
\mathcal{F}_{Sch}\equiv \left(R^{Sch,WKB}_{j_r,\ell}/R^{c,WKB}_{j_r,\ell}\right)\,,
\end{eqnarray}
where $R^{c,WKB}_{j_r,\ell}$ is given in \eqref{eq:WKBKGc}, and we make the identification
\begin{eqnarray}\label{eq:Coulident}
K&\equiv& (G M {\mu})\frac{2\gamma^2_{\infty}-1}{\gamma_{\infty}}~~,~~ \mathcal{N}\equiv\sqrt{J^2-(2GM\mu)^2(3\gamma^2_{\infty}-1)}\,.
\end{eqnarray}
The latter are specifically chosen so that $\sqrt{U^{Sch}_r}$ and $\sqrt{U^{Coul}_r}$ coincide at large $r$. Note that the $R^{Sch}_{j_r,\ell}$ constructed this way does not provide a genuine solution for the KG equation \eqref{eq:KGPhiSch}. Nevertheless, it has the correct WKB limit by construction, which is all we need for the QSM. The advantage of this definition is that, since both $R^{Sch,WKB}_{j_r,\ell}$ and $R^{c,WKB}_{j_r,\ell}$ are known, we can directly compute $\mathcal{F}_{Sch}$ and even expand it in PM/PN (see \cite{Adamo:2023cfp} for a similar PM expansion of the WKB phase in the context of gravitational emission), so that 
\begin{eqnarray}\label{eq:Fsch2}
\mathcal{F}_{Sch}=\left(\frac{\Omega^{Sch}_r}{  \Omega^{Coul}_r}\,\frac{r^2}{\Delta}\sqrt{\frac{U^{Coul}_r(r)}{U^{Sch}_r(r)}}\right)^{\frac{1}{2}}\,\exp\left[-\frac{i}{\hbar}\Delta S(r)\right]\,,
\end{eqnarray}
where 
\begin{eqnarray}
\Delta S(r)=S^{Sch}_r(r)-S^{Coul}_r(r)\,,
\end{eqnarray}
which is given exactly in \eqref{eq:appeqSsch} and \eqref{eq:Srexp}, and depicted in Fig.~\ref{fig:DeltaS}. Importantly, this function is regular in all of the classically allowed region and can be easily Taylor expanded in $1/r$. In a follow-up work \cite{Khalaf:2025}, we will explicitly use this expansion to compute time-dependent geodesic motion and self-force in Schwarzschild to any desired PM/PN order.

\section{Conclusions}\label{section:Conc}
In this paper we showed how to compute bound-unbound universal expressions for conservative motion in Coulomb and Schwarzschild backgrounds. While the analytical continuation between the bound and unbound cases is well known in the Coulomb-Kepler case, we demonstrated it directly also for the Schwarzschild case, by deriving a generalized Kepler equation providing an implicit analytical solution for $r(t)$.  

At the heart of the paper we set up a detailed infrastructure that allows for the calculation of any time-dependent classical observable for relativistic point-particle motion in a Coulomb potential, and for conservative motion in the background of a Schwarzschild black hole -- to any desired order in the PM or PN expansions. In particular, for any observable $\mathcal{O}(t)$, we showed how to compute its corresponding \textit{Laplace observable} using the QSM, as the classical limit of a ``quantum" matrix element. Furthermore, the Laplace observable can be computed for unbound motion but is equally useful for bound motion -- in other words, it provides bound-unbound universality. We demonstrated our method by computing the Laplace observables for relativistic Keplerian motion: the radius $r(t)$, the azimuthal angle $\varphi(t)$, and the {all-order} EM field $A_\mu(t)$ emitted by a relativistic Keplerian electron. The latter computation was explicitly cross-checked with the known lowest-order PL result. Finally, we showed how in the classical limit, Schwarzschild wave functions are proportional to Coulomb ones up to a PM expandable prefactor. This makes the generalization of our method to Schwarzschild automatic, at any desired order in a PM or PN expansion.

\section*{Acknowledgments}
The authors thank Cliff Cheung, Alessandro Lenoci and Rotem Ovadia for stimulating discussions. MK is grateful to Scott Hughes for his generous hospitality at MIT and for many fruitful discussions. OT and MK are supported by the ISF grant No. 3533/24 and by the NSF-BSF grant No. 2022713. MK is grateful for
the support of the Azrieli Foundation Fellows program. The work of MK is also supported by the BSF grant No. 2020220, and an ERC STG grant (``Light-Dark,'' grant No. 101040019). 
CHS is supported by a Yushan Young Scholarship award number 112V1039 from the Ministry of Education (MOE) of Taiwan, and also by the National Science and Technology Council (NSTC) grant number 114L7329. CHS is also grateful for the hospitality of CERN and the Munich Institute for Astro-, Particle and BioPhysics (MIAPbP) which is funded by the Deutsche Forschungsgemeinschaft (DFG, German Research Foundation) under Germany´s Excellence Strategy–EXC-2094–390783311 during the completion of the work. This project has received funding from the European Research Council (ERC) under the European Union’s Horizon Europe research and innovation programme (grant agreement No. 101040019).  Views and opinions expressed are however those of the author(s) only and do not necessarily reflect those of the European Union. The European Union cannot be held responsible for them.

\appendix
\section{Exact Time Dependence for Schwarzschild} \label{KepSch}
In this appendix we solve the differential equation for $t(\varphi)$ by direct integration. This equation is
\begin{eqnarray}\label{eq:tdotphi}
\frac{dt}{d\varphi}=\frac{Er^4(\varphi)}{L\Delta(r(\varphi))}\,,
\end{eqnarray}
where $\Delta(r)=r(r-2GM)$ and
\begin{eqnarray}\label{eq:rphischeqslapp}
r(\varphi)=\frac{p}{1+e\,\cos\left[2{\rm am}(A\,\varphi,k)\right]}\,.
\end{eqnarray}

An explicit solution to \eqref{eq:tdotphi} is
\begin{eqnarray}\label{eq:Fexp}
&&t(\varphi)=\frac{E}{L}\,\left[\frac{p^2}{(1+e)^2}I_{f1}\left(\frac{2e}{1+e}\right)+\frac{2GMp}{1+e}I_{f2}\left(\frac{2e}{1+e}\right)+\frac{2GM}{D}I_{f2}\left(\frac{2e}{pD}\right)\right]\,,\nonumber\\[5pt]
   \end{eqnarray}
where $D\equiv(p/(1+e))^{-1}-(2GM)^{-1}$ and
\begin{eqnarray}\label{eq:Itau}
I_{f1}(\tau)&=&\int\,\frac{1}{\left[1-\tau {\rm sn}^2(A\varphi,k)\right]^2}\,d\varphi\nonumber\\[5pt]
I_{f2}(\tau)&=&\int\,\frac{1}{1-\tau {\rm sn}^2(A\varphi,k)}\,d\varphi=\frac{1}{A}\,\Pi\left(\tau,{\rm am}(A\varphi,k),k\right)\,.
\end{eqnarray}
Here, $\Pi$ is the incomplete elliptic integral of the third kind. Although an explicit expression for $I_{f1}(\tau)$ in terms of elliptic integrals can be derived within a second in \textit{Mathematica}, it is rather unpleasant looking and we omit it here. 

\section{Calculation of $\mathcal{A}_j^{\Delta \ell}(\Delta j_r)$ and  $\mathcal{B}_j^{\Delta \ell}(\Delta j_r)$} \label{AjDeltal}
In this section, we shall calculate $\mathcal{A}_j^{\Delta \ell}(\Delta j_r)$ and $\mathcal{B}_j^{\Delta \ell}(\Delta j_r)$ defined in \eqref{eq:AB} as
\begin{eqnarray}\label{eq:ABapp}
\mathcal{A}_j^{\Delta \ell}\left(\Delta j_r\right)&\equiv&\lim_{\hbar\rightarrow0}I_j=\lim_{\hbar\rightarrow0}\,\left\langle j_r',\ell'\right|\frac{r^j}{1+\frac{K}{Er}}\left|j_r,\ell\right\rangle\nonumber\\[5pt]
\mathcal{B}_{j}^{\Delta \ell}\left(\Delta j_r\right)&\equiv&\lim_{\hbar\rightarrow0}\hbar\left\langle j_r',\ell'\right|\,\frac{r^j}{1+\frac{K}{Er}}\,\partial_{r}\left|j_r,\ell\right\rangle\,.
\end{eqnarray}
We begin with a calculation of $\mathcal{A}_j^{\Delta \ell}$. We note that $I_{j}$ is a Gordon integral \cite{Gordon1929,Matsumoto1991}, which can be expressed using an Appell $F_2$ function as
\begin{eqnarray} \label{eq:IjAppA}
I_j=&&C_{j'_r,\nu'}C_{j_r,\ell}(2k_{j_r,\ell})^{\nu}(2k_{j'_r,\ell'})^{\nu'}\frac{\Gamma(\nu+\nu'+j+3)}{\Gamma(2\nu+2)\Gamma(2\nu'+2)}\left(ik_{j_r,\ell}+ik_{j'_r,\ell'}\right)^{-\nu-\nu'-j-3}\,\times\nonumber\\[5pt]
&&F_2\left(\nu+\nu'+j+3,-j_r,-j'_r,2\nu+2,2\nu'+2;\frac{2k_{j_r,\ell}}{k_{j_r,\ell}+k_{j'_r,\ell'}},\frac{2k_{j'_r,\ell'}}{k_{j_r,\ell}+k_{j'_r,\ell'}}\right)\,,
\end{eqnarray}
where $\Sigma k\equiv k_{j_r,\ell}+k_{j'_r,\ell'}$. The variables of the $F_{2}$ function approach $\left(1,1\right)$
in the classical limit. Hence, in order to compute the classical limit
of~\eqref{eq:IjAppA}, we need the analytical continuation of $F_{2}$ around $\left(1,1\right)$.
This is given by the following expansion \cite{bateman1953higher}
\begin{align}\label{eq:HahneForm}
&F_{2}\left(\alpha,\beta,\beta',\gamma,\gamma';x,y\right) 
=\frac{\Gamma\left(\gamma\right)\Gamma\left(\gamma'\right)\Gamma\left(\alpha-\gamma'+\beta'-\beta\right)}{\Gamma\left(\alpha\right)\Gamma\left(\beta'\right)\Gamma\left(\gamma-\beta\right)} 
\exp\left(i\pi\left(\alpha+\beta'-\gamma'\right)\right)
\left(y-1\right)^{\gamma'-\alpha+\beta-\beta'} \nonumber\\
&\quad \times \sum_{m,s=0}^{\infty} 
\frac{\left(\gamma'+\beta-\beta'+m\right)_{s}\left(\beta\right)_{m}\left(\beta-\beta'+\gamma'-\gamma+1\right)_{s}}{m!s!\left(\beta-\beta'+\gamma'-\alpha+1\right)_{s}} \left(1-x\right)^{m}\left(1-y\right)^{s}\nonumber\\
&\quad \times \,_{3}F_{2}\left(\alpha-\gamma+1+\beta-\beta'+m,-s,\gamma'-\beta'; \gamma'+\beta-\beta'+m,\beta-\beta'+\gamma'-\gamma+1;1\right) \nonumber\\
&\quad + \frac{\Gamma\left(\gamma\right)\Gamma\left(\gamma'\right)\Gamma\left(\gamma-\beta-\gamma'+\beta'\right)\Gamma\left(\gamma'-\alpha+\beta-\beta'\right)}{\Gamma\left(\beta'\right)\Gamma\left(\gamma-\beta\right)\Gamma\left(\gamma'+\beta-\beta'\right)\Gamma\left(\gamma-\alpha\right)}\exp\left(i\pi\left(\alpha+\beta'-\gamma'\right)\right)\nonumber\\
&\quad\times\sum_{m,s=0}^{\infty}\frac{\left(\alpha\right)_{m+s}\left(\beta\right)_{m}\left(\alpha-\gamma+1\right)_{s}}{m!s!\left(\gamma'+\beta-\beta'\right)_{m}\left(\alpha-\gamma'+\beta'-\beta+1\right)_{s}}\left(1-x\right)^{m}\left(1-y\right)^{s}\nonumber\\
&\quad\times\,_{3}F_{2}\left(\alpha-\gamma+1+\beta-\beta'+m,\gamma'-\alpha+\beta-\beta'-s,\gamma'-\beta',\gamma'+\beta-\beta'+m,\beta-\beta'+\gamma'-\gamma+1;1\right)\nonumber\\
&\quad
\hspace{5.5cm}+\left(\beta\leftrightarrow\beta',\gamma\leftrightarrow\gamma',x\leftrightarrow y\right),
\end{align}
where in our case the values of $\alpha,\,\beta,\,\beta',\,\gamma,\,\gamma',\,x$ and $y$ are read off from the arguments of $F_2$ in~\eqref{eq:IjAppA}. Here, $\left(\beta\leftrightarrow\beta',\gamma\leftrightarrow\gamma',x\leftrightarrow y\right)$
means similar terms as the previous two with the exchanges described.

Denote the first and second terms of~\eqref{eq:HahneForm} by $\mathcal{F}^{(1)}$ and $\mathcal{F}^{(2)}$ respectively so that
\begin{align}\label{eq:HahneFormCompact}
&F_{2}\left(\alpha,\beta,\beta',\gamma,\gamma';x,y\right) 
=\mathcal{F}^{(1)}\left(\alpha,\beta,\beta',\gamma,\gamma';x,y\right)+\mathcal{F}^{(2)}\left(\alpha,\beta,\beta',\gamma,\gamma';x,y\right)\nonumber\\
&\quad
\hspace{4.1cm}+\left(\beta\leftrightarrow\beta',\gamma\leftrightarrow\gamma',x\leftrightarrow y\right).
\end{align}
Accordingly, we also define
\begin{eqnarray} \label{eq:IjiAppA}
&I_j^{(i)}\equiv C_{j'_r,\nu'}C_{j_r,\ell}(2k_{j_r,\ell})^{\nu}(2k_{j'_r,\ell'})^{\nu'}\frac{\Gamma(\nu+\nu'+j+3)}{\Gamma(2\nu+2)\Gamma(2\nu'+2)}\left(i\Sigma k\right)^{-\nu-\nu'-j-3}\nonumber\\
&\quad\times\mathcal{F}^{(i)}\left(\nu+\nu'+j+3,-j_r,-j'_r,2\nu+2,2\nu'+2;\frac{2k_{j_r,\ell}}{\Sigma k},\frac{2k_{j'_r,\ell'}}{\Sigma k}\right)\,.
\end{eqnarray}
The classical limit of $I_j^{(i)}$ is calculated in Appendices~\ref{App:ClassLimCCIj1} and~\ref{App:clc}. Explicitly,
\begin{align} \label{eq:CCIj1}
&\lim_{\hbar\to0}\,I_j^{(1)} = \left(\frac{pe}{e^{2}-1}\right)^{j}\left(-1\right)^{j+\Delta \nu+1}2^{-j-2}\frac{i\gamma^2_\infty e}{\sinh\left(\pi\Delta n\right)}\nonumber\\
&\quad
\times\sum_{s,k=0}^{\infty}\left(\begin{array}{c}
j-\Delta \nu+1\\
s
\end{array}\right)\left(\begin{array}{c}
j+\Delta \nu+1\\
k
\end{array}\right)\left(\frac{1-i\sqrt{e^{2}-1}}{e}\right)^{s-k+\Delta \nu}J_{-j-1+k+s+i\Delta n}\left(-ie\gamma_\infty^2\Delta n\right)\,,\nonumber\\
\end{align}
\begin{align}\label{eq:CCIj2}
&\lim_{\hbar\to0}\,I_j^{(2)} = \left(\frac{p e}{e^{2}-1}\right)^{j}2^{-j-2}e^{\pi\left(\Delta n+i\Delta \nu\right)}\frac{\pi \gamma^2_\infty e\left(\frac{1-i\sqrt{e^{2}-1}}{e}\right)^{j+\Delta \nu+1-i\Delta n}}{\Gamma\left(-j+\Delta \nu-1\right)\sinh\left(\pi\Delta n\right)\sinh\left(\pi\left(\Delta n+i\Delta \nu\right)\right)}\nonumber\\
&\quad
\times\sum_{m,s=0}^{\infty}\frac{\left(-1\right)^{s}\left(j-\Delta \nu+2\right)_{s}\left(i\frac{1-i\sqrt{e^{2}-1}}{2}\gamma_\infty^2\Delta n\right)^{s+m}}{m!s!\Gamma\left(j+2-i\Delta n+s\right)}\nonumber\\
&\quad
\times\,_{2}\widetilde{F}_{1}\left(-\left(j+\Delta \nu+1+m\right),-j-s+i\Delta n-1;i\Delta n-\Delta \nu+1;\frac{1+i\sqrt{e^{2}-1}}{1-i\sqrt{e^{2}-1}}\right)\,,
\end{align}
where $\Delta n = i\left(\Delta j_r+\Delta \nu\right),\,\Delta\nu=\nu-\nu'$, and $\gamma_\infty=\frac{E}{\mu}$ is the relativistic factor far away from the source of the potential. Using~\eqref{eq:CCIj1} and~\eqref{eq:CCIj2}, we can obtain $\mathcal{A}_j^{\Delta\ell}$ as
\begin{equation}\label{eq:Aeq}
\mathcal{A}_j^{\Delta\ell}=\lim_{\hbar\to0}\,I_j^{(1)}+\lim_{\hbar\to0}\,I_j^{(2)}+\left(\Delta j_r \to-\Delta j_r,\Delta\nu\to -\Delta\nu\right)\,.
\end{equation}

\subsection{Calculation of $\lim_{\hbar\to0}\,I_j^{(1)}$} \label{App:ClassLimCCIj1}
In this section, we shall calculate the classical limit of $I_j^{(1)}$. To this, we shall consider first the classical limit of $\mathcal{F}^{(1)}$ in~\eqref{eq:HahneFormCompact}. We will eventually set the values of the parameters $\alpha,\,\beta,\,\beta',\,\gamma,\,\gamma',\,x$ and $y$ according to~\eqref{eq:IjiAppA}, but there is no need to do this now since the analysis is more general. We will assume though that the behavior of the parameters
in the classical limit is the same as in~\eqref{eq:IjiAppA}, in that  $\alpha-\gamma,\,\alpha-\gamma'$ and $\beta-\beta'$
are finite as $\hbar\to0$, while $\alpha,\,\beta,\,\beta',\,\gamma,\,\gamma'= \mathcal{O}\left(\hbar^{-1}\right)$. Similarly, we assume that $x+y=2$ and $1-x,1-y=\mathcal{O}\left(\hbar\right)$.
To begin, note that the classical limit of the $\,_{3}F_{2}$ appearing
in $\mathcal{F}^{(1)}$ is easily found to be
\begin{align}
&\,_{3}F_{2}\left(\dots;1\right)\to\,_{2}F_{1}\left(-s,\alpha-\gamma+1+\beta-\beta'+m,,\beta-\beta'+\gamma'-\gamma+1;1-\frac{\beta}{\gamma}\right)\nonumber\\
&\quad
=\left(\frac{\beta}{\gamma}\right)^{s}\,_{2}F_{1}\left(-s,-\left(\alpha-\gamma'+m\right),\beta-\beta'+\gamma'-\gamma+1;1-\frac{\gamma}{\beta}\right)\,,
\end{align}
where we applied an Euler transformation in the second equality. Moreover, by using the classical limit of the Pochhammer symbol \cite{Khalaf:2023ozy},
\begin{eqnarray}
\lim_{\hbar\rightarrow 0}\left(A\hbar^{-1}\right)_n=\left(A\hbar^{-1}\right)^n\,,
\end{eqnarray}
we obtain the following classical
limit of $\mathcal{F}^{(1)}$ 
\begin{align}
\mathcal{F}^{(1)}\left(\alpha,\beta,\beta',\gamma,\gamma';x,y\right)&\to\frac{\Gamma\left(\gamma\right)\Gamma\left(\gamma'\right)\Gamma\left(\alpha-\gamma'+\beta'-\beta\right)}{\Gamma\left(\alpha\right)\Gamma\left(\beta'\right)\Gamma\left(\gamma-\beta\right)}\exp\left(i\pi\left(\alpha+\beta'-\gamma'\right)\right)\left(y-1\right)^{\gamma'-\alpha+\beta-\beta'}\nonumber\\
&\quad
\times\sum_{m,s=0}^{\infty}\frac{\left(\beta-\beta'+\gamma'-\gamma+1\right)_{s}}{m!s!\left(\beta-\beta'+\gamma'-\alpha+1\right)_{s}}\left(\beta\left(1-x\right)\right)^{m}\left(\beta\left(1-y\right)\right)^{s}\nonumber\\
&\quad
\times\,_{2}F_{1}\left(-s,-\left(\alpha-\gamma'+m\right),\beta-\beta'+\gamma'-\gamma+1;1-\frac{\gamma}{\beta}\right)\,.
\end{align}
Using the power series of $\,_{2}F_{1}$, we can carry out the sums
over $m$ and $s$, yielding
\begin{align}
&\mathcal{F}^{(1)}\left(\alpha,\beta,\beta',\gamma,\gamma';x,y\right)\to\frac{\Gamma\left(\gamma\right)\Gamma\left(\gamma'\right)\Gamma\left(\alpha-\gamma'+\beta'-\beta\right)\Gamma\left(\beta-\beta'+\gamma'-\alpha+1\right)\Gamma\left(\alpha-\gamma'+1\right)}{\Gamma\left(\alpha\right)\Gamma\left(\beta'\right)\Gamma\left(\gamma-\beta\right)}\nonumber\\
&
\times e^{i\pi\left(\alpha+\beta'-\gamma'\right)}\left(y-1\right)^{\gamma'-\alpha+\beta-\beta'}\sum_{k=0}^{\infty}\frac{\left(\left(\beta-\gamma\right)\left(1-y\right)\right)^{k}}{k!}\,_{1}\widetilde{F}_{1}\left(\alpha-\gamma'+1,\alpha-\gamma'+1-k;\beta\left(1-x\right)\right)\nonumber\\
&\quad
\times\,_{1}\widetilde{F}_{1}\left(\beta-\beta'+\gamma'-\gamma+k+1,\beta-\beta'+\gamma'-\alpha+k+1;\beta\left(1-y\right)\right)\,.
\end{align}
Here, $\,_{1}\widetilde{F}_{1}$ denotes the regularized Kummer hypergeometric
function. Doing an Euler transformation to both $\,_{1}\widetilde{F}_{1}$'s,
we get
\begin{align}
&\mathcal{F}^{(1)}\left(\alpha,\beta,\beta',\gamma,\gamma';x,y\right)\to\frac{\Gamma\left(\gamma\right)\Gamma\left(\gamma'\right)\Gamma\left(\alpha-\gamma'+\beta'-\beta\right)\Gamma\left(\beta-\beta'+\gamma'-\alpha+1\right)\Gamma\left(\alpha-\gamma'+1\right)}{\Gamma\left(\alpha\right)\Gamma\left(\beta'\right)\Gamma\left(\gamma-\beta\right)}\nonumber\\
&\quad
\times e^{i\pi\left(\alpha+\beta'-\gamma'\right)}\left(y-1\right)^{\gamma'-\alpha+\beta-\beta'}e^{\beta\left(2-x-y\right)}\sum_{k=0}^{\infty}\frac{\left(\left(\beta-\gamma\right)\left(1-y\right)\right)^{k}}{k!}\,_{1}\widetilde{F}_{1}\left(-k,\alpha-\gamma'+1-k;-\beta\left(1-x\right)\right)\nonumber\\
&\quad
\hspace{2cm}\times\,_{1}\widetilde{F}_{1}\left(-\left(\alpha-\gamma\right),\beta-\beta'+\gamma'-\alpha+k+1;-\beta\left(1-y\right)\right)\,.
\end{align}
By using the power expansion of the two $\,_{1}\widetilde{F}_{1}$ polynomials, and shifting $k\to k+m$ where $m$ is the summation index of $\,_{1}\widetilde{F}_{1}\left(-k,\dots\right)$,
we get the following simplification
\begin{align}
&\mathcal{F}^{(1)}\left(\alpha,\beta,\beta',\gamma,\gamma';x,y\right)\to\frac{\Gamma\left(\gamma\right)\Gamma\left(\gamma'\right)\Gamma\left(\alpha-\gamma'+\beta'-\beta\right)\Gamma\left(\beta-\beta'+\gamma'-\alpha+1\right)}{\Gamma\left(\alpha\right)\Gamma\left(\beta'\right)\Gamma\left(\gamma-\beta\right)}\nonumber\\
&\quad
\times e^{i\pi\left(\alpha+\beta'-\gamma'\right)}\left(y-1\right)^{\gamma'-\alpha+\beta-\beta'}\sum_{s,k=0}^{\infty}\left(\begin{array}{c}
\alpha-\gamma\\
s\end{array}\right)\left(\begin{array}{c}
\alpha-\gamma'\\
k\end{array}\right)\left(\beta\left(1-y\right)\right)^{s}\left(\left(\beta-\gamma\right)\left(1-y\right)\right)^{k}\nonumber\\
&\quad
\times\sum_{m=0}^{\infty}\frac{\left(\beta\left(\beta-\gamma\right)\left(1-x\right)\left(1-y\right)\right)^{m}}{m!\Gamma\left(\beta-\beta'+\gamma'-\alpha+k+m+s+1\right)}
\end{align}
where we also used that $2-x-y=0$. For the majority of radial matrix elements we use in this work, we also have $\gamma-\beta=\beta^{*}$, so we can simplify further to get
\begin{align} \label{eq:classlimCurlyF1}
&\mathcal{F}^{(1)}\left(\alpha,\beta,\beta',\gamma,\gamma';x,y\right)\to\frac{\pi\Gamma\left(\gamma\right)\Gamma\left(\gamma'\right)\left|\beta\right|^{\beta'-\beta+\alpha-\gamma'}}{\sin\left(\pi\left(\alpha-\gamma'+\beta'-\beta\right)\right)\Gamma\left(\alpha\right)\Gamma\left(\beta'\right)\Gamma\left(\beta^{*}\right)} e^{i\frac{\pi}{2}\left(\alpha-\gamma'+\beta'+\beta\right)}\nonumber\\
&
\times\sum_{s,k=0}^{\infty}\left(\begin{array}{c}
\alpha-\gamma\\
s\end{array}\right)\left(\begin{array}{c}
\alpha-\gamma'\\
k\end{array}\right)\left(-\frac{i\beta}{\left|\beta\right|}\right)^{s}\,\left(\frac{i\beta^{*}}{\left|\beta\right|}\right)^{k}J_{\beta-\beta'+\gamma'-\alpha+k+s}\left(-2i\left|\beta\right|\left(1-x\right)\right)\,,
\end{align}
where we used the Euler formula $\Gamma\left(z\right)\Gamma\left(1-z\right)=\frac{\pi}{\sin\left(\pi z\right)}$, and $J_\nu$ is the Bessel function of the first kind of order $\nu$. Plugging back~\eqref{eq:classlimCurlyF1} into~\eqref{eq:IjiAppA}, the classical limit of $I_j^{(1)}$ is found to be
\begin{align} \label{eq:CCIj1orig}
&\lim_{\hbar\to0}\,I_j^{(1)} = \left(\frac{p e}{e^{2}-1}\right)^{j}\left(-1\right)^{j+\Delta \nu+1}2^{-j-2}\frac{i\gamma_\infty e}{\sinh\left(\pi\Delta n\right)}\nonumber\\
&\quad
\times\sum_{s,k=0}^{\infty}\left(\begin{array}{c}
j-\Delta \nu+1\\
s
\end{array}\right)\left(\begin{array}{c}
j+\Delta \nu+1\\
k
\end{array}\right)\left(\frac{1-i\sqrt{e^{2}-1}}{e}\right)^{s-k+\Delta \nu}J_{-j-1+k+s+i\Delta n}\left(-i e\gamma_\infty^2\Delta n\right)\,,\nonumber\\
\end{align}
where $\Delta n = i\left(\Delta j_r+\Delta \nu\right)$, and $\gamma_\infty=\frac{E}{\mu}$.

\subsection{Calculation of $\lim_{\hbar\to0}\,I_j^{(2)}$}\label{App:clc}
In this section, we calculate the classical limit of $I_j^{(2)}$. The calculation proceeds similarly to Appendix~\ref{App:ClassLimCCIj1}.
Hence, we first consider the classical limit of $\mathcal{F}^{(2)}$ in~\eqref{eq:HahneFormCompact}. Moreover, we shall keep the parameters general and assume the same assumptions as Appendix~\ref{App:ClassLimCCIj1}.

We first note that in the classical limit,
\begin{align}
&\,_{3}F_{2}\left(\dots;1\right)\to\,_{2}F_{1}\left(\alpha-\gamma+1+\beta-\beta'+m,\gamma'-\alpha+\beta-\beta'-s,\beta-\beta'+\gamma'-\gamma+1;1-\frac{\beta}{\gamma}\right)\nonumber\\
&\quad
=\left(\frac{\beta}{\gamma}\right)^{s+\alpha-\gamma'+\beta'-\beta}\,_{2}F_{1}\left(\gamma'-\alpha-m,\gamma'-\alpha+\beta-\beta'-s,\beta-\beta'+\gamma'-\gamma+1;1-\frac{\gamma}{\beta}\right)\,.
\end{align}
Hence, the classical limit of $\mathcal{F}^{(2)}$ is given by
\begin{align} \label{eq:ClassLimCurlyF2basic}
&\mathcal{F}^{(2)}\left(\alpha,\beta,\beta',\gamma,\gamma';x,y\right)\to\frac{\Gamma\left(\gamma\right)\Gamma\left(\gamma'\right)\Gamma\left(\gamma-\beta-\gamma'+\beta'\right)\Gamma\left(\gamma'-\alpha+\beta-\beta'\right)}{\Gamma\left(\beta'\right)\Gamma\left(\gamma-\beta\right)\Gamma\left(\gamma'+\beta-\beta'\right)\Gamma\left(\gamma-\alpha\right)}\nonumber\\
&\quad
\times\left(\frac{\beta}{\gamma}\right)^{\alpha-\gamma'+\beta'-\beta}e^{i\pi\left(\alpha+\beta'-\gamma'\right)}\sum_{m,s=0}^{\infty}\frac{\left(\alpha-\gamma+1\right)_{s}}{m!s!\left(\alpha-\gamma'+\beta'-\beta+1\right)_{s}}\left(\beta\left(1-x\right)\right)^{m}\left(\beta\left(1-y\right)\right)^{s}\nonumber\\
&\quad
\times\,_{2}F_{1}\left(\gamma'-\alpha-m,\gamma'-\alpha+\beta-\beta'-s,\beta-\beta'+\gamma'-\gamma+1;1-\frac{\gamma}{\beta}\right)\,.
\end{align}
Using the Euler formula $\Gamma\left(z\right)\Gamma\left(1-z\right)=\frac{\pi}{\sin\left(\pi z\right)}$, we can recast~\eqref{eq:ClassLimCurlyF2basic} as 
\begin{align} \label{eq:ClassLimCurlyF2reg}
&\mathcal{F}^{(2)}\left(\alpha,\beta,\beta',\gamma,\gamma';x,y\right)\to\frac{\pi^2 \Gamma\left(\gamma\right)\Gamma\left(\gamma'\right)}{\Gamma\left(\beta'\right)\Gamma\left(\gamma-\beta\right)\Gamma\left(\gamma'+\beta-\beta'\right)\Gamma\left(\gamma-\alpha\right)}\nonumber\\
&\quad
\times\frac{\left(\frac{\beta}{\gamma}\right)^{\alpha-\gamma'+\beta'-\beta}e^{i\pi\left(\alpha+\beta'-\gamma'\right)}}{\sin\left(\pi\left(\gamma'-\alpha+\beta-\beta'\right)\right)\sin\left(\pi\left(\gamma-\beta-\gamma'+\beta'\right)\right)}\sum_{m,s=0}^{\infty}\frac{\left(\alpha-\gamma+1\right)_{s}\left(\beta\left(1-x\right)\right)^{m}\left(\beta\left(1-y\right)\right)^{s}}{m!s!\Gamma\left(\alpha-\gamma'+\beta'-\beta+s+1\right)}\nonumber\\
&\quad
\times\,_{2}\widetilde{F}_{1}\left(\gamma'-\alpha-m,\gamma'-\alpha+\beta-\beta'-s,\beta-\beta'+\gamma'-\gamma+1;1-\frac{\gamma}{\beta}\right)\,.
\end{align}
Plugging back~\eqref{eq:ClassLimCurlyF2reg} into~\eqref{eq:IjiAppA}, the classical limit of $I_j^{(2)}$ is found to be
\begin{align}\label{eq:CCIj2orig}
&\lim_{\hbar\to0}\,I_j^{(2)} = \left(\frac{p e}{e^{2}-1}\right)^{j}2^{-j-2}e^{\pi\left(\Delta n+i\Delta \nu\right)}\frac{\pi \gamma_\infty e\left(\frac{1-i\sqrt{e^{2}-1}}{e}\right)^{j+\Delta \nu+1-i\Delta n}}{\Gamma\left(-j+\Delta \nu-1\right)\sinh\left(\pi\Delta n\right)\sinh\left(\pi\left(\Delta n+i\Delta \nu\right)\right)}\nonumber\\
&\quad
\times\sum_{m,s=0}^{\infty}\frac{\left(-1\right)^{s}\left(j-\Delta \nu+2\right)_{s}\left(i\frac{1-i\sqrt{e^{2}-1}}{2}\gamma_\infty^2\Delta n\right)^{s+m}}{m!s!\Gamma\left(j+2-i\Delta n+s\right)}\nonumber\\
&\quad
\times\,_{2}\widetilde{F}_{1}\left(-\left(j+\Delta \nu+1+m\right),-j-s+i\Delta n-1;i\Delta n-\Delta \nu+1;\frac{1+i\sqrt{e^{2}-1}}{1-i\sqrt{e^{2}-1}}\right)\,,
\end{align}
where $\Delta n = i\left(\Delta j_r+\Delta \nu\right)$, and $\gamma_\infty=\frac{E}{\mu}$.

\subsection{Derivative Matrix Element}
In this section, we calculate 
\begin{equation}
\mathcal{B}_{j}^{\Delta \ell}\left(\Delta j_r\right)\equiv\lim_{\hbar\rightarrow0}\hbar\left\langle j_r',\ell'\right|\,\frac{r^j}{1+\frac{K}{Er}}\,\partial_{r}\left|j_r,\ell\right\rangle,
\end{equation}
that is relevant for the computation of the vector potential. Denoting the position-space projection of $\left|j_r,\ell\right\rangle$ by $R^\Psi_{j_{r},\ell}$, we have
\begin{eqnarray}
R^\Psi_{j_{r},\ell}(r)=\sqrt{1+\frac{K}{E_{j_{r},\ell}\,r}}\,R_{j_{r},\ell}(r)\,,
\end{eqnarray}
where $R_{j_{r},\ell}$ is given by~\eqref{eq:Regsol}.
It is straightforward to obtain that
\begin{eqnarray} \label{eq:DerRjl}
&&\partial_{r}R^{\Psi}_{j_{r},\ell}(r)=-\frac{K}{2E_{j_{r},\ell}r^{2}\sqrt{\frac{K}{E_{j_{r},\ell}r}+1}}\,R_{j_{r},\ell}\left(r\right)+\sqrt{1+\frac{K}{E_{j_{r},\ell}\,r}}\,\partial_{r} R_{j_{r},\ell}\left(r\right)\,,\\
&&\partial_{r}R_{j_{r},\ell}=-ik_{j_{r},\ell}R_{j_{r},\ell}+\frac{\nu}{r}R_{j_{r},\ell}-2iC_{j_{r},\ell}k_{j_{r},\ell}j_{r}r^{\nu}\,e^{-ik_{j_{r},\ell}r}\,\,_{1}\widetilde{F}_{1}\left(1-j_{r},2\nu+3,2ik_{j_{r},\ell}r\right)\,.\nonumber\\
\end{eqnarray}
The first term in~\eqref{eq:DerRjl} vanishes in the classical limit, so we will drop it. Therefore, using the definition of $\mathcal{A}_{j}^{\Delta\ell}$
and making use of the Gordon integral~\eqref{eq:IntIdenKummer}
\begin{align}\label{eq:Beq}
\mathcal{B}_{j}^{\Delta\ell}\left(s\right)=\lim_{\hbar\rightarrow0}\hbar\left\langle j_{r}',\ell'\right|\,\frac{r^j}{1+\frac{K}{Er}}\,\partial_{r}\left|j_{r},\ell\right\rangle =-i\beta_{\infty}E\mathcal{A}_{j}^{\Delta\ell}\left(s\right)+\nu\mathcal{A}_{j-1}^{\Delta\ell}\left(s\right)-2i\beta_{\infty}E\lim_{\hbar\rightarrow0}j_{r} I_{j}^{d}\,,
\end{align}
where
\begin{align}
&I_{j}^{d}\equiv\,C_{j'_{r},\ell'}C_{j_{r},\ell}\,(2k_{j_r,\ell})^{\nu}(2k_{j'_r,\ell'})^{\nu'}\,\frac{\Gamma\left(\nu+\nu'+j+3\right)}{\Gamma\left(2\nu+3\right)\Gamma\left(2\nu'+2\right)}\left(i\Sigma k\right)^{-\nu-\nu'-j-3}\nonumber\\
&\quad \times\,F_{2}\left(\nu+\nu'+j+3,1-j_{r},-j'_{r},2\nu+3,2\nu'+2,\frac{2k_{j_{r},\ell}}{\Sigma k},\frac{2k_{j'_{r},\ell'}}{\Sigma k}\right)\,\,.
\end{align}
The first two terms are obtained from the results of Appendix~\ref{AjDeltal}. The calculation of the third term is similar to $\mathcal{A}_{j}^{\Delta\ell}$, and
results in
\begin{equation}
\lim_{\hbar\rightarrow0}j_{r}I_{j}^{d}=\sum_{i=1}^{4}\lim_{\hbar\rightarrow0}j_{r}I_{j}^{d,\left(i\right)}\,,
\end{equation}
where
\begin{align}
&\lim_{\hbar\rightarrow0}j_{r}I_{j}^{d,\left(1\right)}=-i\frac{J_{r}}{\left|J_{r}\right|}\left(\frac{pe}{e^{2}-1}\right)^{j}\left(-1\right)^{j+\Delta\nu+1}2^{-j-2}\frac{i\gamma_{\infty}e}{\sinh\left(\pi\Delta n\right)}\nonumber\\
&\quad\times\sum_{s,k=0}^{\infty}\left(\begin{array}{c}
j-\Delta\nu\\
s
\end{array}\right)\left(\begin{array}{c}
j+\Delta\nu+1\\
k
\end{array}\right)\left(\frac{1-i\sqrt{e^{2}-1}}{e}\right)^{s-k+\Delta\nu}J_{-j+k+s+i\Delta n}\left(-ie\gamma_{\infty}^{2}\Delta n\right)\,,
\end{align}
\begin{align}
&\lim_{\hbar\rightarrow0}j_{r} I_{j}^{d,\left(2\right)}=\left(\frac{pe}{e^{2}-1}\right)^{j}2^{-j-2}e^{\pi\left(\Delta n+i\Delta\nu\right)}\frac{\pi\gamma_{\infty}e\left(\frac{1-i\sqrt{e^{2}-1}}{e}\right)^{j+\Delta\nu+1-i\Delta n}}{\Gamma\left(-j+\Delta\nu\right)\sinh\left(\pi\Delta n\right)\sinh\left(\pi\left(\Delta n+i\Delta\nu\right)\right)}\nonumber\\
&\times\sum_{m,s=0}^{\infty}\frac{\left(-1\right)^{s}\left(j-\Delta\nu+1\right)_{s}\left(i\frac{1-i\sqrt{e^{2}-1}}{2}\gamma_{\infty}^{2}\Delta n\right)^{s+m}}{m!s!\Gamma\left(j+1-i\Delta n+s\right)}\nonumber\\
&\quad\times\,_{2}\widetilde{F}_{1}\left(-\left(j+\Delta\nu+1+m\right),-j-s+i\Delta n;i\Delta n-\Delta\nu+1;\frac{1+i\sqrt{e^{2}-1}}{1-i\sqrt{e^{2}-1}}\right)\,,
\end{align}

\begin{align}
&\lim_{\hbar\rightarrow0}j_{r} I_{j}^{d,\left(3\right)}=\left(\frac{pe}{e^{2}-1}\right)^{j}\left(-1\right)^{j-\Delta\nu+1}2^{-j-2}\frac{i\gamma_{\infty}e}{\sinh\left(\pi\Delta n\right)}\nonumber\\
&\quad\times\sum_{s,k=0}^{\infty}\left(\begin{array}{c}
j+\Delta\nu+1\\
s
\end{array}\right)\left(\begin{array}{c}
j-\Delta\nu\\
k
\end{array}\right)\left(\frac{1-i\sqrt{e^{2}-1}}{e}\right)^{s-k-\Delta\nu}J_{-j-1+k+s-i\Delta n}\left(ie\gamma_{\infty}^{2}\Delta n\right)\,,
\end{align}

\begin{align}
&\lim_{\hbar\rightarrow0}j_{r} I_{j}^{d,\left(4\right)}=-\left(\frac{pe}{e^{2}-1}\right)^{j}2^{-j-2}e^{-\pi\left(\Delta n+i\Delta\nu\right)}\frac{\pi\gamma_{\infty}e\left(\frac{1-i\sqrt{e^{2}-1}}{e}\right)^{j-\Delta\nu+1+i\Delta n}}{\Gamma\left(-j-\Delta\nu-1\right)\sinh\left(\pi\Delta n\right)\sinh\left(\pi\left(\Delta n+i\Delta\nu\right)\right)}\nonumber\\
&\quad\times\sum_{m,s=0}^{\infty}\frac{\left(-1\right)^{m}\left(j+\Delta\nu+2\right)_{s}\left(i\frac{1-i\sqrt{e^{2}-1}}{2}\gamma_{\infty}^{2}\Delta n\right)^{s+m}}{m!s!\Gamma\left(j+2+i\Delta n+s\right)}\nonumber\\
&\quad\times\,_{2}\widetilde{F}_{1}\left(-\left(j-\Delta\nu+m\right),-j-s-i\Delta n-1;-i\Delta n+\Delta\nu+1;\frac{1+i\sqrt{e^{2}-1}}{1-i\sqrt{e^{2}-1}}\right)\,.
\end{align}
\section{Classical Limit of Spherical Matrix Elements} \label{App:classspherelem}
The classical limit of the spherical matrix elements was calculated in the first QSM paper~\cite{Khalaf:2023ozy}. Here we state the resulting expressions, 
\begin{equation} \label{eq:classYlmelem}
\lim_{\hbar\rightarrow0}\left\langle \ell',m'\right|Y_{\ell_{\gamma}}^{m_{\gamma}}\left(\hat{r}\right)\left|l,l\right\rangle =\delta_{\ell',m'}\delta_{-\Delta \ell,m_{\gamma}}\,f_{\ell_{\gamma},m_{\gamma}}\,,
\end{equation}
\begin{equation}
\lim_{\hbar\rightarrow0}\left\langle \ell',m'\right|Y_{\ell_{\gamma}}^{m_{\gamma}}\left(\hat{r}\right)\left(\hat{r}\right)_{q}\left|l,l\right\rangle =\delta_{\ell',m'}\delta_{-\Delta \ell,m_{\gamma}+q}\,\frac{1}{\sqrt{2}}\left(\delta_{q,1}-\delta_{q,-1}\right)\,f_{\ell_{\gamma},m_{\gamma}}\,,
\end{equation}
\begin{equation}
\lim_{\hbar\rightarrow0}\,\hbar\left\langle \ell',m'\right|Y_{\ell_{\gamma}}^{m_{\gamma}}\left(\hat{r}\right)\left(\vec{\nabla}_{\Omega}\right)_{q}\left|l,l\right\rangle =-\frac{L}{\sqrt{2}}\delta_{\ell',m'}\delta_{-\Delta \ell,m_{\gamma}+q}\,\left(\delta_{q,1}+\delta_{q,-1}\right)\,f_{\ell_{\gamma},m_{\gamma}}\,,
\end{equation}
where
\begin{equation}\label{flm}
f_{\ell_{\gamma},m_{\gamma}}\equiv Y_{\ell_\gamma}^{-m_\gamma}\left(\frac{\pi}{2},0\right)=\frac{\cos\left[\tfrac{\pi(\ell_{\gamma}-m_{\gamma})}{2}\right]}{2\pi}\sqrt{\frac{\left(2\ell_{\gamma}+1\right)\Gamma\left(\frac{\ell_{\gamma}+m_{\gamma}+1}{2}\right)\Gamma\left(\frac{\ell_{\gamma}-m_{\gamma}+1}{2}\right)}{\Gamma\left(\frac{\ell_{\gamma}+m_{\gamma}}{2}+1\right)\Gamma\left(\frac{\ell_{\gamma}-m_{\gamma}}{2}+1\right)}}\,\,.
\end{equation}
\section{Non-Relativistic Case: Resummation of the 1PL Expansion of $\widetilde{A}^t$} \label{AppendixB}
In this section, we first expand the non-relativistic $\widetilde{A}^t$, obtained using the QSM, to first order in the coupling $K$ in Subsection~\ref{App:nonrelAtexp}. We then resum it in Subsection~\ref{App:nonrelAtresum} into the non-relativistic limit of \eqref{eq:CSExp}. 
 Auxiliary calculations are provided in Subsection~\ref{App:nonrelAux}.

\subsection{First-Order Expansion in $K$} \label{App:nonrelAtexp}
We are interested in expanding~\eqref{eq:Atnonrel} up to first order in $K$, while keeping $\beta_\infty$ and $L$ fixed. This is equivalent to expanding~\eqref{eq:AjDeltalnonrel} up to $\mathcal{O}\left(K\right)$, as we do in the two following subsections. It might seem that~\eqref{eq:AjDeltalnonrel} behaves as $1/K^2$ at leading order in $K$, but as we show in Subsection~\ref{App:subsubK-2}, it vanishes as it should, since otherwise $\hat{A}^t$ would diverge as $K\to0$. There is also a $1/K$ term which corresponds to a memory effect; however, since the 1PL computation in \eqref{section:1pll} does not include it, we drop it for an apples-to-apples comparison. As such, the $K^0$ order expression is the one of interest here, and we calculate it in Subsection~\ref{App:subsubK0}.

\subsubsection{$1/K^{2}$ Singularity Cancellation} \label{App:subsubK-2}
The expression for the $1/K^{2}$ term is obtained using~\eqref{eq:AjDeltalnonrel} at leading order, yielding
\begin{align} \label{eq:1/K2term}
&\left(\frac{L}{\mu\beta_\infty}\right)^j\left(-1\right)^{j+\Delta \ell}2^{-j-1}\frac{\mu\beta_\infty^4L}{\pi\omega}\cos\left(\frac{\pi}{2}\left(j-\Delta \ell\right)\right)\nonumber\\
&\quad
\times\sum_{s=0}^{j-\Delta \ell+1}\left(-1\right)^{s}\,\sum_{k=0}^{j+\Delta \ell+1}\left(\begin{array}{c}
j-\Delta \ell+1\\
s
\end{array}\right)\left(\begin{array}{c}
j+\Delta \ell+1\\
k
\end{array}\right)I_{-j-1+k+s}\left(\frac{\omega L}{\mu\beta_\infty^2}\right)\,.
\end{align}
Here, $I_\nu$ is the modified Bessel function of the first kind of order $\nu$.
Note that if $j-\Delta \ell$ is odd, then~\eqref{eq:1/K2term} vanishes.
If otherwise $j-\Delta \ell$ is even, then one can show that (see Subsection~\ref{App:nonrelAux})
\begin{equation}
\sum_{s=0}^{j-\Delta \ell+1}\left(-1\right)^{s}\,\sum_{k=0}^{j+\Delta \ell+1}\left(\begin{array}{c}
j-\Delta \ell+1\\
s
\end{array}\right)\left(\begin{array}{c}
j+\Delta \ell+1\\
k
\end{array}\right)I_{-j-1+k+s}\left(\frac{\omega L}{\mu\beta_\infty^2}\right)=0\,\,.\label{eq:Besseliden1}
\end{equation}
Hence,~\eqref{eq:1/K2term} generally vanishes and the $1/K^{2}$ terms cancels as they should.
\subsubsection{$K^{0}$ Term} \label{App:subsubK0}
A proof analogous to that of~\eqref{eq:Besseliden1} shows that
\begin{equation}
\sum_{s=0}^{j-\Delta \ell+1}\left(-1\right)^{s}\sum_{k=0}^{j+\Delta \ell+1}\left(k-s-\Delta \ell\right)\left(\begin{array}{c}
j-\Delta \ell+1\\
s
\end{array}\right)\left(\begin{array}{c}
j+\Delta \ell+1\\
k
\end{array}\right)I_{-j-1+k+s}\left(\frac{\omega L}{\mu\beta_\infty^2}\right)=0
\end{equation}
when $j-\Delta \ell$ is odd, and
\begin{equation}
\sum_{s=0}^{j-\Delta \ell+1}\left(-1\right)^{s}\sum_{k=0}^{j+\Delta \ell+1}\left(k-s-\Delta \ell\right)^{2}\left(\begin{array}{c}
j-\Delta \ell+1\\
s
\end{array}\right)\left(\begin{array}{c}
j+\Delta \ell+1\\
k
\end{array}\right)I_{-j-1+k+s}\left(\frac{\omega L}{\mu\beta_\infty^2}\right)=0
\end{equation}
when $j-\Delta \ell$ is even. As such, contributions to the $K^{0}$ expression that do not involve a derivative of the order of the Bessel function vanish. Hence, up to order $K^{0}$ and neglecting the memory term, we get
\begin{align}\label{eq:matelemK0}
&\mathcal{A}_{j}^{\Delta \ell}\left(m_\gamma-i\frac{\omega}{\Omega_r}\right) =\left(\frac{L}{\mu\beta_\infty}\right)^{j}\left(-1\right)^{j+\Delta \ell}2^{-j-1}\frac{\mu\beta_\infty^4 L}{4\pi\omega}\sum_{s=0}^{j-\Delta \ell+1}\left(-1\right)^{s}\sum_{k=0}^{j+\Delta \ell+1}\left(\begin{array}{c}
j-\Delta \ell+1\\
s
\end{array}\right)\left(\begin{array}{c}
j+\Delta \ell+1\\
k
\end{array}\right)\nonumber\\
&\quad {\rm Re}\Biggl[\exp\left(\frac{i\pi}{2}\left(j-\Delta \ell\right)\right)\sum_{m=0}^{1}\left(\frac{i \omega}{\mu \beta_\infty^3}\right)^{1+m}\left\{ \left(\frac{2\pi\omega}{\mu \beta_\infty^3}\right)^{1-m}+\left(-\frac{4i\left(k-s-\Delta \ell\right)}{\beta_\infty L}\right)^{1-m}\right\} \nonumber\\
&\quad \hspace{5.5cm}\times\left.\frac{d^{1+m}}{d\nu^{1+m}}I_{\nu}\left(\frac{\omega L}{\mu\beta_\infty^2}\right)\right|_{\nu=-j-1+k+s}\Biggr]\,.
\end{align}
Changing variables from $\left(k,s\right)$ to $\left(u,s\right)\equiv\left(k+s-\left(j+1\right),s\right)$,
it is possible to simplify this expression to (see Subsection~\ref{App:nonrelAux})
)
\begin{align}
&\mathcal{A}_{j}^{\Delta \ell}\left(m_\gamma-i\frac{\omega}{\Omega_r}\right) =\left(\frac{L}{\mu\beta_\infty}\right)^{j} 2^{-j-1}\frac{\omega L}{2\pi \mu\beta_\infty^2}\sum_{u=-\left(j+1\right)}^{j+1}\sum_{s=u-\Delta \ell}^{j+1+u}\left(-1\right)^{s+u}\left(\begin{array}{c}
j-\Delta \ell+1\\
s
\end{array}\right)\left(\begin{array}{c}
j+\Delta \ell+1\\
j+1+u-s
\end{array}\right)\nonumber\\
&\times\sum_{m=0}^{1}\left\{ -\cos\left(\frac{\pi}{2}\left(j-\Delta \ell+1+m\right)\right)\left(-\pi\right)^{1-m}+\cos\left(\frac{\pi}{2}\left(j-\Delta \ell\right)\right)\left(-\frac{2\mu\beta_\infty^2}{\omega L}\left(j+1+u-2s-\Delta \ell\right)\right)^{1-m}\right\} \nonumber\\
&\quad\hspace{5.5cm}\times\left.\frac{d^{m}}{d\nu^{m}}K_{\nu}\left(\frac{\omega L}{\mu\beta_\infty^2}\right)\right|_{\nu=u}\,\,.\label{eq:matelemK0simp1}
\end{align}

For the calculation of the scalar potential below, we will only need
the case where $j-\Delta \ell$ is even (see~\eqref{eq:jlmatelemNR}), for which we can simplify
further
\begin{align}\label{eq:matelemK0simp2}
&\mathcal{A}_{j}^{\Delta \ell}\left(m_\gamma-i\frac{\omega}{\Omega_r}\right) =\left(\frac{L}{2\mu\beta_\infty}\right)^{j}\frac{\omega L}{2\pi \mu\beta_\infty^2}\cos\left(\frac{\pi}{2}\left(j-\Delta \ell\right)\right)\sum_{u=-\left(j+1\right)}^{j+1}\sum_{s=u-\Delta \ell}^{j+1+u}\left(-1\right)^{s+u}\left(\protect\begin{array}{c}
j-\Delta \ell+1\protect\\
s
\protect\end{array}\right)\nonumber\\
&\quad
\times\left(\protect\begin{array}{c}
j+\Delta \ell+1\protect\\
j+1+u-s
\protect\end{array}\right)\left\{ \left.\frac{d}{d\nu}K_{\nu}\left(\frac{\omega L}{\mu\beta_\infty^2}\right)\right|_{\nu=u}-\frac{\mu\beta_\infty^2}{\omega L}\left(j+1+u-2s-\Delta \ell\right)K_{u}\left(\frac{\omega L}{\mu\beta_\infty^2}\right)\right\}\,.
\end{align}
Reverting back to the summation
variables $\left(k,s\right)$ by the transformation $\left(k,s\right)\equiv\left(j+1+u-s,s\right)$ yields
\begin{align}\label{eq:matelemnonrelinks}
&\mathcal{A}_{j}^{\Delta \ell}\left(m_\gamma-i\frac{\omega}{\Omega_r}\right) =\left(\frac{L}{2\mu\beta_\infty}\right)^{j}\frac{\omega L}{2\pi \mu\beta_\infty^2}\cos\left(\frac{\pi}{2}\left(j-\Delta \ell\right)\right)\sum_{k=0}^{j+\Delta \ell+1}\sum_{s=0}^{j-\Delta \ell+1}\left(-1\right)^{k+j+1}\left(\protect\begin{array}{c}
j-\Delta \ell+1\protect\\
s
\protect\end{array}\right)\nonumber\\
&\quad
\times\left(\protect\begin{array}{c}
j+\Delta \ell+1\protect\\
k
\protect\end{array}\right)\left\{ \left.\frac{d}{d\nu}K_{\nu}\left(\frac{\omega L}{\mu\beta_\infty^2}\right)\right|_{\nu=k+s-j-1}-\frac{\mu\beta_\infty^2}{\omega L}\left(k-s-\Delta \ell\right)K_{k+s-j-1}\left(\frac{\omega L}{\mu\beta_\infty^2}\right)\right\}\,.
\end{align}
\subsubsection{First-Order Expression of $\widetilde{A}^t\left(\omega,\vec{x}\right)$}
Using~\eqref{eq:Atnonrel},~\eqref{eq:jlmatelemNR}, and~\eqref{eq:matelemK0simp2}, we obtain the first-order expansion (in $K$) of $\widetilde{A}^t\left(\omega,\vec{x}\right)$
\begin{align} \label{App:AtnonrelOrderK}
&4\pi |\vec{x}|\widetilde{A}^{t}\left(\omega,\vec{x}\right)=-\frac{q K}{ \mu{\pi}\beta_\infty^3}\,\sum_{\ell_{\gamma}=0}^{\infty}\sum_{m_{\gamma}=-\ell_{\gamma}}^{\ell_{\gamma}}(-1)^{m_\gamma}(-i)^{\ell_{\gamma}}\,Y_{\ell_{\gamma}}^{m_{\gamma}}\left(\theta,\varphi\right)\, \cos^2\left(\tfrac{\pi(\ell_{\gamma}-m_{\gamma})}{2}\right)\nonumber\\
&\quad \times
\sqrt{\frac{\left(2\ell_{\gamma}+1\right)\Gamma\left(\frac{\ell_{\gamma}+m_{\gamma}+1}{2}\right)\Gamma\left(\frac{\ell_{\gamma}-m_{\gamma}+1}{2}\right)}{\Gamma\left(\frac{\ell_{\gamma}+m_{\gamma}}{2}+1\right)\Gamma\left(\frac{\ell_{\gamma}-m_{\gamma}}{2}+1\right)}}\,\sum_{\kappa=0}^{\infty}\frac{\left(\kappa+\ell_{\gamma}\right)!}{\kappa!\left(2\kappa+2 \ell_{\gamma}+1\right)!4^\kappa} \,\left(\xi \beta_\infty\right)^{2\kappa+\ell_{\gamma}}\,
\nonumber\\
&\quad
\times\sum_{k=0}^{2\kappa+\ell_{\gamma}-m_\gamma+1}\sum_{s=0}^{2\kappa+\ell_{\gamma}+m_\gamma+1}\left(-1\right)^{k}\left(\protect\begin{array}{c}
2\kappa+\ell_{\gamma}+m_\gamma+1\protect\\
s
\protect\end{array}\right) \left(\protect\begin{array}{c}
2\kappa+\ell_{\gamma}-m_\gamma+1\protect\\
k
\protect\end{array}\right)\nonumber\\
&\quad
\times\left\{ \xi \left.\frac{d}{d\nu}K_{\nu}\left(\xi\right)\right|_{\nu=k+s-2\kappa-\ell_{\gamma}-1}-\left(k-s+m_\gamma\right)K_{k+s-2\kappa-\ell_{\gamma}-1}\left(\xi\right)\right\}\,.
\end{align}
where we defined $\xi\equiv\frac{\omega L}{\mu\beta_\infty^2}$.
\subsection{Resummation} \label{App:nonrelAtresum}
In this section, we show that~\eqref{App:AtnonrelOrderK} resums into the non-relativistic limit of~\eqref{eq:CSExp}. The key identity to do the resummation is 

\begin{align}
&\sum_{k=0}^{1+2j_{-}}\sum_{s=0}^{1+2j_{+}}\left(-1\right)^{k}\left(\begin{array}{c}
1+2j_{+}\\
s
\end{array}\right)\left(\begin{array}{c}
1+2j_{-}\\
k
\end{array}\right)
\Bigg\{ \xi \frac{d}{d\nu}K_{\nu}\left(\xi\right)\Bigg|_{\nu=k+s-\left(1+j_{-}+j_{+}\right)}\nonumber\\
&\quad -\left(k+j_{+}-s-j_{-}\right)K_{k+s-\left(1+j_{-}+j_{+}\right)}\left(\xi\right)\Bigg\} =\left(-1\right)^{j_{-}}2^{1+j_{-}+j_{+}}\frac{\Gamma\left(j_{+}+1\right)\Gamma\left(j_{-}+1\right)}{\Gamma\left(j_{-}+j_{+}\right)}\nonumber\\
&\hspace{1cm}\times\sum_{u=0}^{j_{-}+j_{+}-1}\left(\begin{array}{c}
j_{-}+j_{+}-1\\
u
\end{array}\right) \Bigg\{ f_{u;j_
{-}+j_{+}-u;j_{-}-j_{+}}K_{1}^{\left(1,j_{-}+j_{+}-1-u\right)}\left(\xi\right)\nonumber\\
&\hspace{1cm}-f_{u+1;j_{-}+j_{+}-1-u;j_{-}-j_{+}}\frac{K_{2}^{\left(0,j_{-}+j_{+}-1-u\right)}\left(\xi\right)-K_{0}^{\left(0,j_{-}+j_{+}-1-u\right)}\left(\xi\right)}{2}\Bigg\} \,\,.
\end{align}
where 
\begin{equation}
f_{k;n;m}\equiv\sum_{j=0}^{n}\left(-1\right)^{j}\left(\begin{array}{c}
n\\
j
\end{array}\right)\left(\begin{array}{c}
k\\
j+\frac{k+m-n}{2}
\end{array}\right)\,\,,
\end{equation}
and the notation $K_{i}^{\left(j,l\right)}\left(\xi\right)$ stands
for $\left.\frac{\partial^{l+j}}{\partial\nu^{j}\partial\xi^{l}}K_{\nu}\left(\xi\right)\right|_{\nu=i}$.
We note that the identity is valid for $j_{-}+j_{+}\ge1$. This can be proven via the resummation identity \cite{NIST:DLMFE7}. Applying this identity for $j_{\pm}=\kappa+\frac{\ell_{\gamma}\pm m_{\gamma}}{2}$
(this satisfies the $j_{-}+j_{+}\ge1$ criterion since the monopole
actually vanishes), we can replace~\eqref{App:AtnonrelOrderK} with

\begin{align}
&4\pi |\vec{x}|\widetilde{A}^{t}\left(\omega,\vec{x}\right)=-\frac{q K}{ \mu{\pi}\beta_\infty^3}\sum_{\ell_{\gamma}=0}^{\infty}\sum_{m_{\gamma}=-\ell_{\gamma}}^{\ell_{\gamma}}(-1)^{m_\gamma}\left(-i\right)^{\ell_{\gamma}}Y_{\ell_{\gamma}}^{m_{\gamma}}\left(\theta,\phi\right)\cos^{2}\left(\frac{\pi}{2}\left(\ell_{\gamma}-m_{\gamma}\right)\right)\nonumber\\
&\quad\times\sqrt{\frac{\left(2\ell_{\gamma}+1\right)\Gamma\left(\frac{\ell_{\gamma}+m_{\gamma}+1}{2}\right)\Gamma\left(\frac{\ell_{\gamma}-m_{\gamma}+1}{2}\right)}{\Gamma\left(\frac{\ell_{\gamma}+m_{\gamma}}{2}+1\right)\Gamma\left(\frac{\ell_{\gamma}-m_{\gamma}}{2}+1\right)}}\,\sum_{\kappa=0}^{\infty}\frac{\left(\kappa+\ell_{\gamma}\right)!\,\left(\xi \beta_\infty\right)^{2\kappa+\ell_{\gamma}}}{\kappa!\left(2\kappa+2\ell_{\gamma}+1\right)!4^{\kappa}}\,\left(-1\right)^{\kappa+\frac{\ell_{\gamma}-m_{\gamma}}{2}}\nonumber\\
&\hspace{1cm}\times 2^{2\kappa+\ell_{\gamma}+1}\frac{\Gamma\left(\kappa+\frac{\ell_{\gamma}-m_{\gamma}}{2}+1\right)\Gamma\left(\kappa+\frac{\ell_{\gamma}+m_{\gamma}}{2}+1\right)}{\Gamma\left(2\kappa+\ell_{\gamma}\right)}\,
\sum_{u=0}^{2\kappa+\ell_{\gamma}-1}\left(\begin{array}{c}
2\kappa+\ell_{\gamma}-1\\
u
\end{array}\right)\nonumber\\
&\hspace{1cm}\times\Bigg\{ f_{u;2\kappa+\ell_{\gamma}-u;-m_{\gamma}}K_{1}^{\left(1,2\kappa+\ell_{\gamma}-1-u\right)}\left(\xi\right)\nonumber\\
&\hspace{2cm}-f_{u+1;2\kappa+\ell_{\gamma}-1-u;-m_{\gamma}}\frac{K_{2}^{\left(0,2\kappa+\ell_{\gamma}-1-u\right)}\left(\xi\right)-K_{0}^{\left(0,2\kappa+\ell_{\gamma}-1-u\right)}\left(\xi\right)}{2}\Bigg\} \,\,.
\end{align}

One can show that
\begin{align}
&\left(-1\right)^{\kappa+\frac{\ell_{\gamma}-m_{\gamma}}{2}}2^{2\kappa+\ell_{\gamma}+1}\frac{\Gamma\left(\kappa+\frac{\ell_{\gamma}-m_{\gamma}}{2}+1\right)\Gamma\left(\kappa+\frac{\ell_{\gamma}+m_{\gamma}}{2}+1\right)}{\Gamma\left(2\kappa+\ell_{\gamma}\right)}\sqrt{\frac{\Gamma\left(\frac{\ell_{\gamma}+m_{\gamma}+1}{2}\right)\Gamma\left(\frac{\ell_{\gamma}-m_{\gamma}+1}{2}\right)}{\Gamma\left(\frac{\ell_{\gamma}+m_{\gamma}}{2}+1\right)\Gamma\left(\frac{\ell_{\gamma}-m_{\gamma}}{2}+1\right)}}\nonumber\\
&\hspace{1cm}\times\frac{\left(\kappa+\ell_{\gamma}\right)!}{\kappa!\left(2\kappa+2\ell_{\gamma}+1\right)!}=\frac{\sqrt{\pi}\left(-1\right)^{\kappa+\ell_{\gamma}}}{\left(2\kappa+\ell_{\gamma}-1\right)!}\sqrt{\frac{\left(\ell_{\gamma}+m_{\gamma}\right)!}{\left(\ell_{\gamma}-m_{\gamma}\right)!}}\binom{\frac{1}{2}\left(\ell_{\gamma}-m_{\gamma}-1\right)}{\ell_{\gamma}}\nonumber\\
&\hspace{2cm}\times\frac{2\,\Gamma\left(\kappa+\frac{\ell_{\gamma}-m_{\gamma}}{2}+\frac{3}{2}\right)\Gamma\left(\kappa+1+\frac{\ell_{\gamma}+m_{\gamma}}{2}\right)\ell_{\gamma}!\left(2\kappa+\ell_{\gamma}-m_{\gamma}\right)!!}{\kappa!\Gamma\left(\kappa+\ell_{\gamma}+\frac{3}{2}\right)\Gamma\left(\ell_{\gamma}+m_{\gamma}+1\right)\left(2\kappa+\ell_{\gamma}-m_{\gamma}+1\right)\text{!!}}\,,
\end{align}
using which we can recast the potential as follows,
\begin{align}
&4\pi |\vec{x}|\widetilde{A}^{t}\left(\omega,\vec{x}\right)=-\frac{{2}q K}{ \mu{\pi}\beta_\infty^3}\sum_{\ell_{\gamma}=0}^{\infty}\sum_{m_{\gamma}=-\ell_{\gamma}}^{\ell_{\gamma}}(-1)^{m_\gamma}\left(-i\right)^{\ell_{\gamma}}Y_{\ell_{\gamma}}^{m_{\gamma}}\left(\theta,\phi\right)\cos^{2}\left(\frac{\pi}{2}\left(\ell_{\gamma}-m_{\gamma}\right)\right)\nonumber\\
&\quad\times\sqrt{\frac{\pi\left(2\ell_{\gamma}+1\right)\left(\ell_{\gamma}+m_{\gamma}\right)!}{\left(\ell_{\gamma}-m_{\gamma}\right)!}}\binom{\frac{1}{2}\left(\ell_{\gamma}-m_{\gamma}-1\right)}{\ell_{\gamma}}\nonumber\\
&\quad\quad\times\sum_{\kappa=0}^{\infty}\,\frac{\left(-1\right)^{\kappa+\ell_{\gamma}}\,\left(\xi \beta_\infty\right)^{2\kappa+\ell_{\gamma}}\,\Gamma\left(\kappa+\frac{\ell_{\gamma}-m_{\gamma}}{2}+\frac{3}{2}\right)\Gamma\left(\kappa+1+\frac{\ell_{\gamma}+m_{\gamma}}{2}\right)\ell_{\gamma}!\left(2\kappa+\ell_{\gamma}-m_{\gamma}\right)!!}{2^{2\kappa}\kappa!\Gamma\left(2\kappa+\ell_{\gamma}\right)\Gamma\left(\kappa+\ell_{\gamma}+\frac{3}{2}\right)\Gamma\left(\ell_{\gamma}+m_{\gamma}+1\right)\left(2\kappa+\ell_{\gamma}-m_{\gamma}+1\right)\text{!!}}\nonumber\\
&\quad\quad\quad\times\sum_{u=0}^{2\kappa+\ell_{\gamma}-1}\left(\begin{array}{c}
2\kappa+\ell_{\gamma}-1\\
u
\end{array}\right)
\Bigg\{ f_{u;2\kappa+\ell_{\gamma}-u;-m_{\gamma}}K_{1}^{\left(1,2\kappa+\ell_{\gamma}-1-u\right)}\left(\xi\right)\nonumber\\
&\hspace{2cm}-f_{u+1;2\kappa+\ell_{\gamma}-1-u;-m_{\gamma}}\frac{K_{2}^{\left(0,2\kappa+\ell_{\gamma}-1-u\right)}\left(\xi\right)-K_{0}^{\left(0,2\kappa+\ell_{\gamma}-1-u\right)}\left(\xi\right)}{2}\Bigg\} \,\,.
\end{align}

We now use the identities (which can be shown by explicit calculation)

\begin{align}
&\cos^{2}\left(\frac{\pi}{2}\left(\ell_{\gamma}-m_{\gamma}\right)\right)f_{u;2\kappa+\ell_{\gamma}-u;-m_{\gamma}}=i^{u+m_\gamma}2^{2\kappa+\ell_{\gamma}}\left(-1\right)^{\kappa+\frac{\ell_{\gamma}+m_{\gamma}}{2}}\nonumber\\
&\hspace{2cm}\times\frac{1}{2\pi}\int_{0}^{2\pi}\cos^{u}\phi\sin^{2\kappa+\ell_{\gamma}-u}\phi\,e^{-i m_{\gamma}\phi}\,\,d\phi\,,
\end{align}

\begin{align}
&2^{\ell_{\gamma}}\binom{\frac{1}{2}\left(\ell_{\gamma}-m_{\gamma}-1\right)}{\ell_{\gamma}}\frac{\Gamma\left(\kappa+\frac{\ell_{\gamma}-m_{\gamma}}{2}+\frac{3}{2}\right)\Gamma\left(\kappa+1+\frac{\ell_{\gamma}+m_{\gamma}}{2}\right)\ell_{\gamma}!\left(2\kappa+\ell_{\gamma}-m_{\gamma}\right)!!}{\kappa!\Gamma\left(\kappa+\ell_{\gamma}+\frac{3}{2}\right)\Gamma\left(\ell_{\gamma}+m_{\gamma}+1\right)\left(2\kappa+\ell_{\gamma}-m_{\gamma}+1\right)\text{!!}}\nonumber\\
&\hspace{2cm}=(-1)^{m_\gamma}\frac{\Gamma\left(\ell_\gamma-m_\gamma+1\right)}{2\,\Gamma\left(\ell_\gamma+m_\gamma+1\right)}\int_{0}^{\pi}\sin^{2\kappa+\ell_{\gamma}+1}\theta\,\,P_{\ell_{\gamma}}^{m_{\gamma}}\left(\cos\theta\right)\,\,d\theta\,,
\end{align}
so that
\begin{multline}
4\pi |\vec{x}|\widetilde{A}^{t}\left(\omega,\vec{x}\right)=-\frac{q K}{ \mu{\pi}\beta_\infty^3}\sum_{\ell_{\gamma}=0}^{\infty}\sum_{m_{\gamma}=-\ell_{\gamma}}^{\ell_{\gamma}}Y_{\ell_{\gamma}}^{m_{\gamma}}\left(\theta,\phi\right)\sqrt{\frac{\left(2\ell_{\gamma}+1\right)\left(\ell_{\gamma}-m_{\gamma}\right)!}{4\pi\left(\ell_{\gamma}+m_{\gamma}\right)!}}\\
\times\sum_{\kappa=0}^{\infty}\,\frac{\left(\xi \beta_\infty\right)^{2\kappa+\ell_{\gamma}}}{\Gamma\left(2\kappa+\ell_{\gamma}\right)}\int_{0}^{\pi}\sin^{2\kappa+\ell_{\gamma}+1}\theta'\,\,P_{\ell_{\gamma}}^{m_{\gamma}}\left(\cos\theta'\right)\,\,d\theta'\sum_{u=0}^{2\kappa+\ell_{\gamma}-1}i^{u}\left(\begin{array}{c}
2\kappa+\ell_{\gamma}-1\\
u
\end{array}\right)\\
\times\Biggl\{ K_{1}^{\left(1,2\kappa+\ell_{\gamma}-1-u\right)}\left(\xi\right)\int_{0}^{2\pi}\cos^{u}\phi'\sin^{2\kappa+\ell_{\gamma}-u}\phi'\,e^{-im_{\gamma}\phi'}\,\,d\phi'\\
-i\frac{K_{2}^{\left(0,2\kappa+\ell_{\gamma}-1-u\right)}\left(\xi\right)-K_{0}^{\left(0,2\kappa+\ell_{\gamma}-1-u\right)}\left(\xi\right)}{2}\int_{0}^{2\pi}\cos^{u+1}\phi'\sin^{2\kappa+\ell_{\gamma}-1-u}\phi'\,e^{-im_{\gamma}\phi'}\,\,d\phi'\Biggr\}\,.
\end{multline}
We recast the potential as

\begin{multline}
4\pi |\vec{x}|\widetilde{A}^{t}\left(\omega,\vec{x}\right)=-\frac{q K}{ \mu{\pi}\beta_\infty^3}\sum_{\ell_{\gamma}=0}^{\infty}\sum_{m_{\gamma}=-\ell_{\gamma}}^{\ell_{\gamma}}(-1)^{m_\gamma}Y_{\ell_{\gamma}}^{m_{\gamma}}\left(\theta,\phi\right)\sum_{\kappa=0}^{\infty}\,\frac{\left(\xi \beta_\infty\right)^{2\kappa+\ell_{\gamma}}}{\Gamma\left(2\kappa+\ell_{\gamma}\right)}\\
\times\int\sin^{2\kappa+\ell_{\gamma}}\theta'\sum_{u=0}^{2\kappa+\ell_{\gamma}-1}i^{u}\left(\begin{array}{c}
2\kappa+\ell_{\gamma}-1\\
u
\end{array}\right)\Biggl\{ K_{1}^{\left(1,2\kappa+\ell_{\gamma}-1-u\right)}\left(\xi\right)\cos^{u}\phi'\sin^{2\kappa+\ell_{\gamma}-u}\phi'\\
-i\frac{K_{2}^{\left(0,2\kappa+\ell_{\gamma}-1-u\right)}\left(\xi\right)-K_{0}^{\left(0,2\kappa+\ell_{\gamma}-1-u\right)}\left(\xi\right)}{2}\cos^{u+1}\phi'\sin^{2\kappa+\ell_{\gamma}-1-u}\phi'\Biggr\}\,\,Y_{\ell_{\gamma}}^{m_{\gamma}*}\left(\theta',\phi'\right)\,\,d\Omega'\,\,,
\end{multline}

We now identify the sums over $u$ as follows
\begin{multline}
\sum_{u=0}^{2\kappa+\ell_{\gamma}-1}i^{u}\left(\begin{array}{c}
2\kappa+\ell_{\gamma}-1\\
u
\end{array}\right)K_{1}^{\left(1,2\kappa+\ell_{\gamma}-1-u\right)}\left(\xi\right)\cos^{u}\phi\sin^{2\kappa+\ell_{\gamma}-u}\phi\\
=\frac{\left(-1\right)^{\ell_{\gamma}-1}\sin\phi}{\xi^{2\kappa+\ell_{\gamma}}\sin^{2\kappa+\ell_{\gamma}-1}\theta}\frac{d^{2\kappa+\ell_{\gamma}-1}}{d\beta_\infty^{2\kappa+\ell_{\gamma}-1}}\left\{ \exp\left(-i\xi \beta_\infty\sin\theta\cos\phi\right)\frac{K_{0}\left(\xi\left(1-\beta_\infty\sin\theta\sin\phi\right)\right)}{1-\beta_\infty\sin\theta\sin\phi}\right\} _{v=0}\,,
\end{multline}
and

\begin{multline}
\sum_{u=0}^{2\kappa+\ell_{\gamma}-1}i^{u}\left(\begin{array}{c}
2\kappa+\ell_{\gamma}-1\\
u
\end{array}\right)\frac{K_{2}^{\left(0,2\kappa+\ell_{\gamma}-1-u\right)}\left(\xi\right)-K_{0}^{\left(0,2\kappa+\ell_{\gamma}-1-u\right)}\left(\xi\right)}{2}\cos^{u+1}\phi\sin^{2\kappa+\ell_{\gamma}-1-u}\phi\\
=\frac{\left(-1\right)^{\ell_{\gamma}-1}\cos\phi}{\xi^{2\kappa+\ell_{\gamma}}\sin^{2\kappa+\ell_{\gamma}-1}\theta}\frac{d^{2\kappa+\ell_{\gamma}-1}}{d\beta_\infty^{2\kappa+\ell_{\gamma}-1}}\left\{ \exp\left(-i\xi \beta_\infty\sin\theta\cos\phi\right)\frac{K_{1}\left(\xi\left(1-\beta_\infty\sin\theta\sin\phi\right)\right)}{1-\beta_\infty\sin\theta\sin\phi}\right\} _{\beta_\infty=0}\,,
\end{multline}
hence
\begin{align}
&4\pi |\vec{x}|\widetilde{A}^{t}\left(\omega,\vec{x}\right)=\frac{q K}{ \mu{\pi}\beta_\infty^2}\sum_{\ell_{\gamma}=0}^{\infty}\sum_{m_{\gamma}=-\ell_{\gamma}}^{\ell_{\gamma}}Y_{\ell_{\gamma}}^{m_{\gamma}}\left(\theta,\phi\right)\sum_{\kappa=0}^{\infty}\,\frac{\beta_\infty^{2\kappa+\ell_{\gamma}-1}}{\Gamma\left(2\kappa+\ell_{\gamma}\right)}\nonumber\\
&\quad\times\frac{d^{2\kappa+\ell_{\gamma}-1}}{d\beta_\infty^{2\kappa+\ell_{\gamma}-1}}\Bigg[\int\frac{\sin\theta'\exp\left(-i\xi \beta_\infty\sin\theta'\cos\phi'\right)}{1-\beta_\infty\sin\theta'\sin\phi'}\Big\{ \sin\phi'K_{0}\left(\xi\left(1-\beta_\infty\sin\theta'\sin\phi'\right)\right)\nonumber\\
&\hspace{2cm} -i\cos\phi'\,K_{1}\left(\xi\left(1-\beta_\infty\sin\theta'\sin\phi'\right)\right)\Big\}
\,Y_{\ell_{\gamma}}^{m_{\gamma}*}\left(\theta',\phi'\right)\,\,d\Omega'\Bigg]_{\beta_\infty=0}\,,
\end{align}

It is possible to show that the derivative term above, namely $\frac{d^{k}}{d\beta_\infty^{k}}\left\{ \int\cdots\right\} _{v=0}$
for a generic $k$ actually vanishes unless $k\ge \ell_{\gamma}-1$ and
$k$ have the same parity as $\ell_{\gamma}-1$. In other words, we can
extend the sum over $\kappa$ as follows
\begin{align}
&4\pi |\vec{x}|\widetilde{A}^{t}\left(\omega,\vec{x}\right)=\frac{q K}{ \mu{\pi}\beta_\infty^2}\sum_{\ell_{\gamma}=0}^{\infty}\sum_{m_{\gamma}=-\ell_{\gamma}}^{\ell_{\gamma}}Y_{\ell_{\gamma}}^{m_{\gamma}}\left(\theta,\phi\right)\sum_{\kappa=0}^{\infty}\,\frac{\beta_\infty^{\kappa}}{\kappa!}\nonumber\\
&\quad\times\frac{d^{\kappa}}{d\beta_\infty^{\kappa}}\Bigg[\int\frac{\sin\theta'\exp\left(-i\xi \beta_\infty\sin\theta'\cos\phi'\right)}{1-\beta_\infty\sin\theta'\sin\phi'}\Big\{ \sin\phi'\,K_{0}\left(\xi\left(1-\beta_\infty\sin\theta'\sin\phi'\right)\right)\nonumber\\
&\hspace{2cm}-i\cos\phi'K_{1}\left(\xi\left(1-\beta_\infty\sin\theta'\sin\phi'\right)\right)\Big\}
\,Y_{\ell_{\gamma}}^{m_{\gamma}*}\left(\theta',\phi'\right)\,\,d\Omega'\Bigg]_{\beta_\infty=0}\,.
\end{align}
Finally, we identify the Taylor expansions (in $\beta_\infty$), and the spherical harmonic expansion to conclude the re-summation
\begin{align}
&4\pi |\vec{x}|\widetilde{A}^{t}\left(\omega,\vec{x}\right)=\frac{q K}{ \mu{\pi}\beta_\infty^2}\sum_{\ell_{\gamma}=0}^{\infty}\sum_{m_{\gamma}=-\ell_{\gamma}}^{\ell_{\gamma}}Y_{\ell_{\gamma}}^{m_{\gamma}}\left(\theta,\phi\right)\int d\Omega'\,\frac{\sin\theta'\exp\left(-i\xi \beta_\infty\sin\theta'\cos\phi'\right)}{1-\beta_\infty\sin\theta'\sin\phi'}\nonumber\\
&\quad\times\Big\{ \sin\phi'K_{0}\left(\xi\left(1-\beta_\infty\sin\theta'\sin\phi'\right)\right)-i\cos\phi'K_{1}\left(\xi\left(1-\beta_\infty\sin\theta'\sin\phi'\right)\right)\Big\} \,\,Y_{\ell_{\gamma}}^{m_{\gamma}*}\left(\theta',\phi'\right)\nonumber\\
&\quad
=\frac{q K}{ \mu{\pi}\beta_\infty^2}\frac{\sin\theta\exp\left(-i\xi \beta_\infty\sin\theta\cos\phi\right)}{1-\beta_\infty\sin\theta\sin\phi}\Big\{\sin\phi\,K_{0}\left(\xi\left(1-\beta_\infty\sin\theta\sin\phi\right)\right)\nonumber\\
&\hspace{4cm}-i\cos\phi\, K_{1}\left(\xi\left(1-\beta_\infty\sin\theta\sin\phi\right)\right)\Big\}\,.
\end{align}
\subsection{Auxiliary Calculations} \label{App:nonrelAux}

\subsubsection{Proof of~\eqref{eq:Besseliden1}}

To prove~\eqref{eq:Besseliden1}, we change variables from $\left(k,s\right)$
to $\left(u,s\right)\equiv\left(k+s-\left(j+1\right),s\right)$, yielding

\begin{equation}
\eqref{eq:Besseliden1}=\sum_{u=-\left(j+1\right)}^{j+1}I_{u}\left(\frac{\omega L}{\mu\beta_\infty^2}\right)\sum_{s=u-\Delta \ell}^{j+1+u}\left(-1\right)^{s}\left(\begin{array}{c}
j-\Delta \ell+1\\
s
\end{array}\right)\left(\begin{array}{c}
j+\Delta \ell+1\\
j+1+u-s
\end{array}\right)\,\,.
\end{equation}
To facitilate the proof, it is useful to extend the summation limits
over $s$ to $\left(-\infty,\infty\right)$. This is possible because
of the second binomial coefficient which vanishes outside the correct
limits. Hence, we can write
\begin{equation}
\eqref{eq:Besseliden1}=\sum_{u=-\left(j+1\right)}^{j+1}I_{u}\left(\frac{\omega L}{\mu\beta_\infty^2}\right)\sum_{s=-\infty}^{\infty}\left(-1\right)^{s}\left(\begin{array}{c}
j-\Delta \ell+1\\
s
\end{array}\right)\left(\begin{array}{c}
j+\Delta \ell+1\\
j+1+u-s
\end{array}\right)\,\,.
\end{equation}
Applying the transformation $\left(u,s\right)\to\left(-u,j-\Delta \ell+1-s\right)$,
and using the property $I_{u}=I_{-u}$ yields $~\eqref{eq:Besseliden1}=\left(-1\right)^{j-\Delta \ell+1}~\eqref{eq:Besseliden1}$,
proving the identity.

\subsubsection{Proof of~\eqref{eq:matelemK0simp1}}

Starting from~\eqref{eq:matelemK0}, we pick up the real part and change variables from $\left(k,s\right)$ to $\left(u,s\right)\equiv\left(k+s-\left(j+1\right),s\right)$,
resulting in

\begin{align}\label{eq:proofofmatelemK0simp1:1}
&\left(\frac{L}{\mu \beta_\infty}\right)^{j}\left(-1\right)^{j+\Delta \ell}2^{-j-1}\frac{\omega L}{4\pi \mu\beta_\infty^2}\sum_{u=-\left(j+1\right)}^{j+1}\sum_{s=-\infty}^{\infty}\left(-1\right)^{s}\left(\begin{array}{c}
j-\Delta \ell+1\\
s
\end{array}\right)\left(\begin{array}{c}
j+\Delta \ell+1\\
j+1+u-s
\end{array}\right)\nonumber\\
&\times\sum_{m=0}^{1}\Bigg\{ \cos\left(\frac{\pi}{2}\left(j-\Delta \ell+1+m\right)\right)\left(2\pi\right)^{1-m}-\cos\left(\frac{\pi}{2}\left(j-\Delta \ell\right)\right)\nonumber\\
&\hspace{1.2cm}
\times\left(-\frac{4\mu\beta_\infty^2}{\omega L}\left(j+1+u-2s-\Delta \ell\right)\right)^{1-m}\Bigg\}
\frac{d^{1+m}}{d\nu^{1+m}}I_{\nu}\left(\frac{\omega L}{\mu\beta_\infty^2}\right)\Bigg|_{\nu=u},
\end{align}
where we extended the sum limits over $s$ to $\left(-\infty,\infty\right)$, which is possible thanks to the second binomial coefficient. It is instructive to now perform the transformation $\left(u,s\right)\to\left(-u,j-\Delta \ell+1-s\right)$,
yielding
\begin{align}
&\left(\frac{L}{\mu \beta_\infty}\right)^{j}\left(-1\right)^{j+\Delta \ell}2^{-j-1}\frac{\omega L}{4\pi \mu\beta_\infty^2}\sum_{u=-\left(j+1\right)}^{j+1}\sum_{s=-\infty}^{\infty}\left(-1\right)^{s}\left(\begin{array}{c}
j-\Delta \ell+1\\
s
\end{array}\right)\left(\begin{array}{c}
j+\Delta \ell+1\\
j+1+u-s
\end{array}\right)\nonumber\\
&\times\sum_{m=0}^{1}\left(-1\right)^{j-\Delta \ell+1}\Bigg\{ \cos\left(\frac{\pi}{2}\left(j-\Delta \ell+1+m\right)\right)\left(2\pi\right)^{1-m}+\left(-1\right)^{m}\cos\left(\frac{\pi}{2}\left(j-\Delta \ell\right)\right)\nonumber\\
&\hspace{1cm}\times\left(-\frac{4\mu\beta_\infty^2}{\omega L}\left(j+1+u-2s-\Delta \ell\right)\right)^{1-m}\Bigg\}\,\frac{d^{1+m}}{d\nu^{1+m}}I_{\nu}\left(\frac{\omega L}{\mu\beta_\infty}\right)\Bigg|_{\nu=-u}\,.\label{eq:proofofmatelemK0simp1:2}
\end{align}

Let us denote the RHS of~\eqref{eq:proofofmatelemK0simp1:1} by
$\mathcal{I}$. Using that for $m\in\left\{ 0,1\right\} $
\begin{align}
&\left.\frac{d^{1+m}}{d\nu^{1+m}}I_{\nu}\left(\frac{\omega L}{\mu\beta_\infty^2}\right)\right|_{\nu=-u}=\left(-1\right)^{1+m}\frac{d^{1+m}}{d\nu^{1+m}}I_{\nu}\left(\frac{\omega L}{\mu\beta_\infty^2}\right)\Bigg|_{\nu=u}\nonumber\\
&\hspace{1cm}+\left(-2\right)^{1+m}\cos\left(\pi u\right)\frac{d^{m}}{d\nu^{m}}K_{\nu}\left(\frac{\omega L}{\mu\beta_\infty^2}\right)\Bigg|_{\nu=u}\,,
\end{align}
we plug it in~\eqref{eq:proofofmatelemK0simp1:2}, yielding that $\mathcal{I}=-\mathcal{I}+\mathcal{I}_{2}$, where 
\begin{align}
&\mathcal{I}_{2}\equiv\left(\frac{L}{\mu \beta_\infty}\right)^{j}2^{-j}\frac{\omega L}{2\pi \mu\beta_\infty^2}\sum_{u=-\left(j+1\right)}^{j+1}\sum_{s=-\infty}^{\infty}\left(-1\right)^{s+u}\left(\begin{array}{c}
j-\Delta \ell+1\\
s
\end{array}\right)\left(\begin{array}{c}
j+\Delta \ell+1\\
j+1+u-s
\end{array}\right)\nonumber\\
&\quad\times\sum_{m=0}^{1}\Bigg\{ -\cos\left(\frac{\pi}{2}\left(j-\Delta \ell+1+m\right)\right)\left(-\pi\right)^{1-m}+\cos\left(\frac{\pi}{2}\left(j-\Delta \ell\right)\right)\nonumber\\
&\hspace{1cm}\times\left(-\frac{2\mu\beta_\infty^2}{\omega L}\left(j+1+u-2s-\Delta \ell\right)\right)^{1-m}\Bigg\}\,\frac{d^{m}}{d\nu^{m}}K_{\nu}\left(\frac{\omega L}{\mu\beta_\infty^2}\right)\Bigg|_{\nu=u}\,.
\end{align}
Hence,
\begin{align}
&\eqref{eq:matelemK0}=\frac{\mathcal{I}_{2}}{2}=\left(\frac{L}{\mu \beta_\infty}\right)^{j}2^{-j-1}\frac{\omega L}{2\pi \mu\beta_\infty^2}\sum_{u=-\left(j+1\right)}^{j+1}\sum_{s=-\infty}^{\infty}\left(-1\right)^{s+u}\left(\begin{array}{c}
j-\Delta \ell+1\\
s
\end{array}\right)\nonumber\\
&\quad\times\left(\begin{array}{c}
j+\Delta \ell+1\\
j+1+u-s
\end{array}\right)
\sum_{m=0}^{1}\Bigg\{ -\cos\left(\frac{\pi}{2}\left(j-\Delta \ell+1+m\right)\right)\left(-\pi\right)^{1-m}+\cos\left(\frac{\pi}{2}\left(j-\Delta \ell\right)\right)\nonumber\\
&\hspace{1cm}\times\left(-\frac{2\mu\beta_\infty^2}{\omega L}\left(j+1+u-2s-\Delta \ell\right)\right)^{1-m}\Bigg\}\, 
\frac{d^{m}}{d\nu^{m}}K_{\nu}\left(\frac{\omega L}{\mu\beta_\infty^2}\right)\Bigg|_{\nu=u}\,,
\end{align}
concluding the proof of~\eqref{eq:matelemK0simp1}.
\section{Exact Radial Hamilton-Jacobi Function in Schwarzschild}\label{sec:radialSch}
In this appendix we present the exact solution for $S^{Sch}_r(r)$, given by
\begin{eqnarray}\label{eq:Uschapp}
S^{Sch}_r(r)&=&\int_{r_{min}}^r\,\sqrt{U^{Sch}_r(r')}~dr'\,,\nonumber\\[5pt]
U^{Sch}_r(r)&=&\left(\frac{r^2}{\Delta}\right)^2\,\left[E^2-\frac{\Delta}{r^2}\left(\frac{L^2}{r^2}+\mu^2\right)\right]\equiv\frac{(\mu^2-E^2)}{r-2GM}(r-r_{b})(r-r_{min})(r_{*}-r)\,,\nonumber\\
\end{eqnarray}
where $r_{b}<r_{min}$. In the bound regime, $r_*$ is the maximal radius, whereas in the unbound regime it becomes negative. This integral can be performed exactly, though tediously. The result is
\begin{eqnarray}\label{eq:appeqSsch}
&&S^{Sch}_r(r)=\mu\sqrt{\frac{\left(E^2-\mu^2\right)
   (r-r_{b}) (r-r_{min}) (r-r_{*})}{r}}+\nonumber\\[5pt]
   &&\mu\sqrt{r_{min} \left(E^2-\mu^2\right)
   (r_{*}-r_{b})}\left\{\frac{r_{b} }{r_{*}-r_{b}}\left[F\left(\psi,k\right)-K\left(k\right)\right]+E\left(\psi,k\right)-E\left(k\right)+\right.\nonumber\\[5pt]
   &&
   \left.\frac{r_{b}^2 \left(3
   r_{min}^2+2 r_{min} r_{*}+3 r_{*}^2\right)}{8 r_{min}^2 r_{*} (r_{b}-r_{*})}\,\left[\Pi
   \left(\frac{r_{min}-r_{b}}{r_{min}};\psi,k\right)-\Pi
   \left(\frac{r_{min}-r_{b}}{r_{min}},k\right)\right]\right\}\,.~~~~\nonumber\\[5pt]
   &&-\frac{\mu}{\sqrt{r_{*}
   \left(E^2-\mu^2\right) (r_{b}-r_{min})}}\left\{2 J^2 F\left(\eta,q\right)-\frac{8 E^2 G^2 M^2 r_{min} }{r_{min}-2
   G M}\Pi \left(-\frac{2 G M
   (r_{*}-r_{min})}{(r_{min}-2 G M) r_{*}};\eta,q\right)\right.\nonumber\\[5pt]
   &&\left.-\mathcal{A}\,\Pi \left(\frac{r_{*}-r_{min}}{r_{*}};\eta,q\right)\right\}
\end{eqnarray}
where $F$ ($K$) is a complete (incomplete) integral of the first kind, $E$ is an elliptic integral of the second kind, and $\Pi$ is an elliptic integral of the third kind. Also,
\begin{eqnarray}
k^2&=&\frac{(r_{min}-r_{b}) r_{*}}{r_{min}
   (r_{*}-r_{b})}~~,~~\psi=\sin
   ^{-1}\left(\sqrt{\frac{(r-r_{b}) r_{min}}{r
   (r_{min}-r_{b})}}\right)\nonumber\\[5pt]
   q^2&=&-\frac{r_{b}
   (r_{*}-r_{min})}{(r_{min}-r_{b}) r_{*}}~~,~~\eta=\sin
   ^{-1}\left(\sqrt{\frac{(r-r_{min}) r_{*}}{r (r_{*}-r_{min})}}\right)\nonumber\\[5pt]
   \mathcal{A}&=&\frac{\left(32 E^2 G M r_{min} r_{*}-\left(E^2-\mu^2\right) \left(3 r_{b} (r_{min}-r_{*})^2-8 r_{min} r_{*}
   (r_{min}+r_{*})\right)\right)}{8 r_{*}}\,.\nonumber\\
\end{eqnarray}

\providecommand{\href}[2]{#2}\begingroup\raggedright\endgroup

\end{document}